\documentclass[12pt]{article}

\usepackage{amsmath}
\usepackage{appendix,epic,eepic,amsmath,amsthm,amssymb,epsfig,cite,indentfirst,oldgerm}
\usepackage{multicol,boxedminipage}

\renewcommand{\theequation}{\arabic{equation}}
\newcommand{\EQ}{\begin{equation}}
\newcommand{\EN}{\end{equation}}

\newcommand{\bear}{\begin{eqnarray}}
\newcommand{\ear}{\end{eqnarray}}
\newcommand{\bt} { \begin{tabular} }
\newcommand{\et}{ \end{tabular} }
\newcommand{\bc} { \begin{center} }
\newcommand{\ec}{ \end{center} }

\newcommand{\btb} { \begin{table} }
\newcommand{\etb}{ \end{table} }

\begin{document}

\topmargin 0pt
\oddsidemargin 5mm
\newcommand{\NP}[1]{Nucl.\ Phys.\ {\bf #1}}
\newcommand{\PL}[1]{Phys.\ Lett.\ {\bf #1}}
\newcommand{\NC}[1]{Nuovo Cimento {\bf #1}}
\newcommand{\CMP}[1]{Comm.\ Math.\ Phys.\ {\bf #1}}
\newcommand{\PR}[1]{Phys.\ Rev.\ {\bf #1}}
\newcommand{\PRL}[1]{Phys.\ Rev.\ Lett.\ {\bf #1}}
\newcommand{\MPL}[1]{Mod.\ Phys.\ Lett.\ {\bf #1}}
\newcommand{\JETP}[1]{Sov.\ Phys.\ JETP {\bf #1}}
\newcommand{\TMP}[1]{Teor.\ Mat.\ Fiz.\ {\bf #1}}

\renewcommand{\thefootnote}{\fnsymbol{footnote}}

\newpage
\setcounter{page}{0}
\begin{titlepage}
\begin{flushright}

\end{flushright}
\vspace{0.5cm}
\begin{center}
{\large The factorized F-matrices for arbitrary $U(1)^{(N-1)}$ integrable vertex models} \\
\vspace{1cm}
{\large M.J. Martins, R.A. Pimenta and M. Zuparic} \\
\vspace{0.15cm}
{\em Universidade Federal de S\~ao Carlos\\
Departamento de F\'{\i}sica \\
C.P. 676, 13565-905, S\~ao Carlos (SP), Brazil\\
E-mail Address: {\tt martins, pimenta, zuparic@df.ufscar.br}}\\
\vspace{0.35cm}
\end{center}
\vspace{0.5cm}

\begin{abstract}
We discuss the $F$-matrices associated to the $R$-matrix of
a general $N$-state vertex model whose statistical configurations
encode $N-1$ $U(1)$ symmetries.
The factorization condition is shown for arbitrary weights being based
only on the unitarity property and the Yang-Baxter relation 
satisfied by the $R$-matrix. 
Focusing on the $N=3$ case we are able to conjecture the structure
of some relevant twisted monodromy matrix elements for general weights.
We apply this result providing the algebraic expressions of the 
domain wall partition functions
built up in terms of the creation 
and annihilation monodromy fields. For $N=3$ we also exhibit a $R$-matrix 
whose weights lie
on a del Pezzo surface and have a rather general structure.
\end{abstract}

\vspace{.15cm} \centerline{}
\vspace{.1cm} \centerline{Keywords: N-state Vertex Model, F-Basis, Monodromy Matrix}
\vspace{.15cm} \centerline{November 2011}

\end{titlepage}


\pagestyle{empty}

\newpage

\pagestyle{plain}
\pagenumbering{arabic}

\renewcommand{\thefootnote}{\arabic{footnote}}
\newtheorem{proposition}{Proposition}
\newtheorem{pr}{Proposition}
\newtheorem{remark}{Remark}
\newtheorem{re}{Remark}
\newtheorem{theorem}{Theorem}
\newtheorem{theo}{Theorem}

\def\ll{\left\lgroup}
\def\rr{\right\rgroup}

\newtheorem{Theorem}{Theorem}[section]
\newtheorem{Corollary}[Theorem]{Corollary}
\newtheorem{Proposition}[Theorem]{Proposition}
\newtheorem{Conjecture}[Theorem]{Conjecture}
\newtheorem{Lemma}[Theorem]{Lemma}
\newtheorem{Example}[Theorem]{Example}
\newtheorem{Note}[Theorem]{Note}
\newtheorem{Definition}[Theorem]{Definition}

\section{Introduction}

The R-matrix plays a fundamental role  in the construction of  
two-dimensional integrable systems of statistical mechanics. This operator 
represented here by $R_{ab}(\xi_a,\xi_b)$ 
acts on the tensor product of two $N$-dimensional vectors spaces $V_a \otimes V_b$ depending
on the complex parameters $\xi_a$ and $\xi_b$. The $R$-matrix is required
to satisfy the Yang-Baxter equation \cite{BA},
\begin{equation}
R_{12}(\xi_1,\xi_2)R_{13}(\xi_1,\xi_3)R_{23}(\xi_2,\xi_3)=R_{23}(\xi_2,\xi_3)R_{13}(\xi_1,\xi_3)R_{12}(\xi_1,\xi_2).
\label{bigYB}
\end{equation}

The inverse of the $R$-matrix can be assured by imposing the unitarity condition,
\begin{equation}
R_{12}(\xi_1,\xi_2) R_{21}(\xi_2,\xi_1) = \mathcal{I}_1 \otimes \mathcal{I}_2.
\label{Biguni}
\end{equation}
where 
$\mathcal{I}_a$ is the $N \times N$ identity matrix in $V_a$. 

It turns out that the tensor products of 
the $R$-matrices called monodromy operators are central objects 
in the theory of 
integrable systems \cite{FA,KO}. In recent years, it has been
realized that such monodromy matrix can be decomposed in a suitable
way by means of auxiliary operators that 
have been denominated $F$-matrices \cite{MA1}. This concept
was originally introduced for the six-vertex model motivated
by the notion of twist deformations of quantum groups \cite{DRI}.
Lets us denote by 
$R^{\{\sigma \}}_{1\dots L}(\xi_1,\dots,\xi_L)$ the product 
of $R$-matrices associated to an arbitrary permutation $\sigma$
of the symmetry group $S_L$. 
The factorization
condition for the invertible $F$-matrices defined by any 
element $\sigma(1,\dots,L)=\{\sigma(1),\dots,\sigma(L)\} \in S_L$ reads as \cite{MA1,MA2}, 
\begin{equation}
F_{\sigma(1) \dots \sigma(L)}(\xi_{\sigma(1)}, \dots, \xi_{\sigma(L)}) 
R^{\{\sigma\}}_{1 \dots L}(\xi_1,\dots,\xi_L) =  
F_{1 \dots L}(\xi_1,\dots,\xi_L).
\label{definingFac}
\end{equation}
where the $F$-matrices
$F_{1 \dots L}(\xi_1,\dots,\xi_L)$ act on the tensor product 
spaces $V_1 \otimes \dots \otimes V_L$.

The $F$-matrices can be used as a natural basis to transform the  
monodromy matrix in a way that it becomes totally symmetric with
respect to a general permutation of the indices $1 \dots L$.
This similarity transformation for the six-vertex model
permits the development of an alternative approach \cite{MA2} 
to deal with the combinatorial problem underlying  the general
theory of the scalar product of Bethe states \cite{KO1,IK} and the
respective computation of domain wall partition functions \cite{KO2}.
In some respect this method paved the way for further progress on
the formulation of the correlation functions for the spin-$1/2$ 
Heisenberg chain \cite{MA3,MA4,GHO}. 
It also prompted the search for explicit forms 
of $F$-matrices associated to other
integrable vertex models such as for certain
generalizations of the six-vertex model \cite{TE,BOS} as well as
for multi-state vertex models whose weights are based on the
$SL(n|m)$ superalgebra \cite{BOS1,ZH1,ZH2}. We also remark that 
a diagrammatic interpretation
of the factorization equations for the symmetric six-vertex model
has been discussed in \cite{WEE}.

Recently, it has been argued that the 
existence of the $F$-matrices for an arbitrary 
six-vertex model can be pursued without the need
of using any explicit weights parameterization \cite{MZ}.
The structure of the $F$-matrices depends
basically on the statistical configurations 
encoded in the $R$-matrix and 
the verification of the
factorization condition (\ref{definingFac}) can be done
by using the algebraic weight constraints derived from the 
Yang-Baxter (\ref{bigYB}) and unitarity (\ref{Biguni}) 
relations. We think that this point of view of considering
the formulation of $F$-matrices should not be particular to the
six-vertex model.  In this paper we show that this framework
can indeed be generalized to tackle integrable N-state vertex models 
that are invariant by $N-1$ $U(1)$ symmetries. Recall that for $N=2$
one obtains the standard asymmetric six-vertex model. 
We apply 
the aforementioned construction to the next simplest case $N=3$, presenting
the algebraic expressions of relevant monodromy matrix
elements in the $F$-basis and the corresponding 
domain wall partition
functions. 

This paper is organized as follows. In the next Section we define the
$U(1)^{(N-1)}$ invariant vertex models and write the
algebraic relations (\ref{bigYB},\ref{Biguni}) for the
Boltzmann weights. These explicit relations are required to carry out
simplifications independent of parameterizations. 
We motivate our approach by exhibiting
a solution of the Yang-Baxter equation for $N=3$ which contains a number
of free parameters. In Section \ref{theojust} we discuss a procedure to build up
the $F$-matrices for an arbitrary $U(1)^{(N-1)}$
vertex model. 
It combines, in an effective way, past 
formulations of the $F$-matrices
for specific weights \cite{BOS1} with a 
recent construction devised for the six-vertex model \cite{MZ}. We use the
$F$-basis in Section  
\ref{OPOP} to provide the expressions of
certain relevant monodromy matrix elements for the $N=3$ vertex model  with
general weights. In Sections \ref{DOMAIN} and \ref{DOMAINMIX}  we apply these
results to exhibit the domain wall partition functions
associated to products of creation and annihilation
fields. Our conclusions are presented in 
Section \ref{CONCLU}. In Appendices A-C we summarize
technical details helpful for the understanding
of the main text.

\section{The  $U(1)^{N-1}$ vertex model }\label{VERTEX}
Consider a vertex model 
whose statistical 
configurations on both horizontal and vertical links of a square 
$L \times L$ lattice take 
values on $N$ possible states.  As usual the corresponding 
row-to-row transfer matrix can be written 
as the trace over an auxiliary space 
$\mathcal{A}_a$ of the  monodromy  operator
$\mathcal{T}_{a,1\dots L}(\mu )$. This matrix is constructed  
by the following ordered product
of $R$-matrices,
\begin{equation*}
\mathcal{T}_{a,1 \dots L}(\mu) = R_{aL}(\mu,\xi_L) 
R_{a(L-1)}(\mu,\xi_{L-1})\dots R_{a1}(\mu,\xi_1).
\end{equation*}

From the local Yang-Baxter equation (\ref{bigYB}) 
it follows that the monodromy matrix satisfies the following global 
intertwining relations called Yang-Baxter algebra,
\begin{equation}
R_{ab}(\mu,\nu) \mathcal{T}_{a,1 \dots L}(\mu)\mathcal{T}_{b,1 \dots L}(\nu) = 
\mathcal{T}_{b,1 \dots L}(\nu)\mathcal{T}_{a,1 \dots L}(\mu)R_{ab}(\mu,\nu).
\label{YBalg}\end{equation}

In this paper we shall be considering a 
particular family of $N$-state vertex models whose statistical configurations are invariant by
$N-1$ $U(1)$ symmetries. We shall denote the local generators of such $U(1)$ symmetries by 
$S^{(z,i)}_j$, $i=1,\dots,N-1$. This means that the corresponding $R$-matrix is constrained by the
commutation relations,
\begin{equation}\begin{array}{lll}
[R_{12}(\xi_1,\xi_2) , S^{(z,i)}_1 \otimes \mathcal{I}_2+ 
\mathcal{I}_1 \otimes S^{(z,i)}_2]=0 &, & i=1,\dots,N-1.
\end{array}\label{U(1)sym}
\end{equation}

In terms of the Wely $N \times N$ matrices, $e^{(\alpha \beta)}_j \in V_j$, the expressions for the $N-1$ azimuthal spin operators
$S^{(z,i)}_j$ are, 
\begin{equation}\begin{array}{lll}
S^{(z,i)}_j = e^{(ii)}_j-e^{((i+1)(i+1))}_j&, & i=1,\dots,N-1.
\end{array}\label{U(1)oper}
\end{equation}

Taking into account the property outlined in Eqs.(\ref{U(1)sym},\ref{U(1)oper}) 
one finds that the $R$-matrix $R_{12}(\xi_1 ,\xi_2)$ 
has $N (2N-1)$ non-vanishing weights. For $N=2$, the operator  $S^{(z,1)}_j$ 
reduces to the third component of the spin $1/2$ Pauli matrices and the corresponding 
statistical system is the fully asymmetrical 
six-vertex model. The possible statistical configurations for general
$N$ are given in terms of three distinct classes of weights
denoted here by $a_{i}(\xi_1,\xi_2)$, $b_{ij}(\xi_1,\xi_2)$
and $c_{ij}(\xi_1,\xi_2)$. In Figure \ref{weightsabc} we
show the respective vertex configurations on the square lattice.
\setlength{\unitlength}{3500sp}
\begin{figure}[ht]
\begin{center}
\begin{picture}(2000,1200)(2250,500)
\put(-200,1000){\makebox(0,0){$a_{i}(\xi_1,\xi_2)=$}}
\put(700,1000){\line( 1, 0){1000}}
\put(1200,500){\line( 0, 1){1000}}
\put(1300,550){\makebox(0,0){$i$}}
\put(1300,1450){\makebox(0,0){$i$}}
\put(750,1100){\makebox(0,0){$i$}}
\put(1650,1100){\makebox(0,0){$i$}}
\put(500,1000){\makebox(0,0){$\xi_1$}}
\put(1200,250){\makebox(0,0){$\xi_2$}}
\put(1800,975){\makebox(0,0){,}}

\put(2600,1000){\makebox(0,0){$b_{ij}(\xi_1,\xi_2)=$}}
\put(3500,1000){\line( 1, 0){1000}}
\put(4000,500){\line( 0, 1){1000}}
\put(4100,550){\makebox(0,0){$j$}}
\put(4100,1450){\makebox(0,0){$j$}}
\put(3550,1100){\makebox(0,0){$i$}}
\put(4450,1100){\makebox(0,0){$i$}}
\put(3300,1000){\makebox(0,0){$\xi_1$}}
\put(4000,250){\makebox(0,0){$\xi_2$}}
\put(4600,975){\makebox(0,0){,}}

\put(5400,1000){\makebox(0,0){$c_{ij}(\xi_1,\xi_2)=$}}
\put(6300,1000){\line( 1, 0){1000}}
\put(6800,500){\line( 0, 1){1000}}
\put(6900,550){\makebox(0,0){$j$}}
\put(6900,1450){\makebox(0,0){$i$}}
\put(6350,1100){\makebox(0,0){$i$}}
\put(7250,1100){\makebox(0,0){$j$}}
\put(6100,1000){\makebox(0,0){$\xi_1$}}
\put(6800,250){\makebox(0,0){$\xi_2$}}
\end{picture} \par
\end{center}
\caption{Elementary configuration of Boltzmann weights.}\label{weightsabc}
\end{figure}

From Figure \ref{weightsabc} we see that the expression of the associated $R$-matrix in terms of
of the Weyl matrices is given by,
\begin{equation}\begin{array}{lll}
R_{12}(\xi_1,\xi_2) &=& \displaystyle{\sum^N_{i=1}}a_i(\xi_1,\xi_2)e^{(ii)}_{1}\otimes e^{(ii)}_{2} + 
\sum^N_{\genfrac{}{}{0pt}{}{i,j = 1}{i\ne j}}b_{ij}(\xi_1,\xi_2)e^{(ii)}_{1}\otimes e^{(jj)}_{2}\\
&+&\displaystyle{\sum^N_{\genfrac{}{}{0pt}{}{i,j = 1}{i\ne j}}}c_{ij}(\xi_1,\xi_2)e^{(ij)}_{1}\otimes e^{(ji)}_{2}.
\label{RR}\end{array}\end{equation}

We recall that the integrable vertex model given by Eq.(\ref{RR}) with 
parameterized weights has been 
considered in the literature for some time \cite{DEV,PK}. 
We stress however that 
the main results of this work will be established without the need of any 
specific parameterization of the Boltzmann weights. We shall rely solely on the algebraic 
relations for the weights $a_i(\xi_1,\xi_2)$, $b_{ij}(\xi_1,\xi_2)$ and $c_{ij}(\xi_1,\xi_2)$ 
coming from the Yang-Baxter and unitary properties. For that reason we need to quote them 
here explicitly.
By substituting the expression for the $R$-matrix (\ref{RR}) into Eq.(\ref{Biguni}) we find 
that the Boltzmann weights are required to satisfy the three distinct types of relations,
\begin{eqnarray}
a_i (\xi_1,\xi_2)a_i (\xi_2,\xi_1) = 1 &, & i = 1, \dots N \label{biga} \\
b_{ij} (\xi_1,\xi_2)b_{ji} (\xi_2,\xi_1) + c_{ij} (\xi_1,\xi_2)c_{ij} (\xi_2,\xi_1) =1 &,&  i\ne j=1,\dots,N \label{bigb}\\
b_{ij} (\xi_1,\xi_2)c_{ji} (\xi_2,\xi_1) + c_{ij} (\xi_1,\xi_2)b_{ij} (\xi_2,\xi_1) =0 &,&  i\ne j=1,\dots,N \label{bigc}.
\end{eqnarray}

The Yang-Baxter equation (\ref{bigYB}) generates extra constraints on the Boltzmann weights depending 
now on three independent rapidities. By substituting Eq.(\ref{RR}) into Eq.(\ref{bigYB}) one finds that 
the aforementioned functional relations are much more involved. They can however be written in  a compact
way by the following expressions, 
\begin{eqnarray}
c_{ij}(\xi_1,\xi_2)  c_{ji}(\xi_1,\xi_3)  c_{ij}(\xi_2,\xi_3) =  
c_{ji}(\xi_1,\xi_2)  c_{ij}(\xi_1,\xi_3)  c_{ji}(\xi_2,\xi_3) & i \ne j \label{YB1}\\
b_{ij}(\xi_1,\xi_2)b_{ik}(\xi_1,\xi_3) = b_{ik}(\xi_1,\xi_2)b_{ij}(\xi_1,\xi_3) & i \ne j \ne k \label{YB2} \\
b_{jk}(\xi_1,\xi_3)b_{ik}(\xi_2,\xi_3) = b_{ik}(\xi_1,\xi_3)b_{jk}(\xi_2,\xi_3) & i \ne j \ne k \label{YB3} \\
 b_{ij}(\xi_1,\xi_2)  a_{i}(\xi_1,\xi_3)  c_{ij}(\xi_2,\xi_3) +  c_{ji}(\xi_1,\xi_2)  c_{ij}(\xi_1,\xi_3)  b_{ij}(\xi_2,\xi_3)\nonumber\\
= a_{i}(\xi_1,\xi_2)  b_{ij}(\xi_1,\xi_3)  c_{ij}(\xi_2,\xi_3) & i \ne j \label{YB4}\\
 b_{ij}(\xi_1,\xi_2)  a_{i}(\xi_1,\xi_3)  c_{ji}(\xi_2,\xi_3) +  c_{ij}(\xi_1,\xi_2)  c_{ji}(\xi_1,\xi_3)  b_{ij}(\xi_2,\xi_3)\nonumber\\
= a_{i}(\xi_1,\xi_2)  b_{ij}(\xi_1,\xi_3)  c_{ji}(\xi_2,\xi_3) & i \ne j \label{YB5}\\
 b_{ji}(\xi_1,\xi_2)  c_{ij}(\xi_1,\xi_3)  b_{ij}(\xi_2,\xi_3) +  c_{ij}(\xi_1,\xi_2)  a_{i}(\xi_1,\xi_3)  c_{ij}(\xi_2,\xi_3)\nonumber\\
= a_{i}(\xi_1,\xi_2)  c_{ij}(\xi_1,\xi_3)  a_{i}(\xi_2,\xi_3) & i \ne j \label{YB6}
\end{eqnarray}
\begin{eqnarray}
 b_{ij}(\xi_1,\xi_2)  c_{ij}(\xi_1,\xi_3)  b_{ji}(\xi_2,\xi_3) +  c_{ij}(\xi_1,\xi_2)  a_{j}(\xi_1,\xi_3)  c_{ij}(\xi_2,\xi_3)\nonumber\\
= a_{j}(\xi_1,\xi_2)  c_{ij}(\xi_1,\xi_3)  a_{j}(\xi_2,\xi_3) & i \ne j \label{YB7}\\
 c_{ij}(\xi_1,\xi_2)  a_{i}(\xi_1,\xi_3)  b_{ji}(\xi_2,\xi_3) +  b_{ji}(\xi_1,\xi_2)  c_{ij}(\xi_1,\xi_3)  c_{ji}(\xi_2,\xi_3)\nonumber\\
= c_{ij}(\xi_1,\xi_2)  b_{ji}(\xi_1,\xi_3)  a_{i}(\xi_2,\xi_3) & i \ne j \label{YB8}\\
 c_{ij}(\xi_1,\xi_2)  a_{j}(\xi_1,\xi_3)  b_{ij}(\xi_2,\xi_3) +  b_{ij}(\xi_1,\xi_2)  c_{ij}(\xi_1,\xi_3)  c_{ji}(\xi_2,\xi_3)\nonumber\\
= c_{ij}(\xi_1,\xi_2)  b_{ij}(\xi_1,\xi_3)  a_{j}(\xi_2,\xi_3) & i \ne j \label{YB9}\\
c_{ij}(\xi_1,\xi_2)  c_{jk}(\xi_1,\xi_3)  b_{ij}(\xi_2,\xi_3) +  b_{ij}(\xi_1,\xi_2)  c_{ik}(\xi_1,\xi_3)  c_{ji}(\xi_2,\xi_3)\nonumber\\
= c_{ik}(\xi_1,\xi_2)  b_{ij}(\xi_1,\xi_3)  c_{jk}(\xi_2,\xi_3) & i \ne j \ne k \label{YB10}\\
 c_{kj}(\xi_1,\xi_2)  c_{ik}(\xi_1,\xi_3)  b_{jk}(\xi_2,\xi_3) +  b_{jk}(\xi_1,\xi_2)  c_{ij}(\xi_1,\xi_3)  c_{jk}(\xi_2,\xi_3)\nonumber\\
= c_{ij}(\xi_1,\xi_2)  b_{jk}(\xi_1,\xi_3)  c_{ik}(\xi_2,\xi_3) & i \ne j \ne k \label{YB11}\\
 b_{ij}(\xi_1,\xi_2)  c_{ik}(\xi_1,\xi_3)  b_{ji}(\xi_2,\xi_3) +  c_{ij}(\xi_1,\xi_2)  c_{jk}(\xi_1,\xi_3)  c_{ij}(\xi_2,\xi_3)\nonumber\\
=b_{kj}(\xi_1,\xi_2)  c_{ik}(\xi_1,\xi_3)  b_{jk}(\xi_2,\xi_3)+ c_{jk}(\xi_1,\xi_2)  
c_{ij}(\xi_1,\xi_3)  c_{jk}(\xi_2,\xi_3) & i \ne j \ne k \label{YB12}
\end{eqnarray}
where the indices $i,j,k=1,\dots,N$.

We would like to conclude this section by remarking that  
the Yang-Baxter Eqs.(\ref{YB1}-\ref{YB12}) 
hide integrable models whose weights structure are indeed rather general.
Recall here that the case $N=2$, the asymmetric six-vertex model, has already
been detailed in the literature \cite{BA,BA1}. Therefore, we shall concentrate
on the next simplest system which is the $N=3$ fifteen-vertex
model. The key ingredient to solving integrable models whilst  keeping their weights as
arbitrary as possible is to uncover the main algebraic varieties constraining
the respective Boltzmann weights \cite{BA}. In order to tackle this problem
for the case $N=3$ we have adapted a method, first developed in \cite{PIMA}, 
which handles a large number of functional equations 
associated with three-state vertex
models. The technical details of this analysis have been summarized
in Appendix A and in what follows we will present 
only the main results. It turns out that the underlying algebraic variety
of one possible solution is a homogeneous hypersurface given by the equation,
\begin{equation}
\left(\frac{\Delta_{1} \Delta_{2}-1}{\Delta_1^2}\right)
a(\xi_i)^2b(\xi_i)-\Delta_{2}a(\xi_i)b(\xi_i)\bar{b}(\xi_i)+b(\xi_i) \bar{b}(\xi_i)^2-
\bar{b}(\xi_i)c(\xi_i)\bar{c}(\xi_i)=0
\label{HYPER}
\end{equation}
where $\Delta_1,\Delta_2$ are free constants while $a(\xi_i)$, $b(\xi_i)$, 
$\bar{b}(\xi_i)$, $c(\xi_i)$ and $\bar{c}(\xi_i)$ are arbitrary variables
depending on the spectral parameters.

One possible way to parameterize the hypersurface (\ref{HYPER}) is first to
consider its intersection with the  
hyperplane $c(\xi_i)-\bar{c}(\xi_i)=0$. As a result we obtain 
an algebraic variety in the class of the cubic
del Pezzo surfaces which can be parameterized in
terms of rational functions \cite{HART}. Following an algorithm devised in \cite{SCH}    
we conclude the rational map is attained by just solving Eq.(\ref{HYPER}) for the
linear variable $b(\xi_i)$ in terms of the remaining parameters
$a(\xi_i),\bar{b}(\xi_i)$ and $c(\xi_i)$. Clearly,
the same type of 
procedure also works for the general manifold (\ref{HYPER}) and therefore
the $R$-matrix depends at least on the three free variables
$a(\xi_i),\bar{b}(\xi_i)$ and $c(\xi_i)$. Taking into account
the results of Appendix A we find that the $R$-matrix elements are given
by, 
\begin{equation}
\frac{a_1(\xi_1,\xi_2)}{c_{12}(\xi_1,\xi_2)}=\frac{c(\xi_2)}{c(\xi_1)} \frac{\left[
(\Delta_{1}\Delta_{2}-1) a(\xi_1) a(\xi_2)-\Delta_{1}^2 \bar{b}(\xi_2)(\Delta_{2} a(\xi_1) -
\bar{b}(\xi_1) )\right]}{ \left[a(\xi_2)-\Delta_{1} \bar{b}(\xi_2)\right] \left[
(\Delta_{1}\Delta_{2}-1) a(\xi_2)-\Delta_{1} \bar{b}(\xi_2)\right]}
\end{equation}
\begin{eqnarray}
\frac{b_{12}(\xi_1,\xi_2)}{c_{12}(\xi_1,\xi_2)}&=&\Delta_{1}^2
(-1+\Delta_{1} \Delta_{2}) c(\xi_2)\bar{c}(\xi_1)\left[a(\xi_2) \bar{b}(\xi_1)-a(\xi_1) \bar{b}(\xi_2)\right]\nonumber\\
&&\times\frac{\left[-a(\xi_2)+\Delta_{1} \bar{b}(\xi_2)\right]^{-1}\left[(\Delta_{1}\Delta_{2}-1) a(\xi_2)-\Delta_{1} \bar{b}(\xi_2)\right]^{-1}}
{\left[-a(\xi_1)+\Delta_{1} \bar{b}(\xi_1)\right]
\left[(\Delta_{1}\Delta_{2}-1) a(\xi_1)-\Delta_{1} \bar{b}(\xi_1)\right]}
\end{eqnarray}
\begin{eqnarray}
\frac{b_{13}(\xi_1,\xi_2)}{c_{12}(\xi_1,\xi_2)}&=&\frac{\Delta_{1}^2
(\Delta_{1} \Delta_{2}-1)  c(\xi_2)h_1(\xi_1)h_2(\xi_1)}{\delta_{2} c(\xi_1)\bar{c}(\xi_1)}
\frac{\left[a(\xi_2) \bar{b}(\xi_1)-a(\xi_1)\bar{b}(\xi_2)\right]}
{\left[a(\xi_2)-\Delta_{1}\bar{b}(\xi_2)\right]}
\nonumber\times\\
&&\left[(\Delta_{1} \Delta_{2}-1)a(\xi_1)-\Delta_{1}
\bar{b}(\xi_1)\right]^{-1} \left[(\Delta_{1} \Delta_{2}-1)a(\xi_2)-\Delta_{1} \bar{b}(\xi_2)\right]^{-1}
\end{eqnarray}
\begin{equation}
\frac{b_{21}(\xi_1,\xi_2)}{c_{12}(\xi_1,\xi_2)}=\frac{a(\xi_2) \bar{b}(\xi_1)-a(\xi_1)
\bar{b}(\xi_2)}{c(\xi_1) \bar{c}(\xi_2)}
\end{equation}
\begin{equation}
\frac{a_{2}(\xi_1,\xi_2)}{c_{12}(\xi_1,\xi_2)}=
\frac{\bar{c}(\xi_1)}{\bar{c}(\xi_2)}\frac{\left[(\Delta_{1}
\Delta_{2} -1)a(\xi_1) a(\xi_2)-\Delta_{1}^2 \bar{b}(\xi_2)(\Delta_{2} a(\xi_1) -
\bar{b}(\xi_1) )\right] }
{\left[a(\xi_1)-\Delta_{1} \bar{b}(\xi_1)\right] \left[-a(\xi_1)+\Delta_{1}
\Delta_{2} a(\xi_1)-\Delta_{1} \bar{b}(\xi_1)\right] }
\end{equation}
\begin{equation}
\frac{b_{23}(\xi_1,\xi_2)}{c_{12}(\xi_1,\xi_2)}=
\frac{(\Delta_{1} \Delta_{2}-1) h_1(\xi_1)h_2(\xi_1)}{\delta_{1} c(\xi_1)\bar{c}(\xi_1) \bar{c}(\xi_2)}
\frac{
\left[a(\xi_2) \bar{b}(\xi_1)-a(\xi_1) \bar{b}(\xi_2)\right] }{
\left[(\Delta_{1} \Delta_{2}-1)a(\xi_1)-\Delta_{1} \bar{b}(\xi_1)\right] }
\end{equation}
\begin{equation}
\frac{c_{13}(\xi_1,\xi_2)}{c_{12}(\xi_1,\xi_2)}=\frac{ c(\xi_2) \bar{c}(\xi_2) 
h_1(\xi_1)}{ c(\xi_1) \bar{c}(\xi_1)h_1(\xi_2)}\frac{\left[a(\xi_1)-\Delta_{1} \bar{b}(\xi_1)\right]}
{\left[a(\xi_2)-\Delta_{1} \bar{b}(\xi_2)\right]}
\end{equation}
\begin{equation}
\frac{c_{21}(\xi_1,\xi_2)}{c_{12}(\xi_1,\xi_2)}=\frac{c(\xi_2)
\bar{c}(\xi_1)}{c(\xi_1) \bar{c}(\xi_2)},~~~
\frac{c_{23}(\xi_1,\xi_2)}{c_{12}(\xi_1,\xi_2)}=
\frac{c(\xi_2)h_1(\xi_1)
}{c(\xi_1)h_1(\xi_2) },~~~
\frac{c_{31}(\xi_1,\xi_2)}{c_{12}(\xi_1,\xi_2)}=\frac{c(\xi_2) h_2(\xi_1)}{c(\xi_1)
h_2(\xi_2)}
\end{equation}
\begin{equation}
\frac{b_{31}(\xi_1,\xi_2)}{c_{12}(\xi_1,\xi_2)}=
\frac{\delta_{2} c(\xi_2)\bar{c}(\xi_2)}{h_1(\xi_2) c(\xi_1)h_2(\xi_2)}
\frac{\left[a(\xi_2) \bar{b}(\xi_1)-a(\xi_1) \bar{b}(\xi_2)\right]}
{\left[a(\xi_2)-\Delta_{1} \bar{b}(\xi_2)\right]}
\end{equation}
\begin{equation}
\frac{c_{32}(\xi_1,\xi_2)}{c_{12}(\xi_1,\xi_2)}=
\frac{h_2(\xi_1)}{h_2(\xi_2)}
\frac{\left[(\Delta_{1}\Delta_{2}-1) a(\xi_2)-\Delta_{1} \bar{b}(\xi_2)\right] }
{\left[(\Delta_{1} \Delta_{2}-1)a(\xi_1)-\Delta_{1} \bar{b}(\xi_1)\right] }
\end{equation}
\begin{equation}
\frac{b_{32}(\xi_1,\xi_2)}{c_{12}(\xi_1,\xi_2)}=
\frac{\Delta_{1}^2 \delta_{1} c(\xi_2)\bar{c}(\xi_1) \bar{c}(\xi_2)}{h_1(\xi_2)h_2(\xi_2)}
\frac{
\left[a(\xi_2) \bar{b}(\xi_1)-a(\xi_1) \bar{b}(\xi_2)\right] }
{
\left[(\Delta_{1} \Delta_{2}-1)a(\xi_1)-\Delta_{1} \bar{b}(\xi_1)\right] \left[a(\xi_1)-\Delta_{1}
\bar{b}(\xi_1)\right] \left[a(\xi_2)-\Delta_{1} \bar{b}(\xi_2)\right] }
\end{equation}
\begin{equation}
\frac{a_{3}(\xi_1,\xi_2)}{c_{12}(\xi_1,\xi_2)}=
\frac{h_1(\xi_1) c(\xi_2)\bar{c}(\xi_2) h_2(\xi_1)}{h_1(\xi_2)c(\xi_1)\bar{c}(\xi_1) h_2(\xi_2)}
\frac{
\left[(\Delta_{1} \Delta_{2} -1)a(\xi_1) a(\xi_2)-\Delta_{1}^2\bar{b}(\xi_2)( \Delta_{2}
a(\xi_1)-\bar{b}(\xi_1) )\right] }{
 \left[(\Delta_{1} \Delta_{2}-1)a(\xi_1)-\Delta_{1} \bar{b}(\xi_1)\right] \left[a(\xi_2)-\Delta_{1}
\bar{b}(\xi_2)\right] }
\end{equation}
where 
$c_{12}(\xi_1,\xi_2)$ is an overall normalization, 
$\delta_1$ and $\delta_2$ are extra free constants 
while $h_1(\xi_i)$ and $h_2(\xi_i)$
are free variables. The latter freedom is related to the fact
that the Yang-Baxter equation is preserved under 
local transformations associated to the two $U(1)$ symmetries.  
 
The structure of the $R$-matrix is certainly more general than that of the
the standard $U(1) \otimes U(1)$ invariant vertex models 
containing only one free spectral parameter \cite{DEV,PK}. This rather particular
parameterization for the completely free variables  
$a(\xi_i),\bar{b}(\xi_i)$ and $c(\xi_i)$ is discussed at the end of
Appendix A.
This fact 
highlights the importance to search for explicit results  
in integrable models that are independent
of specific parameterization of the Boltzmann weights.

\section{Factorized F-matrices}\label{theojust}

We begin by providing some basic definitions and notation that are going to be used in
the text. We first recall that 
an element $\sigma$ of the symmetric group $S_L$ can be generated in terms of adjacent permutations. In other
words one can write, 
\begin{equation}
\sigma = \sigma_{\alpha_p (\alpha_p+1)}\dots \sigma_{\alpha_2 (\alpha_2+1)}\sigma_{\alpha_1 (\alpha_1+1)},
\label{sigdecomp}
\end{equation}
where
$\sigma_{\alpha(\alpha+1)}$ denotes the permutation of the indices $\alpha$ and $\alpha+1$, i.e.
$\sigma_{\alpha(\alpha+1)}\left(1,\dots,\alpha,\alpha+1,\dots, L\right) = 
\left(1,\dots, \alpha+1, \alpha,\dots, L\right)$. We shall refer to Eq. (\ref{sigdecomp}) as the minimum decomposition of $\sigma$.

In order to avoid cumbersome notation by the presence of the inhomogeneities $\xi_i$ 
in the $F$-matrices 
we shall omit them when this does not generates confusion. In general, we shall identify 
a given general element $X_{a,\sigma(1),\dots, \sigma(L)} (\mu|\xi_{\sigma(1)}, \dots, \xi_{\sigma(L)})
\in \textrm{End}(\mathcal{A}_a \otimes V_{\sigma(1)} \otimes \dots \otimes V_{\sigma(L)})$ 
to $X_{a,\sigma(1\dots L)} (\mu)$. Taking this notation into account the factorization
condition (\ref{definingFac}) is rewritten as,
\begin{equation}
F_{\sigma(1\dots L)} R^{\{\sigma\}}_{1\dots L} =  F_{1\dots L}.
\label{defining}
\end{equation}

The tensor product of $R$-matrices 
$R^{\{\sigma\}}_{1\dots L}$ entering Eq.(\ref{defining}) is defined in terms of product of auxiliary operators
by the expression,
\begin{equation}
R^{\{\sigma\}}_{1\dots L} = P^{\left\{\sigma\right\}}_{1\dots L} \hat{R}^{\{\sigma^{-1}\}}_{1\dots L}.
\label{decom1}\end{equation}

Through considering the minimum decomposition of $\sigma$ in terms of adjacent permutations, 
the auxiliary operators $P^{\left\{\sigma\right\}}_{1\dots L}$ and $\hat{R}^{\{\sigma^{-1}\}}_{1\dots L}$ can be written
in the following way,
\begin{equation}\begin{array}{lll}
P^{\left\{\sigma\right\}}_{1\dots L} &=& P^{\left\{\sigma_{\alpha_p (\alpha_p+1)}\right\}}_{1\dots L} \dots  P^{\left\{\sigma_{\alpha_2 (\alpha_2 +1)}\right\}}_{1\dots L} P^{\left\{\sigma_{\alpha_1 (\alpha_1+1)}\right\}}_{1\dots L}\\
\hat{R}^{\{\sigma^{-1}\}}_{1\dots L} &=& \hat{R}^{\left\{\sigma_{\alpha_1 (\alpha_1+1)}\right\}}_{1\dots L} \hat{R}^{\left\{\sigma_{\alpha_2 (\alpha_2 +1)}\right\}}_{1\dots L} \dots \hat{R}^{\left\{\sigma_{\alpha_p (\alpha_p+1)}\right\}}_{1\dots L},
\end{array}\label{decom2}\end{equation}
where
$P^{\left\{\sigma_{\alpha (\alpha+1)}\right\}}_{1\dots L} = P_{\alpha (\alpha+1)}$ is the standard permutator while
$\hat{R}^{\left\{\sigma_{\alpha (\alpha+1)}\right\}}_{1\dots L} = 
P_{\alpha (\alpha+1)} R_{\alpha (\alpha+1)}$.

We now list a number of properties that are necessary in our analysis of the factorization condition (\ref{defining}).
We start by mentioning that 
for any given operator $X_{1 \dots L} \in \textrm{End}(V_1 \otimes \dots \otimes V_L)$ we have the following useful relation,
\begin{equation}
X_{\sigma(1\dots L)} = P^{\left\{\sigma\right\}}_{1\dots L} X_{1\dots L} P^{\left\{\sigma^{-1}\right\}}_{1\dots L}.
\label{useful}\end{equation}

Next, given the decomposition laws of the auxiliary 
operators (\ref{decom2}) one can show that for general $\{\sigma,\tau\} \in S_L$ the 
tensor product of permuted $R$-matrices obeys the following decomposition identity,
\begin{equation}
 R^{\{ \sigma \tau \} }_{1 \dots L} = R^{\{ \tau \}}_{\sigma(1\dots L)} R^{\{ \sigma \}}_{1 \dots L}.
\label{combineR}\end{equation}

In order to verify Eq.(\ref{combineR}) we consider the expression $R^{\{ \sigma \tau \} }_{1 \dots L}$ 
and perform the following operations,
\bear
\underbrace{ R^{\{ \sigma \tau \} }_{1 \dots L}}_{\textrm{apply Eq.(\ref{decom1})}} &  =& 
P^{\{ \sigma \tau \} }_{1 \dots L}\hat{R}^{\{ (\sigma \tau)^{-1} \} }_{1 \dots L}
= P^{\{ \sigma  \} }_{1 \dots L} \underbrace{P^{\{  \tau \} }_{1 \dots L} 
\hat{R}^{\{  \tau^{-1} \} }_{1 \dots L} }_{\textrm{apply Eq.(\ref{decom1})}}\hat{R}^{\{ \sigma^{-1} \} }_{1 \dots L}
\nonumber \\
& =& \underbrace{P^{\{ \sigma  \} }_{1 \dots L} R^{\{  \tau \} }_{1 \dots L}  
P^{\{ \sigma^{-1}  \} }_{1 \dots L}}_{\textrm{apply Eq.(\ref{useful})}}  
\underbrace{P^{\{ \sigma  \} }_{1 \dots L}\hat{R}^{\{ \sigma^{-1} \} }_{1 \dots L}}_{\textrm{apply Eq.(\ref{decom1})}} 
= R^{\{ \tau \}}_{\sigma(1\dots L)} R^{\{ \sigma \}}_{1 \dots L}. 
\ear

We end by mentioning  another standard identity regarding  the action of the operator
$ R^{\{ \sigma \}}_{1 \dots L}$ 
on the monodromy matrix, namely
\begin{equation}
R^{\{ \sigma \}}_{1 \dots L} \mathcal{T}_{a,1\dots L}(\mu) = \mathcal{T}_{a,\sigma(1\dots L)}(\mu) R^{\{ \sigma \}}_{1 \dots L}.
\label{RonT}\end{equation}

\subsection{The $L=2$ example} 
We shall start the explicit construction of the $F$-matrices for arbitrary weights 
$a_i(\xi_1,\xi_2)$, $b_{ij}(\xi_1,\xi_2)$ and $c_{ij}(\xi_1,\xi_2)$. Besides the factorization relation (\ref{defining}) we also
require that the $F$-matrices are lower triangular and invertible \cite{MA1}.  In this sense it is instructive to begin with 
the simplest case $L=2$. In this situation 
the only non-trivial permutation is the adjacent permutation $\sigma_{12}$ and Eq.(\ref{defining}) becomes,
\begin{equation}
 F_{21}  R_{12}  = F_{12}.
\label{Ftwo}
\end{equation}

We shall now explicitly show that the solution the expression (\ref{Ftwo}) is given by,
\begin{equation}
F_{12} = \mathcal{N}_{12}\mathcal{F}_{12} 
\label{N=2a}
\end{equation}
where $\mathcal{N}_{12}$ is the diagonal matrix,
\begin{equation}
\mathcal{N}_{12} = \mathcal{I}_{1} \otimes \mathcal{I}_{2} 
-\sum^N_{i=1}e^{(ii)}_1\otimes e^{(ii)}_2 +\sum^N_{i=1}\sqrt{a_i(\xi_1,\xi_2)} e^{(ii)}_1\otimes e^{(ii)}_2,
\label{curlyN}\end{equation}
and $\mathcal{F}_{12}$ is the given by the following operator,
\begin{equation}
\mathcal{F}_{12} = \sum_{1 \le \alpha_1 \le \alpha_2 \le N}e^{(\alpha_1 \alpha_1)}_1\otimes e^{(\alpha_2 \alpha_2)}_2 
\mathcal{I}_{1}\otimes \mathcal{I}_{2} + \sum_{1 \le \alpha_2 < \alpha_1 \le N}e^{(\alpha_1 \alpha_1)}_1\otimes e^{(\alpha_2 \alpha_2)}_2 R_{12}.
\end{equation}

Substituting Eq.(\ref{N=2a}) into the factorization condition and applying Eq.(\ref{biga}) we obtain,
\begin{equation}
\mathcal{F}_{21} R_{12} = \mathcal{R}_{12}\mathcal{F}_{12},
\end{equation}
where the twisted $R$-matrix, $\mathcal{R}_{12}$, is the diagonal matrix,
\begin{equation}
\mathcal{R}_{12} = \mathcal{I}_{1} \otimes \mathcal{I}_{2}
-\sum^N_{i=1}e^{(ii)}_1\otimes e^{(ii)}_2 +\sum^N_{i=1}a_i(\xi_1,\xi_2) e^{(ii)}_1\otimes e^{(ii)}_2. 
\label{twistR2}\end{equation}
Hence, applying unitarity (\ref{Biguni}) the $L=2$ factorization condition becomes,
\begin{equation}\begin{array}{ll}
& \sum_{1 \le \alpha_2 \le \alpha_1 \le N}e^{(\alpha_1 \alpha_1)}_1\otimes 
e^{(\alpha_2 \alpha_2)}_2 {R}_{12} + 
\sum_{1 \le \alpha_1 < \alpha_2 \le N}e^{(\alpha_1 \alpha_1)}_1\otimes 
e^{(\alpha_2 \alpha_2)}_2 \mathcal{I}_1\otimes \mathcal{I}_{2}\\
=& \sum_{1 \le \alpha_1 \le \alpha_2 \le N} \mathcal{R}_{12}e^{(\alpha_1 \alpha_1)}_1\otimes e^{(\alpha_2 \alpha_2)}_2  + \sum_{1 \le \alpha_2 < \alpha_1 \le N}\mathcal{R}_{12} e^{(\alpha_1 \alpha_1)}_1\otimes e^{(\alpha_2 \alpha_2)}_2 R_{12}. 
\end{array}\label{L=2example}\end{equation}
Using the following relation regarding the action of the $\mathcal{R}$-matrix on elementary matrices,
\begin{equation}
\mathcal{R}_{12}e^{(i i)}_1\otimes e^{(jj)}_2 = 
\left\{ \begin{array}{cll} a_i(\xi_1,\xi_2)e^{(i i)}_1\otimes e^{(jj)}_2 & \textrm{for} & i = j\\
                                                         e^{(i i)}_1\otimes e^{(jj)}_2 & \textrm{for} & i \ne j
                                                        \end{array} \right. ,
\label{Rone}\end{equation}
we can immediately see that the $\alpha_1 \ne \alpha_2$ components of 
the summations in Eq.(\ref{L=2example}) cancel. Hence we are left with the expression,
\begin{equation}
 \sum^N_{\alpha=1}e^{(\alpha \alpha)}_1\otimes e^{(\alpha \alpha)}_2 {R}_{12} = \sum^N_{\alpha=1}\mathcal{R}_{12}e^{(\alpha \alpha)}_1\otimes e^{(\alpha \alpha)}_2 ,
\end{equation}
which is true by inspection.

We end by commenting that the form of $\mathcal{F}_{12}$ coincides 
exactly with the solution originally proposed in \cite{BOS1} 
for the rational $SU(N)$ vertex models. This observation provides us a hint on how to
proceed for arbitrary $L$. 

\subsection{The general L case} 
For general $L$ we provide the following ansatz for the form of $F_{1\dots L}$,
\begin{equation}
F_{1\dots L}  = \mathcal{N}_{1\dots L}\mathcal{F}_{1\dots L},
\label{ansatz}\end{equation}
where the definition of $\mathcal{N}_{1\dots L}$ is given by the product of \textit{partial} $\mathcal{N}$-matrices,
\begin{equation}
\mathcal{N}_{1\dots L} = \mathcal{N}_{2\dots L}\mathcal{N}_{1,2\dots L}
= \mathcal{N}_{(L-1)L}\mathcal{N}_{L-2,(L-1)L} \dots \mathcal{N}_{1,2\dots L},
\label{BIGN}
\end{equation}
such that the partial $\mathcal{N}$-matrices $\mathcal{N}_{i,(i+1)\dots L}$, are given by,
\begin{equation}
\mathcal{N}_{i,(i+1)\dots L} = \mathcal{N}_{i L} \mathcal{N}_{i (L-1)} \dots \mathcal{N}_{i(i+1)}.
\label{BIGN2}\end{equation}
As with the $L=2$ case, the terms $\mathcal{F}_{1\dots L}$ coincide with the form of the solution given in \cite{BOS1}:
\begin{equation}
 \mathcal{F}_{1\dots L} = \sum_{\sigma \in S_L} \sum^*_{1\le \alpha_{\sigma(1)} \dots \alpha_{\sigma(L)} \le N} \bigotimes^L_{i=1}e^{(\alpha_{i}\alpha_{i})}_{i} R^{\{\sigma\}}_{1\dots L},
\label{ansatz1a}\end{equation}
where the symbol $*$ in the sum (\ref{ansatz1a}) 
of ordered indices is to be over all non decreasing sequences 
of the indices $\alpha_{\sigma(i)}$. The indices $\alpha_{\sigma(i)}$ satisfy one of the two inequalities for each pair of neighboring indices:
\begin{equation}\begin{array}{lll}
\alpha_{\sigma(i)} \le \alpha_{\sigma(i+1)} & \textrm{if} & \sigma(i) < \sigma(i+1)\\
\alpha_{\sigma(i)} < \alpha_{\sigma(i+1)} & \textrm{if} & \sigma(i) > \sigma(i+1).
\end{array}\label{summ}\end{equation}

The specific choice on the second part of the sum (\ref{ansatz1a}) 
and the form 
of the $R$-matrix ensure that $\mathcal{F}_{1\dots L}$ and $F_{1\dots L}$ are 
lower-triangular (see \cite{BOS1,ZH1,ZH2} for details). Additionally, since each 
diagonal entry is non-zero, the inverse $F^{-1}_{1\dots L}$ is assured to exist. Note also
that the tensor product term in Eq.(\ref{ansatz1a}) is invariant if we apply the permutation to each index $i$,
\begin{equation*}
\bigotimes^L_{i=1}e^{(\alpha_{i}\alpha_{i})}_{i} = \bigotimes^L_{i=1}e^{(\alpha_{\sigma(i)}\alpha_{\sigma(i)})}_{\sigma(i)},
\end{equation*}
and hence an equivalent expression for the $\mathcal{F}$-matrix is,
\begin{equation}
  \mathcal{F}_{1\dots L} = \sum_{\sigma \in S_L} \sum^*_{1\le \alpha_{\sigma(1)} \dots \alpha_{\sigma(L)} \le N} \bigotimes^L_{i=1}e^{(\alpha_{\sigma(i)}\alpha_{\sigma(i)})}_{\sigma(i)} R^{\{\sigma\}}_{1\dots L}.
\label{newansatz1a}\end{equation}

We now are left with the task of verifying the factorization condition (\ref{defining}). This involves a sequence of steps that
we shall now detail.

\subsubsection{Recasting the factorization condition}
In order to show the validity of the factorization condition (\ref{defining}) for all $\sigma \in S_L$ we 
proceed much like \cite{MZ} and take advantage of the 
decomposition of $F_{1\dots L}$ present in Eq.(\ref{ansatz}). In doing so we recast 
Eq.(\ref{defining}) as an equation involving $\mathcal{F}_{1 \dots L}$ and the twisted $R$-matrix. We then adapt the procedure 
first devised in \cite{BOS1, ZH1,ZH2} to tackle the problem. To this end we substitute Eq.(\ref{ansatz}) into Eq.(\ref{defining}) to obtain,
\begin{equation}\begin{array}{lll}
\mathcal{F}_{\sigma(1\dots L)} R^{\{\sigma\}}_{1\dots L}  = \mathcal{N}^{-1}_{\sigma(1\dots L)}\mathcal{N}_{1\dots L} \mathcal{F}_{1\dots L} & \textrm{for all}& \sigma \in S_L.
\end{array}\end{equation}
Recall that all elements of $S_L$ possess a minimal decomposition 
in terms of adjacent permutations (\ref{sigdecomp}). For the expression $\mathcal{N}^{-1}_{\sigma(1\dots L)}\mathcal{N}_{1\dots L}$ we offer the following result.
\begin{proposition}\label{NandR}
\begin{equation}\begin{array}{lll}
\mathcal{N}^{-1}_{\sigma(1\dots L)}\mathcal{N}_{1\dots L}  = \mathcal{R}^{\{\sigma\}}_{1\dots L} &\textrm{for all}& \sigma \in S_L,
\end{array}\label{auxR}\end{equation}
where $\mathcal{R}^{\{\sigma\}}_{1\dots L}$ follows the same decomposition rules as $R^{\{\sigma\}}_{1\dots L}$,
\begin{equation}\begin{array}{llll}
&\mathcal{R}^{\{ \sigma \} }_{1 \dots L} &=& P^{\{ \sigma \} }_{1 \dots L} \hat{\mathcal{R}}^{\{ \sigma^{-1} \} }_{1 \dots L}\\
\textrm{where}&\hat{\mathcal{R}}^{\{ \sigma^{-1} \} }_{1 \dots L} &=& \hat{\mathcal{R}}^{\left\{\sigma_{\alpha_1 (\alpha_1+1)}\right\}}_{1\dots L} \hat{\mathcal{R}}^{\left\{\sigma_{\alpha_2 (\alpha_2 +1)}\right\}}_{1\dots L} \dots \hat{\mathcal{R}}^{\left\{\sigma_{\alpha_p (\alpha_p+1)}\right\}}_{1\dots L}\\
\textrm{and} & \hat{\mathcal{R}}^{\left\{\sigma_{\alpha (\alpha +1)}\right\}}_{1\dots L} &=& P_{\alpha(\alpha+1)}\mathcal{R}_{\alpha(\alpha+1)},
\end{array}\label{combinecurlyR}\end{equation}
and $\mathcal{R}_{12}$ is explicitly given by Eq.(\ref{twistR2}).
\end{proposition}

Because the corresponding auxiliary operators $\hat{\mathcal{R}}^{\{\sigma\}}_{1\dots L}$ 
provide a valid representation for $S_L$, the above result can be verified by showing that 
Eq.(\ref{auxR}) holds for only two permutations: the adjacent permutation $\sigma_{12}$ and 
the cyclic permutation $\sigma_c = \sigma_{12} \sigma_{23} \dots \sigma_{(L-1)L}$. For more 
details refer to Appendix C of \cite{MZ}. 

Hence we recast the factorization condition in the following form,
\begin{equation}\begin{array}{lll}
\mathcal{F}_{\sigma(1\dots L)} R^{\{\sigma\}}_{1\dots L}  = \mathcal{R}^{\{\sigma\}}_{1\dots L} \mathcal{F}_{1\dots L} & \textrm{for all} & \sigma \in S_L.
\end{array}\label{defining2}\end{equation}
We impose that the $\mathcal{R}$-matrices follow the same left-handed and right-handed convention as the $R$-matrices,
\begin{eqnarray}
\mathcal{R}_{1,2\dots,N} = \mathcal{R}_{1N} \mathcal{R}_{1(N-1)} \dots \mathcal{R}_{12},~~\textrm{and}~~
\mathcal{R}_{1\dots N-1,N} = \mathcal{R}_{1N} \mathcal{R}_{2 N} \dots \mathcal{R}_{(N-1)N}.
\end{eqnarray}
Following the above convention the $\mathcal{R}$-matrices obey the same global unitarity condition as the $R$-matrices,
\begin{equation}
 \mathcal{R}_{1,2\dots L}\mathcal{R}_{2\dots L,1}  =\mathcal{I}_{1} \otimes \dots \otimes \mathcal{I}_{L},
\label{unitcurlyR}\end{equation}
and being diagonal the $\mathcal{R}$-matrices commute amongst themselves.

\subsection{Verification of the factorization property} \label{PROOF}
We begin by applying Eq.(\ref{useful}) to Eq.(\ref{defining2}) to obtain,
\begin{equation}
 \hat{R}^{\{\sigma^{-1}\}}_{1\dots L} =\mathcal{F}^{-1}_{1\dots L} 
\hat{\mathcal{R}}^{\left\{\sigma^{-1} \right\}}_{1\dots L} \mathcal{F}_{1\dots L}.
\label{defining3}\end{equation} 
Since both $\hat{R}^{\{\sigma\}}_{1 \dots L}$ and $\hat{\mathcal{R}}^{\{\sigma\}}_{1\dots L}$ provide valid representations of $S_L$, we remark that Eq.(\ref{defining3}) is in a form that one can readily decompose the permutation $\sigma$. To illustrate this consider the permutation $\sigma = \sigma_1 \sigma_2$ on the left-hand side and right-hand side of Eq.(\ref{defining3}) respectively,
\begin{equation}
\begin{array}{lll}
\hat{R}^{\{(\sigma_1\sigma_2)^{-1}\}}_{1\dots L}&=& \hat{R}^{\{\sigma^{-1}_2\}}_{1\dots L}\hat{R}^{\{\sigma^{-1}_1\}}_{1\dots L}\\
\mathcal{F}^{-1}_{1\dots L} \hat{\mathcal{R}}^{\{(\sigma_1\sigma_2)^{-1}\}}_{1 \dots L}\mathcal{F}_{1\dots L} &=& \mathcal{F}^{-1}_{1\dots L} \hat{\mathcal{R}}^{\{\sigma^{-1}_2\}}_{1 \dots L}\mathcal{F}_{1\dots L}\mathcal{F}^{-1}_{1\dots L} \hat{\mathcal{R}}^{\{\sigma^{-1}_1\}}_{1 \dots L}\mathcal{F}_{1\dots L}
\end{array}
\end{equation}
where $\{\sigma_1,\sigma_2\} \in S_L$. 

Since $S_L$ can be constructed entirely from the adjacent permutations $\sigma_{j(j+1)}$, $j = (1,\dots,L-1)$, 
we need only verify Eq.(\ref{defining2}) for the adjacent permutation to guarantee its validity for all $S_L$.
To this end we substitute $\sigma = \sigma_{j(j+1)}$ into the factorization condition (\ref{defining2}) obtaining,
\begin{equation}
\mathcal{F}_{\sigma_{j(j+1)} (1\dots L)} R_{j(j+1)} - \mathcal{R}_{j(j+1)} \mathcal{F}_{1\dots L}=0,
\label{onetoprove}\end{equation}
where,
\begin{equation}\begin{array}{lll}
\mathcal{F}_{\sigma_{j(j+1)} (1\dots L)} R_{j(j+1)} &=& {\displaystyle \sum_{\sigma \in S_L} \sum^*_{1 \le \alpha_{\sigma(1)} \dots \alpha_{\sigma(L)}\le N}} \displaystyle{\bigotimes^L_{i=1}}e^{(\alpha_{\sigma(i)} \alpha_{\sigma(i)})}_{\sigma_{j(j+1)}\sigma(i) } \overbrace{R^{\{\sigma\}}_{\sigma_{j(j+1)}(1\dots L)} R^{\sigma_{j(j+1)}}_{1 \dots L}}^{\textrm{apply Eq.(\ref{combineR})}}\\
&=& {\displaystyle \sum_{\sigma \in S_L} \sum^*_{1 \le \alpha_{\sigma(1)} \dots \alpha_{\sigma(L)}\le N}} \displaystyle{\bigotimes^L_{i=1}}e^{(\alpha_{\sigma(i)} \alpha_{\sigma(i)})}_{\sigma_{j(j+1)}\sigma(i) } R^{\{\sigma_{j(j+1)}\sigma\}}_{1\dots L}.
\end{array}\label{othersum11}\end{equation}
We now perform the change of variables 
$ \alpha_{\sigma(i)} \rightarrow \alpha_{\sigma_{j(j+1)}\sigma(i)}$
to the summation indices $\alpha_{\sigma(i)}$.
In doing so Eq.(\ref{othersum11}) becomes,
\begin{equation}\begin{array}{lll}
\mathcal{F}_{\sigma_{j(j+1)} (1\dots L)} R_{j(j+1)} &=& {\displaystyle \sum_{\sigma \in S_L} \sum^{**}_{1 \le \alpha_{\sigma_{j(j+1)}\sigma(1)} \dots \alpha_{\sigma_{j(j+1)}\sigma(L)} \le N}} \displaystyle{\bigotimes^L_{i=1}}e^{(\alpha_{\sigma_{j(j+1)}\sigma(i)} \alpha_{\sigma_{j(j+1)}\sigma(i)})}_{\sigma_{j(j+1)}\sigma(i) } R^{\{\sigma_{j(j+1)}\sigma\}}_{1\dots L}\\
&=& {\displaystyle \sum_{\tau \in S_L} \sum^{**}_{1 \le \alpha_{\tau(1)} \dots \alpha_{\tau(L)}\le N}} \displaystyle{\bigotimes^L_{i=1}}e^{(\alpha_{\tau(i)} \alpha_{\tau(i)})}_{\tau(i) } R^{\{\tau \}}_{1\dots L},
\end{array}\label{othersum}\end{equation}
where we have applied the relabeling 
$\sigma_{j(j+1)}\sigma = \tau$
of the elements of $S_L$.

The 
summation $\sum^{**}_{1 \le \alpha_{\tau(1)} \dots \alpha_{\tau(L)} \le N} $ in Eq.(\ref{othersum}) for $\mathcal{F}_{\sigma_{j(j+1)} (1\dots L)} R_{j(j+1)}$ is deceptively similar to $\sum^{*}_{1 \le \alpha_{\sigma(1)} \dots \alpha_{\sigma(L)} \le N}$ in Eq.(\ref{newansatz1a}) for $\mathcal{F}_{1\dots L}$, in that its ordered indices are to be summed over all non decreasing sequences of the indices $\alpha_{\tau(i)}$. However the indices $\alpha_{\tau(i)}$ satisfy one of the two inequalities for each pair of neighboring indices:
\begin{equation}\begin{array}{lll}
\alpha_{\tau(i)} \le \alpha_{\tau(i+1)} & \textrm{if} & \sigma_{j(j+1)}\tau(i) < \sigma_{j(j+1)}\tau(i+1)\\
\alpha_{\tau(i)} < \alpha_{\tau(i+1)} & \textrm{if} & \sigma_{j(j+1)}\tau(i) > \sigma_{j(j+1)} \tau(i+1)
\end{array} .
\label{summ2}\end{equation}

Comparing (\ref{summ}) with (\ref{summ2}), one can see that the only 
difference between them is the adjacent permutation $\sigma_{j(j+1)}$ factor 
in the \textit{if} conditions. For any given $\tau \in S_L$, we focus on two 
integers, $k$ and $l$, where $\tau(k)=j$ and $\tau(l) = j+1$. Using these two 
integers we examine how the elementary transposition $\sigma_{j(j+1)}$ will 
affect the inequalities in (\ref{summ2}) compared to (\ref{summ}). There are only two relevant cases, 
given by $|k-l|>1$ and $|k-l|=1$.
When $|k-l|>1$ then the adjacent permutation $\sigma_{j(j+1)}$ does not affect the sequence of indices $\alpha_{\tau}$ at all, meaning that (\ref{summ}) and (\ref{summ2}) are the same. To explicitly show that (\ref{summ}) and (\ref{summ2}) are the same in this case we concentrate on the following four tuples, 
\begin{equation*}\begin{array}{llll}
(\alpha_{\tau(k-1)}, \alpha_{\tau(k)}) &(\alpha_{\tau(k)}, \alpha_{\tau(k+1)}) & (\alpha_{\tau(l-1)}, \alpha_{\tau(l)})& (\alpha_{\tau(l)}, \alpha_{\tau(l+1)}) ,
\end{array}\end{equation*}
which are affected by the permutation $\sigma_{j(j+1)}$.

Focusing on the tuple $(\alpha_{\tau(k-1)}, \alpha_{\tau(k)})$, there are two possible cases for the values of $\tau(k-1)$ and $\tau(k)$ given by,
\begin{equation}\begin{array}{lll}
 \sigma_{j(j+1)} \tau(k-1) > \sigma_{j(j+1)}\tau(k)  & \textrm{or} &  \sigma_{j(j+1)} \tau(k-1) < \sigma_{j(j+1)}\tau(k) 
\end{array}\label{helpme}\end{equation}
Since $|k-l|>1$ we remark that for both cases given in Eq.(\ref{helpme}):
\begin{itemize}
\item{$\tau(k-1)$ is invariant under the permutation $\sigma_{j(j+1)}$.}
\item{$\sigma_{j(j+1)}\tau(k) = \sigma_{j(j+1)}(j) = j+1$ - changing $\tau(k)$ to $\tau(k)+1$ is not going to affect the values of the inequalities.}
\end{itemize}
An equivalently elementary analysis can be performed for the three remaining tuples. Hence with these observations, we are assured that the type of the inequality is unchanged by the addition of the permutation $\sigma_{j(j+1)}$ - meaning that (\ref{summ}) and (\ref{summ2}) lead to the same results when $|k-l|>1$.

We now concentrate on the case $|k-l|=1$.
In this situation the permutation $\sigma_{j(j+1)}$ does affect the inequality, 
leading to a difference between (\ref{summ}) and (\ref{summ2}). To see this consider the case $\tau(k) = j$ and $\tau(k+1)=j+1$, where
\begin{equation*}\begin{array}{clcl}
\tau(k) &<& \tau(k+1)  & \textrm{and} \\
\sigma_{j(j+1)}\tau(k) &>& \sigma_{j(j+1)}\tau(k+1).
\end{array}\end{equation*}
Thus, what is usually a ``$<$'' inequality in (\ref{summ}) changes to a ``$>$'' inequality in (\ref{summ2}). Hence the summation in Eq.(\ref{ansatz1a}),
\begin{equation*}
 \sum^*_{\dots \alpha_{\tau(k)}\alpha_{\tau(k+1)} \dots} =  \sum_{\dots \alpha_{j} \le \alpha_{j+1} \dots},
\end{equation*}
changes when we consider the summation in Eq.(\ref{othersum}),
\begin{equation*}
 \sum^{**}_{\dots \alpha_{\tau(k)}\alpha_{\tau(k+1)} \dots} =  \sum_{\dots \alpha_{j} < \alpha_{j+1} \dots}.
\end{equation*}
Equivalently, consider the case $\tau(k) = j+1$ and $\tau(k+1)=j$, where
\begin{equation*}\begin{array}{clcl}
\tau(k) &>& \tau(k+1)  & \textrm{and} \\
\sigma_{j(j+1)}\tau(k) &<& \sigma_{j(j+1)}\tau(k+1).
\end{array}\end{equation*}
Thus, what is usually a ``$>$'' inequality in (\ref{summ}) changes to a ``$<$'' inequality in (\ref{summ2}). Hence the summation in Eq.(\ref{ansatz1a}),
\begin{equation*}
 \sum^*_{\dots \alpha_{\tau(k)}\alpha_{\tau(k+1)} \dots} =  \sum_{\dots \alpha_{j+1} < \alpha_{j} \dots},
\end{equation*}
changes when we consider the summation in Eq.(\ref{othersum}),
\begin{equation*}
 \sum^{**}_{\dots \alpha_{\tau(k)}\alpha_{\tau(k+1)} \dots} =  \sum_{\dots \alpha_{j+1} \le \alpha_{j} \dots}.
\end{equation*}

\subsubsection{The case $|k-l|>1$}
Applying the above analysis we now consider the left-hand side of Eq.(\ref{onetoprove}),
\begin{equation}
 \sum_{\sigma \in S_L} \sum^{**}_{1 \le \alpha_{\sigma(1)} \dots \alpha_{\sigma(L)} \le N} \bigotimes^L_{i=1}e^{(\alpha_{\sigma(i)} \alpha_{\sigma(i)})}_{\sigma(i) } R^{\{\sigma \}}_{1\dots L} - \mathcal{R}_{j(j+1)}\sum_{\sigma \in S_L} \sum^{*}_{1\le \alpha_{\sigma(1)} \dots \alpha_{\sigma(L)} \le N} \bigotimes^L_{i=1}e^{(\alpha_{\sigma(i)} \alpha_{\sigma(i)})}_{\sigma(i) } R^{\{\sigma \}}_{1\dots L}
\label{BIGEQ}\end{equation}
and in particular, focus on the case where $\sigma(k)=j$, $\sigma(l)=j+1$ and $|k-l|>1$. In this case, 
$\sum^{**}_{1\le \alpha_{\sigma(1)} \dots \alpha_{\sigma(L)} \le N} = 
\sum^{*}_{1 \le \alpha_{\sigma(1)} \dots \alpha_{\sigma(L)} \le N}$, and hence Eq.(\ref{BIGEQ}) becomes,
 \begin{equation}
(\mathcal{I}_{j}\otimes \mathcal{I}_{j+1}- \mathcal{R}_{j(j+1)}) \sum^L_{\genfrac{}{}{0pt}{}{k,l =1}{|k-l|>1}  }\sum_{\genfrac{}{}{0pt}{}{\sigma \in S_L}{\sigma(k)=j,\sigma(l)=j+1} } \sum^{*}_{1\le \alpha_{\sigma(1)} \dots \alpha_{\sigma(L)} \le N} \bigotimes^L_{i=1}e^{(\alpha_{\sigma(i)} \alpha_{\sigma(i)})}_{\sigma(i) } R^{\{\sigma \}}_{1\dots L} .
\label{BIGEQ2}\end{equation}
Through inspection one can see that the condition $|k-l|>1$ means that $\alpha_j \ne \alpha_{j+1}$ in the summation - the easiest way to convince oneself of this statement is to consider the case $|k-l|=2$ (for clarity label $l = k+2$) and look at the inequality conditions between $\alpha_{\sigma(k)}$, $\alpha_{\sigma(k+1)}$ and $\alpha_{\sigma(k+2)}$. Hence, through applying Eq.(\ref{Rone}), Eq.(\ref{BIGEQ2}) becomes,
\begin{equation*}\begin{array}{rll}
 \sum^L_{\genfrac{}{}{0pt}{}{k,l =1}{|k-l|>1}  }\sum_{\genfrac{}{}{0pt}{}{\sigma \in S_L}{\sigma(k)=j,\sigma(l)=j+1} } \sum^{*}_{1 \le \alpha_{\sigma(1)} \dots \alpha_{\sigma(L)} \le N}  \overbrace{(\mathcal{I}_{j} \otimes \mathcal{I}_{j+1}- \mathcal{R}_{j(j+1)})\left\{ e^{(\alpha_{j} \alpha_{j})}_{j} \otimes e^{(\alpha_{j+1} \alpha_{j+1})}_{j+1}  \right\}}^{\textrm{apply Eq.(\ref{Rone})}}\\ \displaystyle{\bigotimes^L_{\genfrac{}{}{0pt}{}{i=1}{i\ne k,l}}}e^{(\alpha_{\sigma(i)} \alpha_{\sigma(i)})}_{\sigma(i) } R^{\{\sigma \}}_{1\dots L} &=&0.
\end{array}\end{equation*}
\subsubsection{The case $|k-l|=1$}
We now focus on the case $\sigma(k)=j$, $\sigma(l)=j+1$ where $|k-l|=1$. In this case $\sum^{**}_{1 \le \alpha_{\sigma(1)} \dots \alpha_{\sigma(L)} \le N}$ $\ne \sum^{*}_{1 \le \alpha_{\sigma(1)} \dots \alpha_{\sigma(L)} \le N}$, hence we look at each term in Eq.(\ref{BIGEQ}) separately,
\begin{eqnarray}
 & \sum^L_{\genfrac{}{}{0pt}{}{k,l =1}{|k-l|=1}  }\sum_{\genfrac{}{}{0pt}{}{\sigma \in S_L}{\sigma(k)=j,\sigma(l)=j+1} } \sum^{**}_{1 \le \alpha_{\sigma(1)} \dots \alpha_{\sigma(L)} \le N} \displaystyle{\bigotimes^L_{i=1}}e^{(\alpha_{\sigma(i)} \alpha_{\sigma(i)})}_{\sigma(i) } R^{\{\sigma \}}_{1\dots L} \nonumber\\
=& \sum^L_{k=1}\sum_{\genfrac{}{}{0pt}{}{\sigma \in S_L}{\sigma(k)=j,\sigma(k+1)=j+1} } \sum^{*}_{1 \le \alpha_{\sigma(1)} \dots \alpha_{j}< \alpha_{j+1} \dots \alpha_{\sigma(L)} \le N} \displaystyle{\bigotimes^L_{i=1}}e^{(\alpha_{\sigma(i)} \alpha_{\sigma(i)})}_{\sigma(i) } R^{\{\sigma \}}_{1\dots L} \label{otherI}\\
&+  \sum^L_{k=1}\sum_{\genfrac{}{}{0pt}{}{\sigma \in S_L}{\sigma(k+1)=j,\sigma(k)=j+1} } \sum^{*}_{1 \le \alpha_{\sigma(1)} \dots \alpha_{j+1}\le \alpha_{j} \dots \alpha_{\sigma(L)}\le N} \displaystyle{\bigotimes^L_{i=1}}e^{(\alpha_{\sigma(i)} \alpha_{\sigma(i)})}_{\sigma(i) } R^{\{\sigma \}}_{1\dots L},
\label{otherII}\end{eqnarray}
and,
\begin{eqnarray}
 & \mathcal{R}_{j(j+1)}\sum^L_{\genfrac{}{}{0pt}{}{k,l =1}{|k-l|=1}  }\sum_{\genfrac{}{}{0pt}{}{\sigma \in S_L}{\sigma(k)=j,\sigma(l)=j+1} } \sum^{*}_{1 \le \alpha_{\sigma(1)} \dots \alpha_{\sigma(L)} \le N} \displaystyle{\bigotimes^L_{i=1}}e^{(\alpha_{\sigma(i)} \alpha_{\sigma(i)})}_{\sigma(i) } R^{\{\sigma \}}_{1\dots L} \nonumber\\
=& \mathcal{R}_{j(j+1)}\sum^L_{k=1}\sum_{\genfrac{}{}{0pt}{}{\sigma \in S_L}{\sigma(k)=j,\sigma(k+1)=j+1} } \sum^{*}_{1 \le \alpha_{\sigma(1)} \dots \alpha_{j}\le \alpha_{j+1} \dots \alpha_{\sigma(L)} \le N} \displaystyle{\bigotimes^L_{i=1}}e^{(\alpha_{\sigma(i)} \alpha_{\sigma(i)})}_{\sigma(i) } R^{\{\sigma \}}_{1\dots L}\label{otherIII}\\
&+  \mathcal{R}_{j(j+1)}\sum^L_{k=1}\sum_{\genfrac{}{}{0pt}{}{\sigma \in S_L}{\sigma(k+1)=j,\sigma(k)=j+1} } \sum^{*}_{1\le \alpha_{\sigma(1)} \dots \alpha_{j+1}< \alpha_{j} \dots \alpha_{\sigma(L)}\le N} \displaystyle{\bigotimes^L_{i=1}}e^{(\alpha_{\sigma(i)} \alpha_{\sigma(i)})}_{\sigma(i) } R^{\{\sigma \}}_{1\dots L}.
\label{otherIV}\end{eqnarray}

We notice that the summations over the $\alpha_i$'s in Eqs. (\ref{otherI}) and (\ref{otherII}) 
are now of the same type as those in Eqs. (\ref{otherIII}) and (\ref{otherIV}) - i.e. they have only one ``$*$''
symbol. This is because the summation indices which cause the difference ($\alpha_j$ and $\alpha_{j+1}$) have been dealt with explicitly. 

We now subtract the $\alpha_j \ne \alpha_{j+1}$ component of Eq.(\ref{otherIII}) from Eq.(\ref{otherI}) to obtain,
\begin{equation}\begin{array}{r}
\sum^L_{k=1}\sum_{\genfrac{}{}{0pt}{}{\sigma \in S_L}{\sigma(k)=j,\sigma(k+1)=j+1} } \sum^{*}_{1 \le \alpha_{\sigma(1)} \dots \alpha_{j}< \alpha_{j+1} \dots \alpha_{\sigma(L)} \le N} \overbrace{(\mathcal{I}_{j} \otimes \mathcal{I}_{j+1}-\mathcal{R}_{j(j+1)})\left\{e^{(\alpha_{j} \alpha_{j})}_{j }\otimes e^{(\alpha_{j+1} \alpha_{j+1})}_{j+1 }\right\}}^{\textrm{\textrm{apply Eq.(\ref{Rone})}}} \\
\displaystyle{\bigotimes^L_{\genfrac{}{}{0pt}{}{i=1}{i\ne k,k+1}}}e^{(\alpha_{\sigma(i)} \alpha_{\sigma(i)})}_{\sigma(i) } R^{\{\sigma \}}_{1\dots L} =0.
\end{array}\end{equation}
Similarly we subtract Eq.(\ref{otherIV}) from the $\alpha_j \ne \alpha_{j+1}$ component of Eq.(\ref{otherII}) to obtain,
\begin{equation}\begin{array}{r}
\sum^L_{k=1}\sum_{\genfrac{}{}{0pt}{}{\sigma \in S_L}{\sigma(k+1)=j,\sigma(k)=j+1} } \sum^{*}_{1 \le \alpha_{\sigma(1)} \dots \alpha_{j+1}< \alpha_{j} \dots \alpha_{\sigma(L)} \le N} \overbrace{(\mathcal{I}_{j} \otimes \mathcal{I}_{j+1}-\mathcal{R}_{j(j+1)})\left\{e^{(\alpha_{j} \alpha_{j})}_{j }\otimes e^{(\alpha_{j+1} \alpha_{j+1})}_{j+1 }\right\}}^{\textrm{\textrm{apply Eq.(\ref{Rone})}}} \\
\displaystyle{\bigotimes^L_{\genfrac{}{}{0pt}{}{i=1}{i\ne k,k+1}}}e^{(\alpha_{\sigma(i)} \alpha_{\sigma(i)})}_{\sigma(i) } R^{\{\sigma \}}_{1\dots L} =0.
\end{array}\end{equation}
Finally, we now subtract the $\alpha_j=\alpha_{j+1}$ component of Eq.(\ref{otherIII}) from the $\alpha_j=\alpha_{j+1}$ component of Eq.(\ref{otherII}) to obtain,
\begin{equation}\begin{array}{r}
\sum^L_{k=1}\sum_{\genfrac{}{}{0pt}{}{\tau \in S_L}{\tau(k+1)=j,\tau(k)=j+1} } \sum^{*}_{1 \le \alpha_{\tau(1)} \dots \alpha_{j+1}= \alpha_{j} \dots \alpha_{\tau(L)}\le N} \displaystyle{\bigotimes^L_{i=1}}e^{(\alpha_{\tau(i)} \alpha_{\tau(i)})}_{\tau(i) } R^{\{\tau \}}_{1\dots L}\\
- \mathcal{R}_{j(j+1)}\sum^L_{k=1}\sum_{\genfrac{}{}{0pt}{}{\sigma \in S_L}{\sigma(k)=j,\sigma(k+1)=j+1} } \sum^{*}_{1 \le \alpha_{\sigma(1)} \dots \alpha_{j}= \alpha_{j+1} \dots \alpha_{\sigma(L)} \le N} \displaystyle{\bigotimes^L_{i=1}}e^{(\alpha_{\sigma(i)} \alpha_{\sigma(i)})}_{\sigma(i) } R^{\{\sigma \}}_{1\dots L}.
\end{array}\label{goomba}\end{equation}
We proceed in Eq.(\ref{goomba}) by first making the change 
$\tau = \sigma \sigma_{k(k+1)}$
in permutation labels.
Next we apply Eq.(\ref{combineR}) to the permuted $R$-matrix $R^{\{\tau \}}_{1\dots L}$ to obtain,
\begin{equation}
R^{\{\tau \}}_{1\dots L} = R^{\{\sigma_{k(k+1)} \}}_{\sigma(1\dots L)} R^{\{\sigma \}}_{1\dots L}
= R_{j(j+1)} R^{\{\sigma \}}_{1\dots L}.
\end{equation}
Taking into account the above change in permutation labels and decomposition of the permuted $R$-matrix, Eq.(\ref{goomba}) becomes,
\begin{equation}\begin{array}{r}
\sum^L_{k=1}\sum_{\genfrac{}{}{0pt}{}{\sigma \in S_L}{\sigma(k)=j,\sigma(k+1)=j+1} } \sum^{*}_{1 \le \alpha_{\sigma(1)} \dots \alpha_{j}= \alpha_{j+1} \dots \alpha_{\sigma(L)} \le N} \displaystyle{\bigotimes^L_{\genfrac{}{}{0pt}{}{i=1}{i\ne k,k+1}}}e^{(\alpha_{\sigma(i)} \alpha_{\sigma(i)})}_{\sigma(i) } \\
\underbrace{\left\{ \left(e^{(\alpha_{j} \alpha_{j})}_{j}\otimes e^{(\alpha_{j+1} \alpha_{j+1})}_{j+1}\right){R}_{j(j+1)} - \mathcal{R}_{j(j+1)}\left(e^{(\alpha_{j} \alpha_{j})}_{j}\otimes e^{(\alpha_{j+1} \alpha_{j+1})}_{j+1}\right) \right\}}_{\textrm{apply Eq.(\ref{Rone})}} R^{\{\sigma \}}_{1\dots L}=0.
\end{array}\end{equation}
Hence we have verified the factorization condition for general $L$ and $N$.

\section{Twisted monodromy operators for $N=3$}\label{OPOP}

The purpose of this section is to  show that the 
constructed $F$-matrices can effectively be used as similarity 
transformation in the simplest situation of the
$N=3$ state vertex model. We shall present an algebraic derivation of the form of some
relevant elements of the monodromy matrix in the $F$-basis 
for arbitrary weights. Here we represent
the $N=3$ monodromy matrix as, 
\begin{equation}
\mathcal{T}_{a,1 \dots L}(\mu) = \left(\begin{array}{ccc}
A^{(11)}_{1\dots L}(\mu) & A^{(12)}_{1\dots L}(\mu) & B^{(1)}_{1\dots L}(\mu)\\
A^{(21)}_{1\dots L} (\mu)& A^{(22)}_{1\dots L}(\mu) & B^{(2)}_{1\dots L}(\mu)\\
C^{(1)}_{1\dots L}(\mu) & C^{(2)}_{1\dots L}(\mu) & D_{1\dots L}(\mu)
                                        \end{array}\right)_a.
\end{equation}

In general, a given monodromy matrix element 
${X}_{1 \dots L}(\mu)$ can be transformed to a new operator    
$\tilde{X}_{1 \dots L}(\mu)$ having a much simpler quasilocal 
form with the help of the $F$-matrices \cite{MA1}. 
Since $F_{1\dots L }$ and $\mathcal{F}_{1 \dots L}$ are related via the 
multiplication of a diagonal matrix (\ref{ansatz}), it is enough to compute the 
following non-trivial twisted operators,
\begin{equation}
\tilde{X}_{1 \dots L}(\mu) = \mathcal{F}_{1 \dots L} X_{1 \dots L}(\mu) \mathcal{F}^{-1}_{1\dots L}.
\label{twisting}
\end{equation}

In what follows we shall present a conjecture for the expressions of the twisted monodromy operators 
$\tilde{D}_{1\dots L}(\mu)$, $\tilde{C}^{(2)}_{1\dots L}(\mu)$, $\tilde{B}^{(2)}_{1\dots L}(\mu)$, 
$\tilde{C}^{(1)}_{1\dots L}(\mu)$ and $\tilde{B}^{(1)}_{1\dots L}(\mu)$. Our results 
are built up from an analysis performed
in the cases $L=2$ and $L=3$ relying only on the identities derived from the 
Yang-Baxter and unitarity relations. Fortunately, this study is sufficient to foresee the main structure
of the mentioned twisted operators for arbitrary $L$ and general Boltzmann weights  
without relying on any specific parameterizations. 

We now list the final expressions for the above mentioned twisted operators,
\begin{equation}
\tilde{D}_{1\dots L}(\mu)= \bigotimes^L_{i=1} diag\left\{b_{31}(\mu,\xi_i),b_{32}(\mu,\xi_i), a_{3}(\mu,\xi_i) \right\}_i 
\label{D}
\end{equation}
\begin{equation}
\begin{array}{lll}
\tilde{C}^{(2)}_{1\dots L}(\mu) 
&=& \displaystyle{\sum^L_{l=1}}c_{32}(\mu,\xi_l)  
e^{(23)}_l \bigotimes^{L}_{\genfrac{}{}{0pt}{}{i=1}{i\ne l}} diag\left\{b_{21}(\mu,\xi_i),\frac{b_{32}(\mu,\xi_i)}{b_{32}(\xi_l,\xi_i)\theta_2(\xi_i,\xi_l)},a_3(\mu,\xi_i)\theta_3(\xi_i,\xi_l)  \right\}_i 
\end{array}
\label{C2}
\end{equation}
\begin{equation}
\begin{array}{lll}
\tilde{B}^{(2)}_{1\dots L}(\mu) 
&=&  \displaystyle{\sum^L_{l=1}}c_{23}(\mu,\xi_l)e^{(32)}_l\bigotimes^{L}_{\genfrac{}{}{0pt}{}{i=1}{\ne l}} 
diag\left\{ b_{31}(\mu,\xi_i),b_{32}(\mu,\xi_i)\theta_2(\xi_l,\xi_i),\frac{a_3(\mu,\xi_i)}{b_{32}(\xi_i,\xi_l)\theta_3(\xi_l,\xi_i)} \right\}_i   
\end{array}
\label{B2}
\end{equation}
\begin{equation}
\begin{array}{lll}
\tilde{C}^{(1)}_{1\dots L}(\mu) 
&=&\displaystyle{\sum^L_{l=1}}c_{31}(\mu,\xi_l)e^{(13)}_l \bigotimes^{L}_{\genfrac{}{}{0pt}{}{i=1}{i\ne l}} diag \left\{ \frac{b_{21}(\mu,\xi_i)}{b_{21}(\xi_l,\xi_i)\theta_1(\xi_i,\xi_l)},\frac{b_{32}(\mu,\xi_i)}{b_{32}(\xi_l,\xi_i)},a_3(\mu,\xi_i)\theta_3(\xi_i,\xi_l) \right\}_i \\  
&&+  \displaystyle{\sum^L_{\genfrac{}{}{0pt}{}{l_1, l_2 = 1}{l_1 \ne l_2}}}\frac{c_{31}(\mu,\xi_{l_1}) b_{32}(\mu, \xi_{l_2}) c_{21}(\xi_{l_1},\xi_{l_2})}{b_{32}(\xi_{l_1},\xi_{l_2})}e^{(23)}_{l_1} \otimes e^{(12)}_{l_2}\\
&&\displaystyle{\bigotimes^{L}_{\genfrac{}{}{0pt}{}{i=1}{i\ne l_1,l_2}}} diag \left\{ \frac{b_{21}(\mu,\xi_i)}{b_{21}(\xi_{l_2},\xi_i)\theta_1(\xi_i,\xi_{l_2})},\frac{b_{32}(\mu,\xi_i)\theta_2(\xi_i,\xi_{l_2})}{b_{32}(\xi_{l_1},\xi_i)\theta_2(\xi_i,\xi_{l_1})},a_3(\mu,\xi_i)\theta_3(\xi_i,\xi_{l_1}) \right\}_i  
\end{array}
\label{C1}
\end{equation}
\begin{equation}
\begin{array}{lll}
\tilde{B}^{(1)}_{1\dots L}(\mu) 
&=&\displaystyle{\sum^L_{l=1}}c_{13}(\mu,\xi_l)e^{(31)}_l \bigotimes^{L}_{\genfrac{}{}{0pt}{}{i=1}{i\ne l}} diag \left\{ b_{31}(\mu,\xi_i)\theta_{1}(\xi_l,\xi_i),\frac{b_{32}(\mu,\xi_i)}{b_{21}(\xi_i,\xi_l)},\frac{a_3(\mu,\xi_i)}{b_{31}(\xi_i,\xi_l)\theta_3(\xi_l,\xi_i)}  \right\}_i \\  
&&+  \displaystyle{\sum^L_{\genfrac{}{}{0pt}{}{l_1, l_2 = 1}{l_1 \ne l_2}}}\frac{c_{13}(\mu,\xi_{l_1}) b_{32}(\mu, \xi_{l_2}) c_{12}(\xi_{l_1},\xi_{l_2})}{ b_{21}(\xi_{l_1},\xi_{l_2})} e^{(32)}_{l_1} \otimes e^{(21)}_{l_2}\\
&&\displaystyle{\bigotimes^{L}_{\genfrac{}{}{0pt}{}{i=1}{i\ne l_1,l_2}}} diag \left\{ b_{31}(\mu,\xi_i)\theta_1(\xi_{l_2},\xi_i) ,\frac{b_{32}(\mu,\xi_i)\theta_2(\xi_{l_1},\xi_i)}{b_{21}(\xi_{i},\xi_{l_2})\theta_2(\xi_{l_2},\xi_i)}, \frac{a_3(\mu,\xi_i)}{b_{31}(\xi_i,\xi_{l_1})\theta_3(\xi_{l_1},\xi_{i})} \right\}_i  
\end{array}
\label{B1}
\end{equation}
where the auxiliary functions 
$\theta_i(\xi_j,\xi_k)$ are given by,
\begin{equation}\begin{array}{ll}
\theta_i(\xi_j,\xi_k) = \left\{ \begin{array}{cll} a_i(\xi_j,\xi_k) 
& \textrm{for} & j < k \\ 1 & \textrm{for} & j \ge k \end{array} \right. & i = 1,2,3 .
\end{array}\label{theta}
\end{equation}

We begin by explicitly verifying that the $L=2$ case holds for each of the twisted operators
given below. 

\subsection{The L=2 case} 
In this section we shall calculate the twisted operators $\tilde{D}_{12}(\mu)$, 
$\tilde{C}^{(2)}_{12}(\mu)$, $\tilde{B}^{(2)}_{12}(\mu)$, 
$\tilde{C}^{(1)}_{12}(\mu)$ and $\tilde{B}^{(1)}_{12}(\mu)$ directly from the similarity transform (\ref{twisting}). 
We shall then detail the necessary Yang-Baxter and unitary relations for the entries obtained 
from Eq.(\ref{twisting}) to match the corresponding entries 
in the $L=2$ operators given by Eqs.(\ref{D}-\ref{B1}).  The expressions derived directly from Eq.(\ref{twisting}) have
the following structure,
\begin{equation}\begin{array}{lll}
\tilde{D}_{12}(\mu) &=& 
diag \left\{ b_{31}(\mu,\xi_1),   b_{32}(\mu,\xi_1), a_{3}(\mu,\xi_1) \right\}_1 \otimes diag \left\{ b_{31}(\mu,\xi_2),   b_{32}(\mu,\xi_2), a_{3}(\mu,\xi_2) \right\}_2 \\
&&+ \kappa^{(D)}_1 e^{(21)}_1 \otimes e^{(12)}_2 +\kappa^{(D)}_2 e^{(31)}_1 \otimes e^{(13)}_2 +\kappa^{(D)}_3 e^{(32)}_1 \otimes e^{(23)}_2\\
\tilde{C}^{(2)}_{12}(\mu)& =& diag \{\kappa^{(C_2)}_1,\kappa^{(C_2)}_2,\kappa^{(C_2)}_3 \}_1\otimes e^{(23)}_2 + e^{(23)}_1 \otimes diag \{\kappa^{(C_2)}_4,\kappa^{(C_2)}_5,\kappa^{(C_2)}_6 \}_2\\
&&+ \kappa^{(C_2)}_7 e^{(21)}_1 \otimes e^{(13)}_2\\
\tilde{B}^{(2)}_{12}(\mu)& =& diag \{\kappa^{(B_2)}_1,\kappa^{(B_2)}_2,\kappa^{(B_2)}_3 \}_1\otimes e^{(32)}_2 + e^{(32)}_1 \otimes diag \{\kappa^{(B_2)}_4,\kappa^{(B_2)}_5,\kappa^{(B_2)}_6 \}_2\\
&&+ \kappa^{(B_2)}_7 e^{(31)}_1 \otimes e^{(12)}_2
\end{array}
\label{kappaL21}
\end{equation}
\begin{equation}\begin{array}{lll}
\tilde{C}^{(1)}_{12}(\mu)& =& diag \{\kappa^{(C_1)}_1,\kappa^{(C_1)}_2,\kappa^{(C_1)}_3 \}_1\otimes e^{(13)}_2 + e^{(13)}_1 \otimes diag \{\kappa^{(C_1)}_4,\kappa^{(C_1)}_5,\kappa^{(C_1)}_6 \}_2\\
&&+ \kappa^{(C_1)}_7 e^{(12)}_1 \otimes e^{(23)}_2+ \kappa^{(C_1)}_8 e^{(23)}_1 \otimes e^{(12)}_2\\
\tilde{B}^{(1)}_{12}(\mu)& =& diag \{\kappa^{(B_1)}_1,\kappa^{(B_1)}_2,\kappa^{(B_1)}_3 \}_1\otimes e^{(31)}_2 + e^{(31)}_1 \otimes diag \{\kappa^{(B_1)}_4,\kappa^{(B_1)}_5,\kappa^{(B_1)}_6 \}_2\\
&&+ \kappa^{(B_1)}_7 e^{(21)}_1 \otimes e^{(32)}_2+ \kappa^{(B_1)}_8 e^{(32)}_1 \otimes e^{(21)}_2
\end{array}
\label{kappaL22}
\end{equation}

All the $\kappa$ entries of the above matrices do not immediately match 
the corresponding entries calculated from the $L=2$ expressions of Eqs.(\ref{D})-(\ref{B1}). It is possible to simplify
these entries using only certain relations coming from the Yang-Baxter (\ref{bigYB}) and unitarity (\ref{Biguni}) relations.
The technical details are quite cumbersome and thus have been deferred to Appendix B. 
In Table (\ref{tab:template}) we provide a summary  of the unitarity and Yang-Baxter relations 
that are required to simplify 
such non-trivial entries in order to bring 
the operators in the form given by Eqs.(\ref{D})-(\ref{B1}). 
\begin{table}\begin{center}
\begin{tabular}{| c | c | c |}
  \hline                       
  Operator &  Y-B equations & Unitarity equations \\\hline
  $\tilde{D}_{12}$  & (\ref{YB2})-\{3,1,2\} (\ref{YB4})-\{3,1\}-\{3,2\} & \\\hline
  $\tilde{C}^{(2)}_{12}$ & (\ref{YB3})-\{2,3,1\} (\ref{YB6})-\{3,2\} (\ref{YB9})-\{3,2\} (\ref{YB11})-\{3,2,1\} & (\ref{bigc})-\{3,2\} \\\hline
 $\tilde{B}^{(2)}_{12}$  & (\ref{YB3})-\{3,2,1\} (\ref{YB7})-\{2,3\} (\ref{YB8})-\{2,3\} (\ref{YB11})-\{2,3,1\} & (\ref{bigc})-\{2,3\} \\\hline
 $\tilde{C}^{(1)}_{12}$  & (\ref{YB3})-\{1,3,2\} (\ref{YB6})-\{3,1\} (\ref{YB9})-\{3,1\} (\ref{YB10})-\{3,2,1\} & (\ref{bigb})-\{2,1\} (\ref{bigc})-\{3,2\} \\\hline 
 $\tilde{B}^{(1)}_{12}$  & (\ref{YB3})-\{1,3,2\} (\ref{YB7})-\{1,3\} (\ref{YB8})-\{1,3\} (\ref{YB11})-\{1,3,2\} & (\ref{bigb})-\{1,2\} (\ref{bigc})-\{2,1\}-\{3,1\} \\\hline 
\end{tabular}
\caption{Required Yang-Baxter and unitarity relations for $L=2$.}
\label{tab:template}\end{center}\end{table}

We remark that in this table a given algebraic relation
among weights is referred to the equation number 
together with its respective indices $\{i\}$, $\{i,j\}$ or $\{i,j,k\}$.   
For example, the symbol (\ref{bigb})-$\{1,3\}$ refers to the equation
$ b_{13} (\xi_1,\xi_2)b_{31} (\xi_2,\xi_1) + c_{13} (\xi_1,\xi_2)c_{31} (\xi_2,\xi_1) =1$ whilst 
(\ref{YB2})-$\{1,3,2\}$ means the relation
$ b_{13}(\xi_1,\xi_2)b_{12}(\xi_1,\xi_3) = b_{12}(\xi_1,\xi_2)b_{13}(\xi_1,\xi_3)$.
We observe that there exists obvious equivalences for some equations such as
for Eq.(\ref{YB2}) the indices $\{i,j,k\}$ or  $\{i,k,j\}$ leads to the same relation.

At this point we remark that the $L=2$ case does not allow one to 
verify the entries of the diagonal matrices of the double summation 
from $\tilde{C}^{(1)}_{1\dots L}$ and $\tilde{B}^{(1)}_{1\dots L}$, we have only verified the coefficients 
of the double summation terms, given by $\frac{c_{31}(\mu,\xi_{l_1}) b_{32}(\mu, \xi_{l_2} 
c_{21}(\xi_{l_1},\xi_{l_2})}{b_{32}(\xi_{l_1},\xi_{l_2})}$ and $\frac{c_{13}(\mu,\xi_{l_1}) 
b_{32}(\mu, \xi_{l_2}) c_{12}(\xi_{l_1},\xi_{l_2})}{ b_{21}(\xi_{l_1},\xi_{l_2})}$ respectively. In order to 
obtain the entries of the aforementioned diagonal matrices we need to consider the $L=3$ case.

\subsection{The case L=3} 
We now consider explicitly calculating the $L=3$ case using Eq.(\ref{twisting}). Here it is only necessary to
consider the operators $\tilde{C}^{(1)}_{123}(\mu)$ and $\tilde{B}^{(1)}_{123}(\mu)$ 
and compute the entries that are completely missed by the $L=2$ case. We begin with 
the operator $\tilde{C}^{(1)}_{123}(\mu)$ whose entries are proportional to,
\begin{equation}\begin{array}{ll}
& e^{(12)}_1 \otimes e^{(23)}_2 \otimes diag\{\alpha^{(C_1)}_1,\alpha^{(C_1)}_2,\alpha^{(C_1)}_3 \}_3 + e^{(12)}_1 \otimes diag\{\beta^{(C_1)}_1,\beta^{(C_1)}_2,\beta^{(C_1)}_3 \}_2 \otimes e^{(23)}_3\\
+& diag\{\gamma^{(C_1)}_1,\gamma^{(C_1)}_2,\gamma^{(C_1)}_3 \}_1 \otimes e^{(12)}_2 \otimes e^{(23)}_3 + e^{(23)}_1 \otimes e^{(12)}_2 \otimes diag\{\delta^{(C_1)}_1,\delta^{(C_1)}_2,\delta^{(C_1)}_3 \}_3\\
+& e^{(23)}_1 \otimes diag\{\phi^{(C_1)}_1,\phi^{(C_1)}_2,\phi^{(C_1)}_3 \}_2 \otimes e^{(12)}_3 +diag\{\omega^{(C_1)}_1,\omega^{(C_1)}_2,\omega^{(C_1)}_3 \}_1 \otimes e^{(23)}_2 \otimes e^{(12)}_3.
\end{array}
\label{gammaL3}
\end{equation}
As with the $L=2$ case, we shall detail the Yang-Baxter and unitary relations necessary for the entries 
obtained from Eq.(\ref{twisting}) to match the $L=3$ case of Eq.(\ref{C1}). In Table (\ref{tab:template2}) 
we provide a summary  of the unitarity and Yang-Baxter relations that are required to simplify 
all the entries in order to bring the operators in the form given by Eq.(\ref{C1}). The technicalities of this calculation are
rather  involved and thus have been deferred to Appendix C.
\begin{table}\begin{center}
\begin{tabular}{| c | c | c |}
  \hline                       
  Entry &  Y-B equations & Unitarity equations \\\hline
  $\alpha^{(C_1)}_1$  & (\ref{YB3})-\{3,2,1\} (\ref{YB10})-\{3,2,1\} & (\ref{bigc})-\{3,2\} \\\hline
  $\alpha^{(C_1)}_2$ & (\ref{YB10})-\{3,2,1\} & (\ref{bigc})-\{3,2\} \\\hline
 $\alpha^{(C_1)}_3$  & (\ref{YB10})-\{3,2,1\} & (\ref{bigc})-\{3,2\} \\\hline
  $\beta^{(C_1)}_1$  & (\ref{YB3})-\{3,2,1\} (\ref{YB10})-\{3,2,1\} & (\ref{bigc})-\{3,2\} \\\hline
  $\beta^{(C_1)}_2$ & (\ref{YB9})-\{3,2\} (\ref{YB10})-\{3,2,1\} & (\ref{bigc})-\{3,2\} \\\hline
 $\beta^{(C_1)}_3$  & (\ref{YB4})-\{3,2\} (\ref{YB5})-\{3,2\} (\ref{YB6})-\{3,2\} (\ref{YB10})-\{3,2,1\} & (\ref{bigc})-\{3,2\} \\\hline
  $\gamma^{(C_1)}_1$  & (\ref{YB3})-\{3,2,1\} (\ref{YB9})-\{2,1\}-\{3,1\}  (\ref{YB10})-\{3,2,1\} & (\ref{bigc})-\{2,1\}-\{3,1\}-\{3,2\} \\\hline
  $\gamma^{(C_1)}_2$ & (\ref{YB3})-\{1,3,2\} (\ref{YB6})-\{2,1\} (\ref{YB9})-\{3,2\} (\ref{YB10})-\{3,2,1\} & (\ref{bigc})-\{3,2\} \\\hline
 $\gamma^{(C_1)}_3$  & (\ref{YB2})-\{3,2,1\} (\ref{YB5})-\{3,2\} (\ref{YB6})-\{3,1\} (\ref{YB10})-\{3,2,1\} & (\ref{bigc})-\{3,2\} \\\hline
  $\delta^{(C_1)}_1$  & (\ref{YB3})-\{3,2,1\} & \\\hline
  $\delta^{(C_1)}_2$ & &\\\hline
 $\delta^{(C_1)}_3$  & & \\\hline
  $\phi^{(C_1)}_1$  & (\ref{YB2})-\{3,2,1\} (\ref{YB3})-\{3,2,1\} (\ref{YB9})-\{2,1\} & (\ref{bigc})-\{2,1\} \\\hline
  $\phi^{(C_1)}_2$ & &\\\hline
 $\phi^{(C_1)}_3$  & & \\\hline
  $\omega^{(C_1)}_1$  & (\ref{YB2})-\{3,2,1\} (\ref{YB3})-\{2,3,1\} (\ref{YB9})-\{2,1\}-\{3,1\} (\ref{YB10})-\{3,2,1\} & (\ref{bigc})-\{2,1\}-\{3,1\} \\\hline
  $\omega^{(C_1)}_2$ & (\ref{YB3})-\{1,3,2\} (\ref{YB6})-\{2,1\} (\ref{YB10})-\{3,2,1\} & (\ref{bigc})-\{3,2\} \\\hline
 $\omega^{(C_1)}_3$  & (\ref{YB2})-\{3,2,1\} (\ref{YB6})-\{3,1\} (\ref{YB10})-\{3,2,1\} & \\\hline
\end{tabular}
\caption{Required Yang-Baxter and unitarity relations for the entries of $\tilde{C}^{(1)}_{123}$.}
\label{tab:template2}\end{center}\end{table}

We now turn our attention to the operator $\tilde{B}^{(1)}_{123}(\mu)$ whose entries are, 
completely missed by the $L=2$ case are given by,
\begin{equation*}\begin{array}{ll}
& e^{(32)}_1 \otimes e^{(21)}_2 \otimes diag\{\alpha^{(B_1)}_1,\alpha^{(B_1)}_2,\alpha^{(B_1)}_3 \}_3 + e^{(32)}_1 \otimes diag\{\beta^{(B_1)}_1,\beta^{(B_1)}_2,\beta^{(B_1)}_3 \}_2 \otimes e^{(21)}_3\\
+& diag\{\gamma^{(B_1)}_1,\gamma^{(B_1)}_2,\gamma^{(B_1)}_3 \}_1 \otimes e^{(32)}_2 \otimes e^{(21)}_3 + e^{(21)}_1 \otimes e^{(32)}_2 \otimes diag\{\delta^{(B_1)}_1,\delta^{(B_1)}_2,\delta^{(B_1)}_3 \}_3\\
+& e^{(21)}_1 \otimes diag\{\phi^{(B_1)}_1,\phi^{(B_1)}_2,\phi^{(B_1)}_3 \}_2 \otimes e^{(32)}_3 +diag\{\omega^{(B_1)}_1,\omega^{(B_1)}_2,\omega^{(B_1)}_3 \}_1 \otimes e^{(21)}_2 \otimes e^{(32)}_3.
\end{array}\end{equation*}

The technical details entering in the simplifications of these entries are similar to that conducted for the
operator $\tilde{C}^{(1)}_{123}(\mu)$. We therefore restrict ourselves in presenting 
only the required Yang-Baxter and unitarity relations that are necessary for the raw entries 
obtained from the $L=3$ case of Eq.(\ref{twisting}) to 
become the corresponding $L=3$ entries in Eq.(\ref{B1}). This is summarized in Table (\ref{tab:template3}).

\begin{table}\begin{center}
\begin{tabular}{| c | c | c |}
  \hline                       
  Entry &  Y-B equations & Unitarity equations \\\hline
  $\alpha^{(B_1)}_1$  & (\ref{YB3})-\{2,3,1\} (\ref{YB8})-\{1,2\}-\{1,3\} (\ref{YB11})-\{1,3,2\}  & (\ref{biga})-\{1\} \\\hline
  $\alpha^{(B_1)}_2$ & (\ref{YB3})-\{1,3,2\} (\ref{YB7})-\{1,2\} (\ref{YB8})-\{2,3\} (\ref{YB11})-\{1,3,2\} & (\ref{biga})-\{2\} (\ref{bigc})-\{2,1\} \\\hline
 $\alpha^{(B_1)}_3$  & $\begin{array}{c} (\ref{YB2})-\{3,1,2\} (\ref{YB3})-\{2,3,1\}\\ (\ref{YB4})-\{3,2\} (\ref{YB7})-\{1,3\} (\ref{YB11})-\{1,3,2\}\end{array}$   & (\ref{bigc})-\{2,1\}-\{3,1\}-\{3,2\} \\\hline
  $\beta^{(B_1)}_1$  & (\ref{YB3})-\{2,3,1\} (\ref{YB11})-\{1,3,2\} & (\ref{bigc})-\{3,2\} \\\hline
  $\beta^{(B_1)}_2$ & (\ref{YB8})-\{2,3\} (\ref{YB11})-\{1,3,2\} & \\\hline
 $\beta^{(B_1)}_3$  & $\begin{array}{c} (\ref{YB2})-\{3,1,2\} (\ref{YB3})-\{2,3,1\} (\ref{YB4})-\{3,2\}\\ (\ref{YB5})-\{3,2\} (\ref{YB7})-\{2,3\} (\ref{YB11})-\{1,3,2\}\end{array}$  & (\ref{bigc})-\{2,1\} \\\hline
  $\gamma^{(B_1)}_1$  & (\ref{YB11})-\{1,3,2\} &  \\\hline
  $\gamma^{(B_1)}_2$ & (\ref{YB11})-\{1,3,2\} & \\\hline
 $\gamma^{(B_1)}_3$  & (\ref{YB2})-\{3,1,2\} (\ref{YB11})-\{1,3,2\}  &  \\\hline
  $\delta^{(B_1)}_1$  & (\ref{YB2})-\{3,1,2\} (\ref{YB8})-\{1,2\}-\{1,3\} (\ref{YB11})-\{1,3,2\} & (\ref{biga})-\{1\} (\ref{bigc})-\{2,1\} \\\hline
  $\delta^{(B_1)}_2$ & (\ref{YB3})-\{1,3,2\} (\ref{YB4})-\{2,1\} (\ref{YB7})-\{1,2\} (\ref{YB11})-\{1,3,2\} & (\ref{biga})-\{2\} (\ref{bigc})-\{2,1\}\\\hline
 $\delta^{(B_1)}_3$  & (\ref{YB7})-\{1,3\} & (\ref{bigc})-\{2,1\}-\{3,1\} \\\hline
  $\phi^{(B_1)}_1$  & (\ref{YB2})-\{3,1,2\}  & (\ref{bigc})-\{2,1\} \\\hline
  $\phi^{(B_1)}_2$ & (\ref{YB4})-\{2,1\} & (\ref{bigc})-\{2,1\} \\\hline
 $\phi^{(B_1)}_3$  & & (\ref{bigc})-\{2,1\} \\\hline
  $\omega^{(B_1)}_1$  &  & (\ref{bigc})-\{2,1\} \\\hline
  $\omega^{(B_1)}_2$ &  & (\ref{bigc})-\{2,1\} \\\hline
 $\omega^{(B_1)}_3$  & (\ref{YB2})-\{3,1,2\}  & (\ref{bigc})-\{2,1\} \\\hline
\end{tabular}
\caption{Required Yang-Baxter and unitarity relations for the entries of $\tilde{B}^{(1)}_{123}$.}
\label{tab:template3}\end{center}\end{table}

\subsection{The general $L$ case}
At present we just have a  proof for arbitrary $L$ in the case of simplest 
the twisted operator $\tilde{D}_{1\dots L}(\mu)$. 
We now present the proof of Eq.(\ref{D}) for general $L$. Our verification is an adaptation  
of an argument first given in \cite{BOS1} for the equivalent operator. 
To begin we note that the operator $D_{1 \dots L}(\mu)$ can be expressed through,
\begin{equation}
 e^{(33)}_a D_{1 \dots L}(\mu) = e^{(33)}_a \mathcal{T}_{a,1\dots L} (\mu) e^{(33)}_a,
\end{equation}
where the Weyl matrices $e^{(33)}_a$ project out the operator 
$D_{1 \dots L}(\mu)$ from the $3\times 3$ matrix expression in auxiliary space $\mathcal{A}_a$. 

Using the above expression we now consider the action of the $\mathcal{F}$-matrix on $D_{1 \dots L}(\mu)$,
\begin{equation}\begin{array}{lll}
 e^{(33)}_a  \mathcal{F}_{1\dots L} D_{1 \dots L}(\mu)&=& \sum_{\sigma \in S_L}\sum^*_{1 \le \alpha_{\sigma(1)} \dots \alpha_{\sigma(L)} \le 3 } \displaystyle{\bigotimes^L_{i=1}}e^{(\alpha_{\sigma(i)}\alpha_{\sigma(i)})}_{\sigma(i)} e^{(33)}_a \overbrace{R^{\{\sigma \}}_{1\dots L} \mathcal{T}_{a,1\dots L} (\mu)}^{\textrm{apply Eq.(\ref{RonT})}} e^{(33)}_a\\
&=& \sum_{\sigma \in S_L}\sum^*_{1 \le \alpha_{\sigma(1)} \dots \alpha_{\sigma(L)} \le 3 } \displaystyle{\bigotimes^L_{i=1}}e^{(\alpha_{\sigma(i)}\alpha_{\sigma(i)})}_{\sigma(i)} e^{(33)}_a \mathcal{T}_{a,\sigma(1\dots L)}(\mu)e^{(33)}_a R^{\{\sigma \}}_{1\dots L}.
\end{array}\end{equation}
In what follows, we separate the sum over the indices $\alpha_{\sigma(i)}$ according to the number of occurrences where $\alpha_{\sigma(i)}=3$,
\begin{equation}
e^{(33)}_a \mathcal{F}_{1\dots L} D_{1 \dots L}(\mu)= \sum_{\sigma \in S_L}\sum^L_{k=0}\sum^{* (k)}_{1 \le \alpha_{\sigma(1)} \dots \alpha_{\sigma(L)} \le 3 } \bigotimes^L_{i=1}e^{(\alpha_{\sigma(i)}\alpha_{\sigma(i)})}_{\sigma(i)} e^{(33)}_a \mathcal{T}_{a,\sigma(1\dots L)}(\mu)e^{(33)}_a R^{\{\sigma \}}_{1\dots L},
\label{hat2}\end{equation}
where,
\begin{equation}
 \sum^{* (k)}_{1 \le \alpha_{\sigma(1)} \dots \alpha_{\sigma(L)} \le 3 } =  \sum^{*}_{\genfrac{}{}{0pt}{}{1 \le \alpha_{\sigma(1)} \dots \alpha_{\sigma(L)} \le 3}{ (\alpha_{\sigma(1)},\dots,\alpha_{\sigma(L-k)})\in\{1,2\} \textrm{  ,  } (\alpha_{\sigma(L-k+1)},\dots,\alpha_{\sigma(L)})=3 }  }.
\label{hat1}\end{equation}
Hence the tensor product of the Weyl matrices $\displaystyle{\bigotimes^L_{i=1}}e^{(\alpha_{\sigma(i)}\alpha_{\sigma(i)})}_{\sigma(i)}$ for each value of the index $k$ becomes,
\begin{equation*}
 \bigotimes^L_{i=1}e^{(\alpha_{\sigma(i)}\alpha_{\sigma(i)})}_{\sigma(i)} =  \bigotimes^{L-k}_{i=1}e^{(\alpha_{\sigma(i)}\alpha_{\sigma(i)})}_{\sigma(i)} \bigotimes^{L}_{i=L-k+1}e^{(33)}_{\sigma(i)},
\end{equation*}
leading to the following form for Eq.(\ref{hat2}),
\begin{equation}\begin{array}{lll}
e^{(33)}_a \mathcal{F}_{1\dots L} D_{1 \dots L}(\mu) &=& \sum_{\sigma \in S_L}\sum^L_{k=0}\sum^{* (k)}_{1 \le \alpha_{\sigma(1)} \dots \alpha_{\sigma(L)} \le 3 } \displaystyle{\bigotimes^L_{i=1}}e^{(\alpha_{\sigma(i)}\alpha_{\sigma(i)})}_{\sigma(i)}\\
&&\times e^{(33)}_a e^{(33)}_{\sigma(L)}R_{a \sigma(L)} (\mu) \dots e^{(33)}_{\sigma(L-k+1)}R_{a \sigma(L-k+1)} (\mu) \mathcal{T}_{a,\sigma(1\dots (L-k))}(\mu)e^{(33)}_a R^{\{\sigma \}}_{1\dots L}.
\end{array}\label{hat3}\end{equation}

In the above expression we realize that the action of the Weyl operators on the $R$-matrices leads to,
\begin{equation*}
e^{(33)}_a e^{(33)}_{\sigma(L)}R_{a \sigma(L)} (\mu) \dots e^{(33)}_{\sigma(L-k+1)}R_{a \sigma(L-k+1)} (\mu) = \prod^L_{i=L-k+1}a_3(\mu,\xi_{\sigma(i)}) e^{(33)}_a \bigotimes^{L}_{j=L-k+1}e^{(33)}_{\sigma(j)},
\end{equation*}
and hence Eq.(\ref{hat3}) becomes,
\begin{equation*}\begin{array}{l}
e^{(33)}_a \mathcal{F}_{1\dots L} D_{1 \dots L}(\mu) = \sum_{\sigma \in S_L}\sum^L_{k=0}\sum^{* (k)}_{1 \le \alpha_{\sigma(1)} \dots \alpha_{\sigma(L)} \le 3 } \prod^{L}_{i=L-k+1}a_3(\mu,\xi_{\sigma(i)})  e^{(33)}_a \\
\times e^{(\alpha_{\sigma(L-k)}\alpha_{\sigma(L-k)})}_{\sigma(L-k)}R_{a \sigma(L-k)} (\mu) \dots e^{(\alpha_{\sigma(1)}\alpha_{\sigma(1)})}_{\sigma(1)}R_{a \sigma(1)} (\mu) e^{(33)}_a  \displaystyle{\bigotimes^{L}_{j=L-k+1}}e^{(33)}_{\sigma(j)}R^{\{\sigma \}}_{1\dots L}.
\end{array}\end{equation*}
Since $\alpha_{\sigma(i)}\in \{1,2\}$, $i=(1,\dots,L-k)$, in the above expression we realize that the action of the Weyl operators on the $R$-matrices leads to the following simplified expression,
\begin{equation*}\begin{array}{ll}
& e^{(33)}_a e^{(\alpha_{\sigma(L-k)}\alpha_{\sigma(L-k)})}_{\sigma(L-k)}R_{a \sigma(L-k)} (\mu) \dots e^{(\alpha_{\sigma(1)}\alpha_{\sigma(1)})}_{\sigma(1)}R_{a \sigma(1)} (\mu) e^{(33)}_a\\
=& e^{(33)}_a \prod^{L-k}_{i=1}b_{3 \alpha_{\sigma(i)}}(\mu,\xi_{\sigma(i)}) \displaystyle{\bigotimes^{L-k}_{j=1}}e^{(\alpha_{\sigma(j)}\alpha_{\sigma(j)})}_{\sigma(j)}.
\end{array}\end{equation*}
The above considerations lead the following result,
\begin{equation*}\begin{array}{lll}
\mathcal{F}_{1\dots L} D_{1 \dots L}(\mu) &=&  \sum_{\sigma \in S_L}\sum^L_{k=0}\sum^{* (k)}_{1 \le \alpha_{\sigma(1)} \dots \alpha_{\sigma(L)} \le 3 } \prod^{L-k}_{i=1}b_{3 \alpha_{\sigma(i)}}(\mu,\xi_{\sigma(i)}) \\
&&  \prod^{L}_{j=L-k+1}a_3(\mu,\xi_{\sigma(j)}) \displaystyle{\bigotimes^{L}_{l=1}}e^{(\alpha_l \alpha_l)}_{l}R^{\{\sigma \}}_{1\dots L}\\
&=&  \displaystyle{\bigotimes^L_{i=1}} diag\left\{b_{31}(\mu,\xi_i),b_{32}(\mu,\xi_i), a_{3}(\mu,\xi_i) \right\}_i \mathcal{F}_{1 \dots L},
\end{array}\end{equation*}
ultimately verifying Eq.(\ref{D}) for general $L$.

We do not have a proof for general $L$ in the cases of the twisted $B$ and $C$ operators. The necessary underlying
recurrence relations to carry out such demonstrations have thus far eluded us. However, with the help of the Yang-Baxter
solution presented in Section  \ref{VERTEX} we have been able to verify Eqs.(\ref{C2}-\ref{B1}) for $L=4$ explicitly. 
Considering the generality of such solution, this is strong evidence 
supporting our conjectured expressions Eqs.(\ref{C2}-\ref{B1}) for general $L$.

\section{Basic domain wall partition functions}\label{DOMAIN}

The purpose of this section is to start the formulation
to compute certain domain wall partition functions
(DWPF's) associated to the $N=3$ vertex model
with arbitrary Boltzmann weights. The domain
wall boundary conditions correspond to certain fixed statistical configurations for the
horizontal and vertical edges at the top
and bottom of the square lattice, see for instance
\cite{KO2,KARA,DOW}. From an algebraic perspective
these objects can be expressed in terms of
the expectation values of combinations of
the creation and annihilation operators
${B}^{(i)}_{1\dots L}(\mu)$ and
${C}^{(i)}_{1\dots L}(\nu)$ on some pseudo-vacuum states \cite{KO2,MA2,PRO,ZH3}.
In Figures \ref{DWPFC} and \ref{DWPFB} we have depicted
the graphical representation of such domain wall partition
functions.
\setlength{\unitlength}{3500sp}
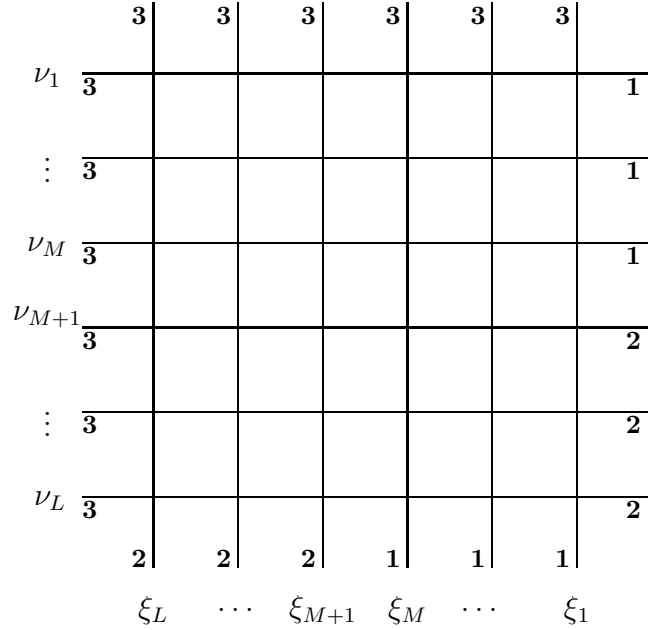
\begin{figure}[ht]
\begin{center}
\begin{picture}(6537,3912)(2000,-6361)
\put(3800,-2350){\makebox(0,0){\fontsize{10}{12}\selectfont \textbf{3}}}
\put(4400,-2350){\makebox(0,0){\fontsize{10}{12}\selectfont \textbf{3}}}
\put(5000,-2350){\makebox(0,0){\fontsize{10}{12}\selectfont \textbf{3}}}
\put(5600,-2350){\makebox(0,0){\fontsize{10}{12}\selectfont \textbf{3}}}
\put(6200,-2350){\makebox(0,0){\fontsize{10}{12}\selectfont \textbf{3}}}
\put(6800,-2350){\makebox(0,0){\fontsize{10}{12}\selectfont \textbf{3}}}
\put(3800,-6200){\makebox(0,0){\fontsize{10}{12}\selectfont \textbf{2}}}
\put(4400,-6200){\makebox(0,0){\fontsize{10}{12}\selectfont \textbf{2}}}
\put(5000,-6200){\makebox(0,0){\fontsize{10}{12}\selectfont \textbf{2}}}
\put(5600,-6200){\makebox(0,0){\fontsize{10}{12}\selectfont \textbf{1}}}
\put(6200,-6200){\makebox(0,0){\fontsize{10}{12}\selectfont \textbf{1}}}
\put(6800,-6200){\makebox(0,0){\fontsize{10}{12}\selectfont \textbf{1}}}
\put(3450,-5850){\makebox(0,0){\fontsize{10}{12}\selectfont \textbf{3}}}
\put(3450,-5250){\makebox(0,0){\fontsize{10}{12}\selectfont \textbf{3}}}
\put(3450,-4650){\makebox(0,0){\fontsize{10}{12}\selectfont \textbf{3}}}
\put(3450,-4050){\makebox(0,0){\fontsize{10}{12}\selectfont \textbf{3}}}
\put(3450,-3450){\makebox(0,0){\fontsize{10}{12}\selectfont \textbf{3}}}
\put(3450,-2850){\makebox(0,0){\fontsize{10}{12}\selectfont \textbf{3}}}
\put(7300,-5850){\makebox(0,0){\fontsize{10}{12}\selectfont \textbf{2}}}
\put(7300,-5250){\makebox(0,0){\fontsize{10}{12}\selectfont \textbf{2}}}
\put(7300,-4650){\makebox(0,0){\fontsize{10}{12}\selectfont \textbf{2}}}
\put(7300,-4050){\makebox(0,0){\fontsize{10}{12}\selectfont \textbf{1}}}
\put(7300,-3450){\makebox(0,0){\fontsize{10}{12}\selectfont \textbf{1}}}
\put(7300,-2850){\makebox(0,0){\fontsize{10}{12}\selectfont \textbf{1}}}
\put(3401,-2761){\line( 1, 0){4000}}
\put(3401,-3361){\line( 1, 0){4000}}
\put(3401,-3961){\line( 1, 0){4000}}
\put(3401,-4561){\line( 1, 0){4000}}
\put(3401,-5161){\line( 1, 0){4000}}
\put(3401,-5761){\line( 1, 0){4000}}
\put(3901,-6261){\line( 0, 1){4000}}
\put(4501,-6261){\line( 0, 1){4000}}
\put(5101,-6261){\line( 0, 1){4000}}
\put(6301,-6261){\line( 0, 1){4000}}
\put(6901,-6261){\line( 0, 1){4000}}
\put(5701,-6261){\line( 0, 1){4000}}
\put(3170,-5786){\makebox(0,0){$\nu_L$}}
\put(3150,-5186){\makebox(0,0){$\vdots$}}
\put(3150,-4486){\makebox(0,0){$\nu_{M+1}$}}
\put(3150,-3986){\makebox(0,0){$\nu_M$}}
\put(3150,-3386){\makebox(0,0){$\vdots$}}
\put(3150,-2786){\makebox(0,0){$\nu_1$}}
\put(3901,-6561){\makebox(0,0){$\xi_L$}}
\put(4501,-6561){\makebox(0,0){$\dots$}}
\put(5101,-6561){\makebox(0,0){$\xi_{M+1}$}}
\put(5701,-6561){\makebox(0,0){$\xi_M$}}
\put(6226,-6561){\makebox(0,0){$\dots$}}
\put(6901,-6561){\makebox(0,0){$\xi_1$}}
\end{picture} \par
\end{center}
\caption{Graphical representation of a mixed DWPF of type C.}\label{DWPFC}
\end{figure}
\setlength{\unitlength}{3500sp}
\begin{figure}[ht]
\begin{center}
\begin{picture}(6537,3912)(2000,-6361)
\put(3800,-2350){\makebox(0,0){\fontsize{10}{12}\selectfont \textbf{1}}}
\put(4400,-2350){\makebox(0,0){\fontsize{10}{12}\selectfont \textbf{1}}}
\put(5000,-2350){\makebox(0,0){\fontsize{10}{12}\selectfont \textbf{1}}}
\put(5600,-2350){\makebox(0,0){\fontsize{10}{12}\selectfont \textbf{2}}}
\put(6200,-2350){\makebox(0,0){\fontsize{10}{12}\selectfont \textbf{2}}}
\put(6800,-2350){\makebox(0,0){\fontsize{10}{12}\selectfont \textbf{2}}}
\put(3800,-6200){\makebox(0,0){\fontsize{10}{12}\selectfont \textbf{3}}}
\put(4400,-6200){\makebox(0,0){\fontsize{10}{12}\selectfont \textbf{3}}}
\put(5000,-6200){\makebox(0,0){\fontsize{10}{12}\selectfont \textbf{3}}}
\put(5600,-6200){\makebox(0,0){\fontsize{10}{12}\selectfont \textbf{3}}}
\put(6200,-6200){\makebox(0,0){\fontsize{10}{12}\selectfont \textbf{3}}}
\put(6800,-6200){\makebox(0,0){\fontsize{10}{12}\selectfont \textbf{3}}}
\put(3450,-5850){\makebox(0,0){\fontsize{10}{12}\selectfont \textbf{1}}}
\put(3450,-5250){\makebox(0,0){\fontsize{10}{12}\selectfont \textbf{1}}}
\put(3450,-4650){\makebox(0,0){\fontsize{10}{12}\selectfont \textbf{1}}}
\put(3450,-4050){\makebox(0,0){\fontsize{10}{12}\selectfont \textbf{2}}}
\put(3450,-3450){\makebox(0,0){\fontsize{10}{12}\selectfont \textbf{2}}}
\put(3450,-2850){\makebox(0,0){\fontsize{10}{12}\selectfont \textbf{2}}}
\put(7300,-5850){\makebox(0,0){\fontsize{10}{12}\selectfont \textbf{3}}}
\put(7300,-5250){\makebox(0,0){\fontsize{10}{12}\selectfont \textbf{3}}}
\put(7300,-4650){\makebox(0,0){\fontsize{10}{12}\selectfont \textbf{3}}}
\put(7300,-4050){\makebox(0,0){\fontsize{10}{12}\selectfont \textbf{3}}}
\put(7300,-3450){\makebox(0,0){\fontsize{10}{12}\selectfont \textbf{3}}}
\put(7300,-2850){\makebox(0,0){\fontsize{10}{12}\selectfont \textbf{3}}}
\put(3401,-2761){\line( 1, 0){4000}}
\put(3401,-3361){\line( 1, 0){4000}}
\put(3401,-3961){\line( 1, 0){4000}}
\put(3401,-4561){\line( 1, 0){4000}}
\put(3401,-5161){\line( 1, 0){4000}}
\put(3401,-5761){\line( 1, 0){4000}}
\put(3901,-6261){\line( 0, 1){4000}}
\put(4501,-6261){\line( 0, 1){4000}}
\put(5101,-6261){\line( 0, 1){4000}}
\put(6301,-6261){\line( 0, 1){4000}}
\put(6901,-6261){\line( 0, 1){4000}}
\put(5701,-6261){\line( 0, 1){4000}}
\put(3100,-5786){\makebox(0,0){$\mu_1$}}
\put(3100,-5186){\makebox(0,0){$\vdots$}}
\put(3100,-4486){\makebox(0,0){$\mu_{M}$}}
\put(3100,-3986){\makebox(0,0){$\mu_{M+1}$}}
\put(3100,-3386){\makebox(0,0){$\vdots$}}
\put(3100,-2786){\makebox(0,0){$\mu_L$}}
\put(3901,-6561){\makebox(0,0){$\xi_L$}}
\put(4501,-6561){\makebox(0,0){$\dots$}}
\put(5101,-6561){\makebox(0,0){$\xi_{M+1}$}}
\put(5701,-6561){\makebox(0,0){$\xi_M$}}
\put(6226,-6561){\makebox(0,0){$\dots$}}
\put(6901,-6561){\makebox(0,0){$\xi_1$}}
\end{picture} \par
\end{center}
\caption{Graphical representation of a mixed DWPF of type B.}\label{DWPFB}
\end{figure}
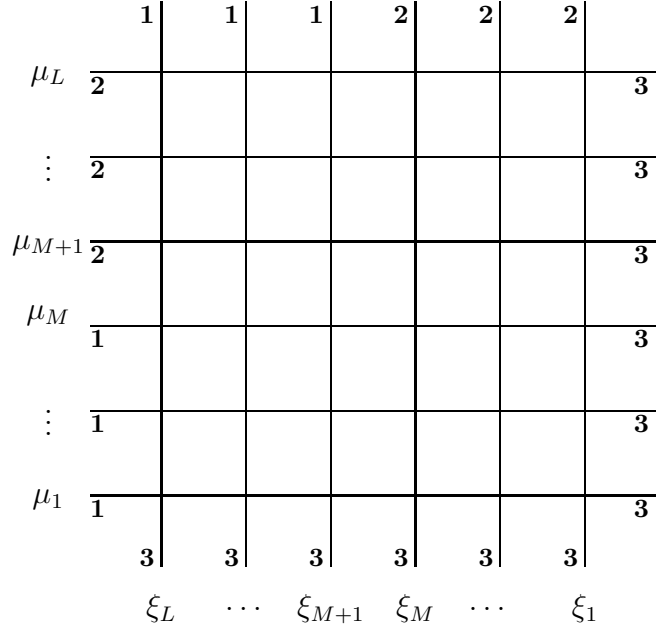

In order to compute  such partition functions we need to first define
the corresponding reference states as well as their properties under the action of the $F$-matrix.
The $N=3$ state vertex model has three possible pseudovacuum states 
which we label as follows,
\begin{equation}\begin{array}{lllll}
|1 \rangle_{\alpha} = \left(\begin{array}{l} 1\\0\\0\end{array} \right)_{\alpha}&, & |2 \rangle_{\alpha} = \left(\begin{array}{l} 0\\1\\0\end{array} \right)_{\alpha}&,&|3 \rangle_{\alpha} = \left(\begin{array}{l} 0\\0\\1\end{array} \right)_{\alpha},
\end{array}\label{reference}\end{equation}
for $\alpha=1,\dots,L$ while the corresponding transpose states are denoted by,
\begin{equation}\begin{array}{lllll}
\,_{\alpha}\langle 1 | = \left(1,0,0 \right)_{\alpha} &,& \,_{\alpha}\langle 2 | = \left(0,1,0 \right)_{\alpha}&,& \,_{\alpha}\langle 3 | = \left(0,0,1 \right)_{\alpha}.
\end{array}\label{treference}\end{equation}

We also use the following convenient notational conventions with the outer products of reference states,
\begin{equation*} \begin{array}{ll}
 |i\rangle_{\alpha_1 \dots \alpha_p} = \displaystyle{\bigotimes^p_{j=1}}|i\rangle_{\alpha_j} & i = (1,2,3),
\end{array}\end{equation*}
and similarly with the transpose reference states.

One then finds that the action of a Weyl basis element on a reference state and a transpose reference state respectively is given by,
\begin{equation}\begin{array}{lll}
 e^{(ij)}_{\alpha}|k \rangle_{\alpha}  = |i \rangle_{\alpha}  \delta_{jk} &,& \,_{\alpha}\langle i | e^{(jk)}_{\alpha} = \,_{\alpha}\langle k | \delta_{ij}.
\end{array}\label{action}\end{equation}
Using the above observation we give the following results:
\begin{proposition}
\begin{equation}\begin{array}{lll}
 \mathcal{F}_{1\dots L} |i \rangle_{1\dots L}  = |i \rangle_{1 \dots L} &,&  \,_{1\dots L}\langle i |\mathcal{F}_{1\dots L} =  \,_{1 \dots L}\langle i | \\
 \mathcal{F}^{-1}_{1\dots L} |i \rangle_{1\dots L}  = |i \rangle_{1 \dots L} &,&  \,_{1\dots L}\langle i |\mathcal{F}^{-1}_{1\dots L} =  \,_{1 \dots L}\langle i |
\end{array}\label{actref}\end{equation}
\end{proposition}
\textbf{Proof.} From Eq.(\ref{action}) we notice that the only element of 
the expression $\displaystyle{\bigotimes^{L}_{j=1}}e^{(\alpha_j \alpha_j)}_j$ 
in Eq.(\ref{ansatz}) which acts non trivially on the reference state 
$|i \rangle_{1\dots L}$ and its transpose $\,_{1 \dots L}\langle i |$ 
is when $\alpha_1 = \alpha_2 = \dots = \alpha_L = i$. From our detailed analysis of 
Eq.(\ref{ansatz}) in Section (\ref{theojust}) we know that this particular set of $\alpha$ values 
in the sum of Eq.(\ref{summ}) only occurs when 
$\sigma = \mathcal{I}$  is the trivial permutation. 
Since $R^{\{ \mathcal{I}\}}_{1 \dots L} = \mathcal{I}_{1} \otimes \dots \otimes \mathcal{I}_L$, 
this proves the first two relations in Eq.(\ref{actref}).

Focusing on the final two results, from the above analysis we 
find that the 1st, $\frac{3^L+1}{2}$th and $3^L$th row and column of 
$\mathcal{F}_{1 \dots L}$ are given explicitly by 1 in the respective diagonal 
position, and zero everywhere else. Noting the following cofactor values concerning the 1st row and column of $\mathcal{F}_{1\dots L}$,
\begin{equation*}\begin{array}{llcl}
\textrm{det}\left\{  \left( \mathcal{F}_{1\dots L} \right)_{kl} \right\}_{\genfrac{}{}{0pt}{}{k=2 \dots 3^L}{l=2 \dots 3^L}} &=& \textrm{det}\left\{  \mathcal{F}_{1\dots L}  \right\}\\
\textrm{det}\left\{  \left( \mathcal{F}_{1\dots L} \right)_{kl} \right\}_{\genfrac{}{}{0pt}{}{k=2 \dots 3^L}{l=1\dots (i-1)(i+1) \dots 3^L}} &=& 0 &\textrm{for $i \in \{2,\dots, 3^L \}$}\\
\textrm{det}\left\{  \left( \mathcal{F}_{1\dots L} \right)_{kl} \right\}_{\genfrac{}{}{0pt}{}{k=1\dots (j-1)(j+1) \dots 3^L}{l=2 \dots 3^L}} &=& 0 &\textrm{for $j \in \{2,\dots, 3^L \}$},
\end{array}
\end{equation*}
and similarly equivalent values for the $\frac{3^L+1}{2}$th and $3^L$th row and column of $\mathcal{F}_{1\dots L}$. 

We now recall Cramer's rule for the inverse of a matrix,
\begin{equation}
\left( \mathcal{F}^{-1}_{1\dots L} \right)_{ij} = \frac{(-1)^{i+j}}{\textrm{det}\left\{ \mathcal{F}_{1\dots L} \right\}}\textrm{det}\left\{  \left( \mathcal{F}_{1\dots L} \right)_{kl} \right\}_{\genfrac{}{}{0pt}{}{k=1\dots j-1,j+1 \dots 3^L}{l=1\dots i-1,i+1\dots 3^L} }.
\label{Cramer}\end{equation}
Hence applying this equation we obtain,
\begin{equation*}\begin{array}{llcl}
\left( \mathcal{F}^{-1}_{1\dots L} \right)_{11} &=& 1\\
\left( \mathcal{F}^{-1}_{1\dots L} \right)_{i1} &=& 0 &\textrm{for $i \in \{2,\dots, 3^L \}$}\\
\left( \mathcal{F}^{-1}_{1\dots L} \right)_{1j} &=& 0 &\textrm{for $j \in \{2,\dots, 3^L \}$},
\end{array}
\end{equation*}
and similarly equivalent results for the $\frac{3^L+1}{2}$th and $3^L$th row and column of $\mathcal{F}_{1 \dots L}$ - thus verifying the last two results of this proposition. $\square$

We now start by computing the four possible basic non-trivial DWPF's consisting of 
the expectation values of tensor product of operators carrying a single charge index. These are the building
blocks required to construct the domain wall partition functions involving mixed fields such as  
$C^{(1)}_{1\dots L}(\nu)$ and $C^{(2)}_{1\dots L}(\nu)$ 
or
$B^{(1)}_{1\dots L}(\mu)$ and $B^{(2)}_{1\dots L}(\mu)$. 

\subsection{Single DWPF for  $C^{(2)}$}\label{useall}
We now consider our first basic DWPF, labeled $Z^{(C,2)}_L$, which is given explicitly by,
\begin{equation}
 Z^{(C,2)}_L (\{ \nu \},\{\xi\}) = \,_{1\dots L}\langle 2 | C^{(2)}_{1\dots L}(\nu_L) \dots C^{(2)}_{1\dots L}(\nu_1) |3 \rangle_{1\dots L}.
\label{ZC2}\end{equation}
Applying the $\mathcal{F}$-matrix similarity transform to each of the $C^{(2)}$ operators in the above expression, and using the results of Eq.(\ref{actref}), Eq.(\ref{ZC2}) immediately becomes,
\begin{equation}
 Z^{(C,2)}_L (\{ \nu \},\{\xi\}) = \,_{1\dots L}\langle 2 | \tilde{C}^{(2)}_{1\dots L}(\nu_L) \dots \tilde{C}^{(2)}_{1\dots L}(\nu_1) |3 \rangle_{1\dots L}.
\label{ZC2a}\end{equation}
We now are going to obtain a recurrence relation for the above expression. This is done by inserting a 
complete set of states between the operators $\tilde{C}^{(2)}_{1\dots L}(\nu_2)$ and $\tilde{C}^{(2)}_{1\dots L}(\nu_1)$ to obtain,
\begin{equation}\begin{array}{lll}
 Z^{(C,2)}_L (\{ \nu \},\{\xi\}) &= & \sum^L_{p=1} \,_{1\dots L}\langle 2 | \tilde{C}^{(2)}_{1\dots L}(\nu_L) \dots \tilde{C}^{(2)}_{1\dots L}(\nu_2) |2 \rangle_{p} |3 \rangle_{\genfrac{}{}{0pt}{}{ 1\dots L}{\ne p}} \\
&& \times\,_{p}\langle 2 | \,_{\genfrac{}{}{0pt}{}{ 1\dots L}{\ne p}}\langle 3 |\tilde{C}^{(2)}_{1\dots L}(\nu_1) |3 \rangle_{1\dots L}.
\end{array}\label{ZC2b}\end{equation}
It is elementary to show that only the above terms of the complete 
set of states are non zero by considering the action of $e^{(23)}$ on the reference states.
Through the property 
$ \,_{p}\langle 2 | \,_{\genfrac{}{}{0pt}{}{ 1\dots L}{\ne p}}\langle 3 | e^{(23)}_l  =  \,_{1 \dots L}\langle 3 |\delta_{p l}$,
one is able to derive the following identity, 
\begin{equation}
 \,_{p}\langle 2 | \,_{\genfrac{}{}{0pt}{}{ 1\dots L}{\ne p}}\langle 3 |\tilde{C}^{(2)}_{1\dots L}(\nu_1) |3 \rangle_{1\dots L} = c_{32}(\nu_1,\xi_p) \prod^L_{\genfrac{}{}{0pt}{}{ i=1}{\ne p}} a_3(\nu_1,\xi_i) \theta_3(\xi_i, \xi_p).
\end{equation}
We now focus on the expression $\,_{1\dots L}\langle 2 | \tilde{C}^{(2)}_{1\dots L}(\nu_L) \dots \tilde{C}^{(2)}_{1\dots L}(\nu_2) |2 \rangle_{p} |3 \rangle_{\genfrac{}{}{0pt}{}{ 1\dots L}{\ne p}}$. Using the fact that $e^{(23)}_p|2 \rangle_{p}=0$, we can discard the value $p$ in the summations of the $\tilde{C}^{(2)}_{1\dots L}(\nu)$ operators. Hence applying this fact, and elementary matrix multiplication, the expression now becomes,
\begin{equation}\begin{array}{ll}
&\,_{1\dots L}\langle 2 | \tilde{C}^{(2)}_{1\dots L}(\nu_L) \dots \tilde{C}^{(2)}_{1\dots L}(\nu_2) |2 \rangle_{p} |3 \rangle_{\genfrac{}{}{0pt}{}{ 1\dots L}{\ne p}}\\
=&\sum^L_{\genfrac{}{}{0pt}{}{ l_2, \dots, l_L=1}{l_2 \ne \dots \ne l_L \ne p}} \prod^{L}_{j=2}\frac{b_{32}(\nu_j,\xi_p)}{b_{32}(\xi_{l_j},\xi_p)\theta_2(\xi_p,\xi_{l_j})}\,_{1\dots L}\langle 2 | \tilde{C}^{(2,l_L)}_{\genfrac{}{}{0pt}{}{ 1 \dots L}{\ne p}}(\nu_L) \dots \tilde{C}^{(2,l_2)}_{\genfrac{}{}{0pt}{}{ 1 \dots L}{\ne p}}(\nu_2) |2 \rangle_{p} |3 \rangle_{\genfrac{}{}{0pt}{}{ 1\dots L}{\ne p}},
\end{array}
\label{ZC2f}\end{equation}
where,
\begin{equation} 
\tilde{C}^{(2,l)}_{\genfrac{}{}{0pt}{}{ 1 \dots L}{\ne p}}(\nu)  = 
c_{32}(\nu,\xi_l)  e^{(23)}_l \bigotimes^{L}_{\genfrac{}{}{0pt}{}{i=1}{i\ne l,p}} diag\left\{b_{21}(\nu,\xi_i),
\frac{b_{32}(\nu,\xi_i)}{b_{32}(\xi_l,\xi_i)\theta_2(\xi_i,\xi_l)},a_3(\nu,\xi_i)\theta_3(\xi_i,\xi_l)  \right\}_i .
\label{auxC2L}
\end{equation}
We now offer the following comments on Eq.(\ref{ZC2f}). Firstly, the product 
$\prod^{L}_{j=2}\frac{b_{32}(\nu_j,\xi_p)}{b_{32}(\xi_{l_j},\xi_p)\theta_2(\xi_p,\xi_{l_j})}$ is 
independent of the value of $l_2, \dots, l_L$ due to the condition $l_2 \ne \dots \ne l_L \ne p$. 
Secondly, the operators 
$\tilde{C}^{(2,l)}_{\genfrac{}{}{0pt}{}{ 1 \dots L}{\ne p}}(\nu)$ act trivially on 
the vector space $V_p$, meaning that we can decrease the number of relevant vector spaces 
in the reference states by one. Thirdly, when the aforementioned coefficient is taken out of 
the sum over $l_2, \dots, l_L$, the following simplification occurs,
\begin{equation*}\begin{array}{lll}
\sum^L_{\genfrac{}{}{0pt}{}{ l_2, \dots, l_L=1}{l_2 \ne \dots \ne l_L \ne p}} \,_{\genfrac{}{}{0pt}{}{ 1\dots L}{\ne p}}\langle 2 | \tilde{C}^{(2,l_L)}_{\genfrac{}{}{0pt}{}{ 1 \dots L}{\ne p}}(\nu_L) \dots \tilde{C}^{(2,l_2)}_{\genfrac{}{}{0pt}{}{ 1 \dots L}{\ne p}}(\nu_2)  |3 \rangle_{\genfrac{}{}{0pt}{}{ 1\dots L}{\ne p}}&=& \,_{\genfrac{}{}{0pt}{}{ 1\dots L}{\ne p}}\langle 2 | \tilde{C}^{(2)}_{\genfrac{}{}{0pt}{}{ 1 \dots L}{\ne p}}(\nu_L) \dots \tilde{C}^{(2)}_{\genfrac{}{}{0pt}{}{ 1 \dots L}{\ne p}}(\nu_2)  |3 \rangle_{\genfrac{}{}{0pt}{}{ 1\dots L}{\ne p}}\\
&=&  Z^{(C,2)}_{L-1} (\{ \nu \},\{\xi\}|\hat{\nu}_1,\hat{\xi}_p) .
\end{array}
\end{equation*}
Hence Eq.(\ref{ZC2b}) becomes the following recurrence relation,
\begin{equation}\begin{array}{lll}
 Z^{(C,2)}_L (\{ \nu \},\{\xi\}) &= & \sum^L_{p=1} c_{32}(\nu_1,\xi_p) \prod^L_{\genfrac{}{}{0pt}{}{ i=1}{\ne p}} \frac{a_3(\nu_1,\xi_i) \theta_3(\xi_i, \xi_p)}{b_{32}(\xi_i,\xi_p) \theta_2(\xi_p,\xi_i)}  \prod^L_{j=2}b_{32}(\nu_j,\xi_p)\\
&& \times Z^{(C,2)}_{L-1} (\{ \nu \},\{\xi\}|\hat{\nu}_1,\hat{\xi}_p).
\end{array}\label{ZC2rec}\end{equation}
\subsubsection{Exact solution}\label{secC2DW}
We now can apply induction to verify the complete algebraic expression 
for the $C^{(2)}$ type DWPF in terms of Boltzmann weights. To begin for $L=1$ and $L=2$ we have,
\begin{equation*}\begin{array}{lll}
Z^{(C,2)}_1 ( \nu_1, \xi_1) &=& c_{32}(\nu_1,\xi_1)\\
Z^{(C,2)}_2 (\{ \nu \},\{\xi\}) &=& \frac{c_{32}(\nu_1,\xi_1)c_{32}(\nu_2,\xi_2)a_{3}(\nu_1,\xi_2)\theta_{3}(\xi_2,\xi_1)b_{32}(\nu_2,\xi_1)}{b_{32}(\xi_2,\xi_1)\theta_{2}(\xi_1,\xi_2)}\\
&&+\frac{c_{32}(\nu_1,\xi_2)c_{32}(\nu_1,\xi_2)a_{3}(\nu_1,\xi_1)\theta_{3}(\xi_1,\xi_2)b_{32}(\nu_2,\xi_2)}{b_{32}(\xi_1,\xi_2)\theta_{2}(\xi_2,\xi_1)}
\end{array}\end{equation*}
We now offer the following general result. 
\begin{proposition}\label{propexpDWPF3}
 \begin{equation}\begin{array}{l}
 Z^{(C,2)}_L(\{\nu\},\{\xi\})=   \sum_{\sigma \in S_L} \prod^L_{i=1}c_{32}(\nu_i,\xi_{\sigma(i)})   \prod_{1 \le j < k \le L} \frac{a_3(\nu_j,\xi_{\sigma(k)})b_{32}(\nu_k,\xi_{\sigma(j)}) \theta_3(\xi_{\sigma(k)},\xi_{\sigma(j)}) }{b_{32}(\xi_{\sigma(k)},\xi_{\sigma(j)})\theta_2(\xi_{\sigma(j)},\xi_{\sigma(k)})}
\end{array}\label{explicitZC2} \end{equation}
\end{proposition}
\textbf{Proof.} Noting that the above formula is correct for $L = 1,2$, we assume 
that it holds for some $L$, and focus on the $L+1$ case of Eq.(\ref{ZC2rec}),
\begin{equation*}
\begin{array}{lll}
Z^{(C,2)}_{L+1}&=&   \sum^{L+1}_{p=1} c_{32}(\nu_1,\xi_p) \prod^{L+1}_{\genfrac{}{}{0pt}{}{ m=1}{\ne p}} \frac{a_3(\nu_1,\xi_m) \theta_3(\xi_m, \xi_p)}{b_{32}(\xi_m,\xi_p) \theta_2(\xi_p,\xi_m)}  \prod^{L+1}_{n=2}b_{32}(\nu_n,\xi_p)\\
&& \times    \sum_{\sigma \in S^{(p)}_L} \prod^{L+1}_{i=2}c_{32}(\nu_i,\xi_{\sigma(i)})   \prod_{2 \le j < k \le L+1} \frac{a_3(\nu_j,\xi_{\sigma(k)})b_{32}(\nu_k,\xi_{\sigma(j)}) \theta_3(\xi_{\sigma(k)},\xi_{\sigma(j)}) }{b_{32}(\xi_{\sigma(k)},\xi_{\sigma(j)}) \theta_2(\xi_{\sigma(j)},\xi_{\sigma(k)})},
\end{array}\end{equation*}
where the sum over the permutations with superscript $p$ is given by,
\begin{equation*}\begin{array}{lll}
 {\displaystyle \sum_{\sigma \in S^{(p)}_L} \equiv \sum^{L+1}_{\genfrac{}{}{0pt}{}{\sigma_2=1}{\ne p} } \sum^{L+1}_{\genfrac{}{}{0pt}{}{\sigma_3=1}{\ne p} } \dots \sum^{L+1}_{\genfrac{}{}{0pt}{}{\sigma_{L+1}=1}{\ne p} }} &\textrm{for}& \sigma_2 \ne \sigma_3 \ne \dots \ne \sigma_{L+1}  .
\end{array}\end{equation*}
The verification of the proposition follows immediately through the change in label, $ p \rightarrow \sigma_{1}$. $\square$

\subsection{Single DWPF for $B^{(2)}$}
We now consider the equivalent DWPF expression for the $B^{(2)}_{1\dots L}$ operators which is given explicitly by,
\begin{equation}\begin{array}{lll}
 Z^{(B,2)}_L (\{ \mu \},\{\xi\})&=& \,_{1\dots L}\langle 3 | B^{(2)}_{1\dots L}(\mu_1) \dots B^{(2)}_{1\dots L}(\mu_L) |2 \rangle_{1\dots L}\\
&=& \,_{1\dots L}\langle 3 | \tilde{B}^{(2)}_{1\dots L}(\mu_1) \dots \tilde{B}^{(2)}_{1\dots L}(\mu_L) |2 \rangle_{1\dots L},
\end{array}\label{ZB2}\end{equation}
where we have applied the results of Eq.(\ref{actref}) to twist the operators.
As with the previous DWPF, we now obtain a recurrence relation for the above expression. We insert a complete set of states between the operators $\tilde{B}^{(2)}_{1\dots L}(\mu_1)$ and $\tilde{B}^{(2)}_{1\dots L}(\mu_2)$ to obtain,
\begin{equation}\begin{array}{lll}
 Z^{(B,2)}_L (\{ \mu \},\{\xi\}) &= & \sum^L_{p=1} \,_{1\dots L}\langle 3 | \tilde{B}^{(2)}_{1\dots L}(\mu_1) |2 \rangle_{p} |3 \rangle_{\genfrac{}{}{0pt}{}{ 1\dots L}{\ne p}} \\
&& \times\,_{p}\langle 2 | \,_{\genfrac{}{}{0pt}{}{ 1\dots L}{\ne p}}\langle 3 |\tilde{B}^{(2)}_{1\dots L}(\mu_2)\dots \tilde{B}^{(2)}_{1\dots L}(\mu_L) |2 \rangle_{1\dots L}.
\end{array}\label{ZB2a}\end{equation}

Once again one can show that only the above terms of the complete set of states 
are non zero by considering the action of $e^{(32)}$ on the reference states.
Taking into account the property
$ e^{(32)}_l |2 \rangle_{p} |3 \rangle_{\genfrac{}{}{0pt}{}{ 1\dots L}{\ne p}}  =  \delta_{p l} |3 \rangle_{ 1\dots L}$ one finds
that,
\begin{equation*}
\,_{1\dots L}\langle 3 | \tilde{B}^{(2)}_{1\dots L}(\mu_1) |2 \rangle_{p} |3 \rangle_{\genfrac{}{}{0pt}{}{ 1\dots L}{\ne p}} = c_{23}(\mu_1,\xi_p) \prod^L_{\genfrac{}{}{0pt}{}{ i=1}{\ne p}} \frac{a_3(\mu_1,\xi_i)}{b_{32}(\xi_i,\xi_p) \theta_3(\xi_p, \xi_i)}.
\end{equation*}

Let us now consider the expression 
$\,_{p}\langle 2 | \,_{\genfrac{}{}{0pt}{}{ 1\dots L}{\ne p}}\langle 3 |\tilde{B}^{(2)}_{1\dots L}(\mu_2)\dots \tilde{B}^{(2)}_{1\dots L}(\mu_L) |2 \rangle_{1\dots L}$. Using the fact that $\,_{p}\langle 2 |e^{(32)}_p=0$, we can discard the value $p$ in the summations of the $\tilde{B}^{(2)}_{1\dots L}(\mu)$ operators. Applying this fact, and elementary matrix multiplication, the expression becomes,
\begin{equation}\begin{array}{ll}
&\,_{p}\langle 2 | \,_{\genfrac{}{}{0pt}{}{ 1\dots L}{\ne p}}\langle 3 |\tilde{B}^{(2)}_{1\dots L}(\mu_2)\dots \tilde{B}^{(2)}_{1\dots L}(\mu_L) |2 \rangle_{1\dots L}\\
=&\sum^L_{\genfrac{}{}{0pt}{}{ l_2, \dots, l_L=1}{l_2 \ne \dots \ne l_L \ne p}} \prod^{L}_{j=2}b_{32}(\mu_j,\xi_p)\theta_2(\xi_{l_j},\xi_{p}) \,_{p}\langle 2 | \,_{\genfrac{}{}{0pt}{}{ 1\dots L}{\ne p}}\langle 3 |\tilde{B}^{(2,l_2)}_{1\dots L}(\mu_2)\dots \tilde{B}^{(2,l_L)}_{1\dots L}(\mu_L) |2 \rangle_{1\dots L},
\end{array}
\label{ZB2f}\end{equation}
where,
\begin{equation*} 
\tilde{B}^{(2,l)}_{\genfrac{}{}{0pt}{}{ 1 \dots L}{\ne p}}(\mu)  =c_{23}(\mu,\xi_l)e^{(32)}_l\bigotimes^{L}_{\genfrac{}{}{0pt}{}{i=1}{i\ne l,p}} diag\left\{ b_{31}(\mu,\xi_i),b_{32}(\mu,\xi_i)\theta_2(\xi_l,\xi_i),\frac{a_3(\mu,\xi_i)}{b_{32}(\xi_i,\xi_l)\theta_3(\xi_l,\xi_i)} \right\}_i.
\end{equation*}

As before the product $\prod^{L}_{j=2}b_{32}(\mu_j,\xi_p)\theta_2(\xi_{l_j},\xi_{p})$ is independent of 
the value of $l_2, \dots, l_L$ and the operators $\tilde{B}^{(2,l)}_{\genfrac{}{}{0pt}{}{ 1 \dots L}{\ne p}}(\mu)$ 
act trivially on the vector space $V_p$, meaning that we can 
decrease the number of relevant vector spaces in the reference states 
by one. Applying these simplifications we find that,
\begin{equation*}\begin{array}{lll}
\sum^L_{\genfrac{}{}{0pt}{}{ l_2, \dots, l_L=1}{l_2 \ne \dots \ne l_L \ne p}} \,_{\genfrac{}{}{0pt}{}{ 1\dots L}{\ne p}}\langle 3 | \tilde{B}^{(2,l_2)}_{\genfrac{}{}{0pt}{}{ 1 \dots L}{\ne p}}(\mu_2) \dots \tilde{B}^{(2,l_L)}_{\genfrac{}{}{0pt}{}{ 1 \dots L}{\ne p}}(\mu_L)  |2 \rangle_{\genfrac{}{}{0pt}{}{ 1\dots L}{\ne p}}&=& \,_{\genfrac{}{}{0pt}{}{ 1\dots L}{\ne p}}\langle 3 | \tilde{B}^{(2)}_{\genfrac{}{}{0pt}{}{ 1 \dots L}{\ne p}}(\mu_2) \dots \tilde{B}^{(2)}_{\genfrac{}{}{0pt}{}{ 1 \dots L}{\ne p}}(\mu_L)  |2 \rangle_{\genfrac{}{}{0pt}{}{ 1\dots L}{\ne p}}\\
&=&  Z^{(B,2)}_{L-1} (\{ \mu \},\{\xi\}|\hat{\mu}_1,\hat{\xi}_p) .
\end{array}
\end{equation*}
Consequently, Eq.(\ref{ZB2a}) becomes the following recurrence relation,
\begin{equation}\begin{array}{lll}
 Z^{(B,2)}_L (\{ \mu \},\{\xi\}) &= & \sum^L_{p=1} c_{23}(\mu_1,\xi_p) \prod^L_{\genfrac{}{}{0pt}{}{ i=1}{\ne p}} \frac{a_3(\mu_1,\xi_i) \theta_2(\xi_i, \xi_p)}{b_{32}(\xi_i,\xi_p) \theta_3(\xi_p,\xi_i)}  \prod^L_{j=2}b_{32}(\mu_j,\xi_p)\\
&& \times Z^{(B,2)}_{L-1} (\{ \mu \},\{\xi\}|\hat{\mu}_1,\hat{\xi}_p).
\end{array}\label{ZB2rec}\end{equation}
\subsubsection{Exact solution}
We now offer the following general result, 
 \begin{equation}\begin{array}{l}
 Z^{(B,2)}_L(\{\mu\},\{\xi\})=   \sum_{\sigma \in S_L} \prod^L_{i=1}c_{23}(\mu_i,\xi_{\sigma(i)})   \prod_{1 \le j < k \le L} \frac{a_3(\mu_j,\xi_{\sigma(k)})b_{32}(\mu_k,\xi_{\sigma(j)}) \theta_2(\xi_{\sigma(k)},\xi_{\sigma(j)}) }{b_{32}(\xi_{\sigma(k)},\xi_{\sigma(j)})\theta_3(\xi_{\sigma(j)},\xi_{\sigma(k)})},
\end{array}\label{explicitZB2} \end{equation}
where we note that the verification of Eq.(\ref{explicitZB2}) follows exactly from the verification of Eq.(\ref{explicitZC2}).

In what follows we shall  consider the basic DWPF's that are constructed from considerably 
more complicated twisted operators. Nevertheless, the task of 
obtaining the explicit forms of the DWPF's of type $C^{(1)}$ and $B^{(1)}$ poses no greater 
challenge than what we have experienced thus far.

\subsection{Single DWPF for $C^{(1)}$}\label{arcolad}

The DWPF expression for the $C^{(1)}_{1\dots L}$ operators is given explicitly by,
\begin{equation}\begin{array}{lll}
Z^{(C,1)}_L (\{ \nu \},\{\xi\})&=& \,_{1\dots L}\langle 1 | C^{(1)}_{1\dots L}(\nu_L) \dots C^{(1)}_{1\dots L}(\nu_1) |3 \rangle_{1\dots L}\\
&=& \,_{1\dots L}\langle 1 | \tilde{C}^{(1)}_{1\dots L}(\nu_L) \dots \tilde{C}^{(1)}_{1\dots L}(\nu_1) |3 \rangle_{1\dots L},
\end{array}\label{ZC1}\end{equation}
where we have applied the results of Eq.(\ref{actref}) to twist the operators.

Before inserting a complete set of states between 
operators we shall first consider the expression, 
$\tilde{C}^{(1)}_{1\dots L}(\nu_1) |3 \rangle_{1\dots L}$, and in particular 
we notice that the action of the matrices $e^{(12)}_{l_1} \otimes e^{(23)}_{l_2}$ on the 
reference states $|3\rangle_{l_1} \otimes |3\rangle_{l_2}$ is zero. Hence if we insert a complete set of states in between the operators $\tilde{C}^{(1)}_{1\dots L}(\nu_2)$ and $\tilde{C}^{(1)}_{1\dots L}(\nu_1)$ we obtain,
\begin{equation}\begin{array}{lll}
 Z^{(C,1)}_L (\{ \nu \},\{\xi\}) &= & \sum^L_{p=1} \,_{1\dots L}\langle 1 | \tilde{C}^{(1)}_{1\dots L}(\nu_L) \dots \tilde{C}^{(1)}_{1\dots L}(\nu_2) |1 \rangle_{p} |3 \rangle_{\genfrac{}{}{0pt}{}{ 1\dots L}{\ne p}} \\
&& \times\,_{p}\langle 1 | \,_{\genfrac{}{}{0pt}{}{ 1\dots L}{\ne p}}\langle 3 |\tilde{C}^{(1)}_{1\dots L}(\nu_1) |3 \rangle_{1\dots L},
\end{array}\label{ZC1a}\end{equation}
which is very similar to the corresponding expression for $\tilde{C}^{(2)}$.

We now use the identity 
$ \,_{p}\langle 1 | \,_{\genfrac{}{}{0pt}{}{ 1\dots L}{\ne p}}\langle 3 | e^{(13)}_l  =  \,_{1 \dots L}\langle 3 |\delta_{p l}$
and by elementary matrix multiplication we obtain,
\begin{equation*}
 \,_{p}\langle 1 | \,_{\genfrac{}{}{0pt}{}{ 1\dots L}{\ne p}}\langle 3 |\tilde{C}^{(1)}_{1\dots L}(\nu_1) |3 \rangle_{1\dots L} = c_{31}(\nu_1,\xi_p) \prod^L_{\genfrac{}{}{0pt}{}{ i=1}{\ne p}} a_3(\nu_1,\xi_i) \theta_3(\xi_i, \xi_p).
\end{equation*}

Focusing on the expression 
$\,_{1\dots L}\langle 1 | \tilde{C}^{(1)}_{1\dots L}(\nu_L) \dots \tilde{C}^{(1)}_{1\dots L}(\nu_2) 
|1 \rangle_{p} |3 \rangle_{\genfrac{}{}{0pt}{}{ 1\dots L}{\ne p}}$ we note 
that due to the elementary relations in Eq.(\ref{action}), there is no possibility of the 
Wely matrices $e^{(12)}_{l_1} \otimes e^{(23)}_{l_2}$ in the expressions for the $\tilde{C}^{(1)}_{1\dots L}$ operators to produce anything but zero, hence they can be discarded from the calculations. Additionally, using the fact that $e^{(13)}_p|1 \rangle_{p}=0$, we can discard the value $p$ in the summations of the $\tilde{C}^{(1)}_{1\dots L}(\nu)$ operators. Applying the above facts, and elementary matrix multiplication, the expression now becomes,
\begin{equation}\begin{array}{ll}
&\,_{1\dots L}\langle 1 | \tilde{C}^{(1)}_{1\dots L}(\nu_L) \dots \tilde{C}^{(1)}_{1\dots L}(\nu_2) |1 \rangle_{p} |3 \rangle_{\genfrac{}{}{0pt}{}{ 1\dots L}{\ne p}}\\
=&\sum^L_{\genfrac{}{}{0pt}{}{ l_2, \dots, l_L=1}{l_2 \ne \dots \ne l_L \ne p}} \prod^{L}_{j=2}\frac{b_{21}(\nu_j,\xi_p)}{b_{21}(\xi_{l_j},\xi_p)\theta_1(\xi_p,\xi_{l_j})}\,_{1\dots L}\langle 1 | \tilde{C}^{(1,l_L)}_{\genfrac{}{}{0pt}{}{ 1 \dots L}{\ne p}}(\nu_L) \dots \tilde{C}^{(1,l_2)}_{\genfrac{}{}{0pt}{}{ 1 \dots L}{\ne p}}(\nu_2) |1 \rangle_{p} |3 \rangle_{\genfrac{}{}{0pt}{}{ 1\dots L}{\ne p}},
\end{array}
\label{ZC1f}\end{equation}
where,
\begin{equation} 
\tilde{C}^{(1,l)}_{\genfrac{}{}{0pt}{}{ 1 \dots L}{\ne p}}(\nu)  = c_{31}(\nu,\xi_l)e^{(13)}_l \bigotimes^{L}_{\genfrac{}{}{0pt}{}{i=1}{i\ne l,p}} diag \left\{ \frac{b_{21}(\nu,\xi_i)}{b_{21}(\xi_l,\xi_i)\theta_1(\xi_i,\xi_l)},\frac{b_{32}(\nu,\xi_i)}{b_{32}(\xi_l,\xi_i)},a_3(\nu,\xi_i)\theta_3(\xi_i,\xi_l) \right\}_i.
\label{H22}\end{equation}

We note that the product $\prod^{L}_{j=2}\frac{b_{21}(\nu_j,\xi_p)}{b_{21}(\xi_{l_j},\xi_p)\theta_1(\xi_p,\xi_{l_j})}$ 
is independent of the indices $l_2, \dots, l_L$, and the operators 
$\tilde{C}^{(1,l)}_{\genfrac{}{}{0pt}{}{ 1 \dots L}{\ne p}}(\nu)$ act trivially on the vector 
space $V_p$, allowing us to decrease the number of relevant vector spaces 
in the reference states by one. Applying such simplifications one finds the relation,
\begin{equation*}\begin{array}{lll}
\sum^L_{\genfrac{}{}{0pt}{}{ l_2, \dots, l_L=1}{l_2 \ne \dots \ne l_L \ne p}} \,_{\genfrac{}{}{0pt}{}{ 1\dots L}{\ne p}}\langle 1 | \tilde{C}^{(1,l_L)}_{\genfrac{}{}{0pt}{}{ 1 \dots L}{\ne p}}(\nu_L) \dots \tilde{C}^{(1,l_2)}_{\genfrac{}{}{0pt}{}{ 1 \dots L}{\ne p}}(\nu_2)  |3 \rangle_{\genfrac{}{}{0pt}{}{ 1\dots L}{\ne p}}&=& \,_{\genfrac{}{}{0pt}{}{ 1\dots L}{\ne p}}\langle 1 | \tilde{C}^{(1)}_{\genfrac{}{}{0pt}{}{ 1 \dots L}{\ne p}}(\nu_L) \dots \tilde{C}^{(1)}_{\genfrac{}{}{0pt}{}{ 1 \dots L}{\ne p}}(\nu_2)  |3 \rangle_{\genfrac{}{}{0pt}{}{ 1\dots L}{\ne p}}\\
&=&  Z^{(C,1)}_{L-1} (\{ \nu \},\{\xi\}|\hat{\nu}_1,\hat{\xi}_p) .
\end{array}
\end{equation*}
Again, we can discard the terms involving $e^{(12)}_{l_1} \otimes e^{(23)}_{l_2}$ in the 
above $\tilde{C}^{(1)}_{\genfrac{}{}{0pt}{}{ 1 \dots L}{\ne p}}$ 
operators because they always produce zero. As a consequence of that, 
Eq.(\ref{ZC1a}) becomes the following recurrence relation,
\begin{equation}\begin{array}{lll}
 Z^{(C,1)}_L (\{ \nu \},\{\xi\}) &= & \sum^L_{p=1} c_{31}(\nu_1,\xi_p) \prod^L_{\genfrac{}{}{0pt}{}{ i=1}{\ne p}} \frac{a_3(\nu_1,\xi_i) \theta_3(\xi_i, \xi_p)}{b_{21}(\xi_i,\xi_p) \theta_1(\xi_p,\xi_i)}  \prod^L_{j=2}b_{21}(\nu_j,\xi_p)\\
&& \times Z^{(C,1)}_{L-1} (\{ \nu \},\{\xi\}|\hat{\nu}_1,\hat{\xi}_p).
\end{array}\label{ZC1rec}\end{equation}

The same type of arguments presented in Section \ref{secC2DW} can be used to provide  
the general solution to the above recurrence relation. It is given explicitly by, 
\begin{equation}\begin{array}{l}
Z^{(C,1)}_L(\{\nu\},\{\xi\})=   \sum_{\sigma \in S_L} 
\prod^L_{i=1}c_{31}(\nu_i,\xi_{\sigma(i)})   \prod_{1 \le j < k \le L} 
\frac{a_3(\nu_j,\xi_{\sigma(k)})b_{21}(\nu_k,\xi_{\sigma(j)}) 
\theta_3(\xi_{\sigma(k)},\xi_{\sigma(j)}) }{b_{21}(\xi_{\sigma(k)},\xi_{\sigma(j)})\theta_1(\xi_{\sigma(j)},\xi_{\sigma(k)})}.
\end{array}\label{explicitZC1} \end{equation}

\subsection{Single DWPF for type $B^{(1)}$}\label{BBB}
We now consider the equivalent DWPF expression for the $B^{(1)}_{1\dots L}$ operators which given explicitly by,
\begin{equation}\begin{array}{lll}
 Z^{(B,1)}_L (\{ \mu \},\{\xi\})&=& \,_{1\dots L}\langle 3 | B^{(1)}_{1\dots L}(\mu_1) \dots B^{(1)}_{1\dots L}(\mu_L) |1 \rangle_{1\dots L}\\
&=& \,_{1\dots L}\langle 3 | \tilde{B}^{(1)}_{1\dots L}(\mu_1) \dots \tilde{B}^{(1)}_{1\dots L}(\mu_L) |1 \rangle_{1\dots L},
\end{array}\label{ZB1}\end{equation}
where we have applied the results of Eq.(\ref{actref}) to twist the operators.

As with the case for $C^{(1)}$, before we insert a complete set of states between operators we shall first consider the expression $\,_{1\dots L}\langle 3 | \tilde{B}^{(1)}_{1\dots L}(\mu_1)$, and in particular we notice that the action of the matrices $e^{(21)}_{l_1} \otimes e^{(32)}_{l_2}$ on the transpose reference states $\,_{l_1}\langle 3 | \otimes \,_{l_2}\langle 3 |$ is zero. Hence if we insert a complete set of states in between the operators $\tilde{B}^{(1)}_{1\dots L}(\mu_1)$ and $\tilde{B}^{(1)}_{1\dots L}(\mu_2)$ we obtain,
\begin{equation}\begin{array}{lll}
 Z^{(B,1)}_L (\{ \mu \},\{\xi\}) &= & \sum^L_{p=1} \,_{1\dots L}\langle 3 | \tilde{B}^{(1)}_{1\dots L}(\mu_1)  |1 \rangle_{p} |3 \rangle_{\genfrac{}{}{0pt}{}{ 1\dots L}{\ne p}} \\
&& \times\,_{p}\langle 1 | \,_{\genfrac{}{}{0pt}{}{ 1\dots L}{\ne p}}\langle 3 |\tilde{B}^{(1)}_{1\dots L}(\mu_2) \dots \tilde{B}^{(1)}_{1\dots L}(\mu_L)|1 \rangle_{1\dots L},
\end{array}\label{ZB1a}\end{equation}
Using the fact that
$ e^{(31)}_l  |1 \rangle_{p} |3 \rangle_{\genfrac{}{}{0pt}{}{ 1\dots L}{\ne p}}  = \delta_{p l}  |3 \rangle_{1\dots L}$
one is bale to write the following identity,
\begin{equation*}
\,_{1\dots L}\langle 3 | \tilde{B}^{(1)}_{1\dots L}(\mu_1)  |1 \rangle_{p} |3 \rangle_{\genfrac{}{}{0pt}{}{ 1\dots L}{\ne p}} = c_{13}(\mu_1,\xi_p) \prod^L_{\genfrac{}{}{0pt}{}{ i=1}{\ne p}} \frac{a_3(\mu_1,\xi_i)}{ b_{31}(\xi_i,\xi_p)\theta_3(\xi_p, \xi_1)}.
\end{equation*}

In parallel to Section \ref{arcolad}, due to the elementary relations in Eq.(\ref{action}), there is no 
possibility of the Wely matrices $e^{(21)}_{l_1} \otimes e^{(32)}_{l_2}$ in the expression 
for the $\tilde{B}^{(1)}_{1\dots L}$ operators to produce anything 
but zero, hence they can be discarded from the calculations. In addition, using the 
fact that $\,_{p}\langle 1 |e^{(31)}_p=0$, we can discard the value $p$ in the summations 
of the $\tilde{B}^{(1)}_{1\dots L}$ operators. Considering these facts together with elementary matrix 
multiplication, the expression now becomes,
\begin{equation}\begin{array}{ll}
&\,_{p}\langle 1 | \,_{\genfrac{}{}{0pt}{}{ 1\dots L}{\ne p}}\langle 3 |\tilde{B}^{(1)}_{1\dots L}(\mu_2) \dots \tilde{B}^{(1)}_{1\dots L}(\mu_L)|1 \rangle_{1\dots L}\\
=&\sum^L_{\genfrac{}{}{0pt}{}{ l_2, \dots, l_L=1}{l_2 \ne \dots \ne l_L \ne p}} \prod^{L}_{j=2}b_{31}(\mu_j,\xi_p)\theta_1(\xi_{l_j},\xi_{p})\,_{p}\langle 1 | \,_{\genfrac{}{}{0pt}{}{ 1\dots L}{\ne p}}\langle 3 |  \tilde{B}^{(1,l_2)}_{\genfrac{}{}{0pt}{}{ 1 \dots L}{\ne p}}(\mu_2) \dots \tilde{B}^{(1,l_L)}_{\genfrac{}{}{0pt}{}{ 1 \dots L}{\ne p}}(\mu_L)  |1 \rangle_{1\dots L},
\end{array}
\label{ZB1f}\end{equation}
where,
\begin{equation} 
\tilde{B}^{(1,l)}_{\genfrac{}{}{0pt}{}{ 1 \dots L}{\ne p}}(\mu)  =c_{13}(\mu,\xi_l)e^{(31)}_l \bigotimes^{L}_{\genfrac{}{}{0pt}{}{i=1}{i\ne l,p}} diag \left\{ b_{31}(\mu,\xi_i)\theta_{1}(\xi_l,\xi_i),\frac{b_{32}(\mu,\xi_i)}{b_{21}(\xi_i,\xi_l)},\frac{a_3(\mu,\xi_i)}{b_{31}(\xi_i,\xi_l)\theta_3(\xi_l,\xi_i)} \right\}.
\label{H11}\end{equation}

Note that the  product $\prod^{L}_{j=2}b_{31}(\mu_j,\xi_p)\theta_1(\xi_{l_j},\xi_{p})$ 
does not depend on the indices $l_2, \dots, l_L$ and the operators 
$\tilde{B}^{(1,l)}_{\genfrac{}{}{0pt}{}{ 1 \dots L}{\ne p}}$ act trivially on the vector space $V_p$, 
allowing us to decrease the number of relevant vector spaces 
in the reference states by one. These simplifications lead to, 
\begin{equation*}\begin{array}{lll}
\sum^L_{\genfrac{}{}{0pt}{}{ l_2, \dots, l_L=1}{l_2 \ne \dots \ne l_L \ne p}} \,_{\genfrac{}{}{0pt}{}{ 1\dots L}{\ne p}}\langle 3 | \tilde{B}^{(1,l_2)}_{\genfrac{}{}{0pt}{}{ 1 \dots L}{\ne p}}(\mu_2) \dots \tilde{B}^{(1,l_L)}_{\genfrac{}{}{0pt}{}{ 1 \dots L}{\ne p}}(\mu_L)  |1 \rangle_{\genfrac{}{}{0pt}{}{ 1\dots L}{\ne p}}&=& \,_{\genfrac{}{}{0pt}{}{ 1\dots L}{\ne p}}\langle 3 | \tilde{B}^{(1)}_{\genfrac{}{}{0pt}{}{ 1 \dots L}{\ne p}}(\mu_2) \dots \tilde{B}^{(1)}_{\genfrac{}{}{0pt}{}{ 1 \dots L}{\ne p}}(\mu_L)  |1 \rangle_{\genfrac{}{}{0pt}{}{ 1\dots L}{\ne p}}\\
&=&  Z^{(B,1)}_{L-1} (\{ \mu \},\{\xi\}|\hat{\mu}_1,\hat{\xi}_p) .
\end{array}
\end{equation*}
Again, we can discard the terms involving $e^{(21)}_{l_1} \otimes e^{(32)}_{l_2}$ in the above $\tilde{B}^{(1)}_{\genfrac{}{}{0pt}{}{ 1 \dots L}{\ne p}}$ operators because they always produce zero. Hence Eq.(\ref{ZB1a}) becomes the following recurrence relation,
\begin{equation}\begin{array}{lll}
 Z^{(B,1)}_L (\{ \mu \},\{\xi\}) &= & \sum^L_{p=1} c_{13}(\mu_1,\xi_p) \prod^L_{\genfrac{}{}{0pt}{}{ i=1}{\ne p}} \frac{a_3(\mu_1,\xi_i) \theta_1(\xi_i, \xi_p)}{b_{31}(\xi_i,\xi_p) \theta_3(\xi_p,\xi_i)}  \prod^L_{j=2}b_{31}(\mu_j,\xi_p)\\
&& \times Z^{(B,1)}_{L-1} (\{ \mu \},\{\xi\}|\hat{\mu}_1,\hat{\xi}_p).
\end{array}\label{ZB1rec}\end{equation}
The exact solution to the above recurrence relation is, 
\begin{equation}\begin{array}{l}
Z^{(B,1)}_L(\{\mu\},\{\xi\})=   \sum_{\sigma \in S_L} \prod^L_{i=1}c_{13}(\mu_i,\xi_{\sigma(i)})  
 \prod_{1 \le j < k \le L} \frac{a_3(\mu_j,\xi_{\sigma(k)})b_{31}(\mu_k,\xi_{\sigma(j)}) 
\theta_1(\xi_{\sigma(k)},\xi_{\sigma(j)}) }{b_{31}(\xi_{\sigma(k)},\xi_{\sigma(j)})\theta_3(\xi_{\sigma(j)},\xi_{\sigma(k)})}.
\end{array}\label{explicitZB1} 
\end{equation}
\section{Mixed DWPF}\label{DOMAINMIX}
We start by discussing the preliminary steps to compute the DWPF's involving mixed operators. 
The elements of the Bethe state vectors built from $C$ fields for the fifteen-vertex model are given by,
\begin{equation}
 C^{(i_1)}_{1\dots L}(\nu_L) C^{(i_2)}_{1\dots L}(\nu_{L-1}) \dots  
C^{(i_L)}_{1\dots L}(\nu_1)|3\rangle_{1\dots L},
\label{BetheC}\end{equation}
where $i_1,\dots,i_L\in \{1,2\}$. Here we are  considering the vector 
$|3\rangle_{1\dots L}$ as our starting ferromagnetic reference state.

We now introduce the integer $M$, $M \le L$, which indicates how many type ``$(1)$'' operators we have in our state vector element. Given $M$, every element of Eq.(\ref{BetheC}) will generally consist of an expression of $C^{(1)}$ and $C^{(2)}$ operators in no particular order. Applying the following Yang-Baxter algebra expression generated from Eq.(\ref{YBalg}),
\begin{equation}
 C^{(1)}_{1\dots L}(\nu)C^{(2)}_{1\dots L}(\mu) = \frac{a_3}{b_{21}}(\mu,\nu)C^{(2)}_{1\dots L}(\mu)C^{(1)}_{1\dots L}(\nu)- \frac{c_{12}}{b_{21}}(\mu,\nu)C^{(2)}_{1\dots L}(\nu)C^{(1)}_{1\dots L}(\mu),
\label{twoC}\end{equation}
it is possible to commute all the $C^{(2)}$ operators to the left, leading to the general expression,
\begin{equation}\begin{array}{ll}
&C^{(i_1)}_{1\dots L}(\nu_L) C^{(i_2)}_{1\dots L}(\nu_{L-1}) \dots  C^{(i_L)}_{1\dots L}(\nu_1)|3\rangle_{1\dots L}\\
=& \sum_{\sigma \in S_L} \phi^{(i_1,\dots,i_L)}_{M,\{\sigma\}} C^{(2)}_{1\dots L}(\nu_{\sigma(L)})\dots C^{(2)}_{1\dots L}(\nu_{\sigma(M+1)}) C^{(1)}_{1\dots L}(\nu_{\sigma(M)})\dots  C^{(1)}_{1\dots L}(\nu_{\sigma(1)})|3\rangle_{1\dots L},
\end{array}\label{generalCC}\end{equation}
where the coefficient $\phi^{(i_1,\dots,i_L)}_{M,\{\sigma\}}$, which is constructed 
from the Boltzmann weights $a_3$, $b_{21}$ and $c_{12}$, generally depends on the initial 
value of the indices $i_1, \dots, i_L$, the integer $M$ 
and the particular permutation $\sigma$.

From Eq.(\ref{generalCC}) we have that the most fundamental 
mixed DWPF expression of type $C$ will be of the form,
\begin{equation}
 Z^{(C)}_{L,M}(\{\nu\},\{\xi\}) = \,_{q_1 \dots q_M}\langle 1 |\,_{\genfrac{}{}{0pt}{}{1\dots L}{\ne q_1 \dots q_M}}\langle 2 | {C}^{(2)}_{1\dots L}(\nu_L)\dots {C}^{(2)}_{1\dots L}(\nu_{M+1}) {C}^{(1)}_{1\dots L}(\nu_{M})\dots {C}^{(1)}_{1\dots L}(\nu_{1}) |3\rangle_{1\dots L},
\label{mixedC}\end{equation}
where for clarity we assume that $q_1 < q_2 < \dots < q_M$. We briefly note that the configuration in Figure \ref{DWPFC} corresponds to Eq.(\ref{mixedC}) for $q_1=1, q_2=2, \dots, q_M=M$.

Similarly for $B$, the elements of the transpose Bethe state vectors are given by,
\begin{equation}
\,_{1 \dots L}\langle 3 | B^{(i_1)}_{1\dots L}(\mu_1) B^{(i_2)}_{1\dots L}(\mu_{2}) \dots  B^{(i_L)}_{1\dots L}(\mu_L),
\label{BetheB}\end{equation}
where again $i_1,\dots,i_L \in \{1,2\}$.

Using $M$, $M \le L$, to indicate how many ``$(1)$'' operators we have in our transpose state vector element, we apply the following Yang-Baxter algebra expression generated from Eq.(\ref{YBalg}),
\begin{equation}
 B^{(2)}_{1\dots L}(\mu)B^{(1)}_{1\dots L}(\nu) = \frac{a_3}{b_{21}}(\mu,\nu)B^{(1)}_{1\dots L}(\nu)B^{(2)}_{1\dots L}(\mu)- \frac{c_{21}}{b_{21}}(\mu,\nu)B^{(1)}_{1\dots L}(\mu)B^{(2)}_{1\dots L}(\nu),
\label{twoB}\end{equation}
to commute all the $B^{(2)}$ operators to the right, leading to the expression,
\begin{equation}\begin{array}{ll}
&\,_{1 \dots L}\langle 3 | B^{(i_1)}_{1\dots L}(\mu_1) B^{(i_2)}_{1\dots L}(\mu_{2}) \dots  B^{(i_L)}_{1\dots L}(\mu_L)\\
=& \sum_{\sigma \in S_L} \psi^{(i_1,\dots,i_L)}_{M,\{\sigma\}} \,_{1 \dots L}\langle 3 | B^{(1)}_{1\dots L}(\mu_{\sigma(1)})\dots B^{(1)}_{1\dots L} (\mu_{\sigma(M)}) B^{(2)}_{1\dots L}(\mu_{\sigma(M+1)})\dots  B^{(2)}_{1\dots L}(\mu_{\sigma(L)}),
\end{array}\label{generalBB}\end{equation}
where the coefficient $\psi^{(i_1,\dots,i_L)}_{M,\{\sigma\}}$, 
is constructed from the Boltzmann weights $a_3$, $b_{21}$ and $c_{21}$ and parallels $\phi^{(i_1,\dots,i_L)}_{M,\{\sigma\}}$ from Eq. (\ref{generalCC}) in structure.

From Eq.(\ref{generalBB}) we see that the most fundamental mixed DWPF expression of type $B$ will be of the form,
\begin{equation}
 Z^{(B)}_{L,M}(\{\mu\},\{\xi\}) = \,_{1 \dots L}\langle 3 | {B}^{(1)}_{1\dots L}(\mu_1)\dots {B}^{(1)}_{1\dots L}(\mu_M) {B}^{(2)}_{1\dots L}(\mu_{M+1})\dots {B}^{(2)}_{1\dots L}(\mu_{L}) |1\rangle_{q_1 \dots q_M} | 2 \rangle_{\genfrac{}{}{0pt}{}{1\dots L}{\ne q_1 \dots q_M}}.
\label{mixedB}\end{equation}
For clarity we briefly note that the configuration in Figure \ref{DWPFB} corresponds to Eq.(\ref{mixedB}) for $q_1=1, q_2=2, \dots, q_M=M$.

We now devote separate subsections for the explicit evaluation of Eqs. (\ref{mixedC}) and (\ref{mixedB}). 
\subsection{Mixed DWPF for  B}
Applying the $\mathcal{F}$-matrix to Eq.(\ref{mixedB}) to twist the monodromy operators we obtain,
\begin{equation}
 Z^{(B)}_{L,M}(\{\mu\},\{\xi\}) = \,_{1 \dots L}\langle 3 | \tilde{B}^{(1)}_{1\dots L}(\mu_1)\dots \tilde{B}^{(1)}_{1\dots L}(\mu_M) \tilde{B}^{(2)}_{1\dots L}(\mu_{M+1})\dots \tilde{B}^{(2)}_{1\dots L}(\mu_{L})\mathcal{F}_{1\dots L} |1\rangle_{q_1 \dots q_M} | 2 \rangle_{\genfrac{}{}{0pt}{}{1\dots L}{\ne q_1 \dots q_M}}.
\label{mixedB1}\end{equation}
We note that there is an $\mathcal{F}_{1\dots L}$ operator on the far right of Eq.(\ref{mixedB1}) as there is no equivalent expression such as Eq.(\ref{actref}) for mixed reference states.
We proceed by inserting two complete sets of states, but for clarity we shall do this in separate stages. Consider first placing a complete set of states in between the operators $\tilde{B}^{(1)}_{1\dots L}(\mu_M)$ and $\tilde{B}^{(2)}_{1\dots L}(\mu_{M+1})$ to obtain,
\begin{equation}\begin{array}{lll}
 Z^{(B)}_{L,M}(\{\mu\},\{\xi\}) &=& \sum_{1 \le p_1 < \dots < p_M \le L}\,_{1 \dots L}\langle 3 | \tilde{B}^{(1)}_{1\dots L}(\mu_1)\dots \tilde{B}^{(1)}_{1\dots L}(\mu_M)|1\rangle_{p_1 \dots p_M} | 3 \rangle_{\genfrac{}{}{0pt}{}{1\dots L}{\ne p_1 \dots p_M}}\\
&&\times \,_{p_1 \dots p_M}\langle 1 |\,_{\genfrac{}{}{0pt}{}{1\dots L}{\ne p_1 \dots p_M}}\langle 3 | \tilde{B}^{(2)}_{1\dots L}(\mu_{M+1})\dots \tilde{B}^{(2)}_{1\dots L}(\mu_{L})\mathcal{F}_{1\dots L} |1\rangle_{q_1 \dots q_M} | 2 \rangle_{\genfrac{}{}{0pt}{}{1\dots L}{\ne q_1 \dots q_M}}.
\end{array}\label{mixedB2}\end{equation}
In order to verify that the above terms of the complete set of states are the only ones which are non zero, consider the expression, 
\begin{equation*}
 \,_{1 \dots L}\langle 3 | \tilde{B}^{(1)}_{1\dots L}(\mu_1)\dots \tilde{B}^{(1)}_{1\dots L}(\mu_M),
\end{equation*}
and recall from Section \ref{BBB} that only the Weyl matrix $e^{31}_l$ (as opposed to $e^{32}_{l_1}\otimes e^{21}_{l_2}$) gives non zero terms when applied to the reference state $\,_{1 \dots L}\langle 3 |$. Hence since $\,_{l}\langle 3 |e^{31}_l = \,_{l}\langle 1 |$, this means the only terms in the complete set of states which are non zero are,
\begin{equation*} 
 \sum_{1 \le p_1 < \dots < p_M \le L} |1\rangle_{p_1 \dots p_M} | 3 \rangle_{\genfrac{}{}{0pt}{}{1\dots L}{\ne p_1 \dots p_M}} \,_{p_1 \dots p_M}\langle 1 |\,_{\genfrac{}{}{0pt}{}{1\dots L}{\ne p_1 \dots p_M}}\langle 3 |.
\end{equation*}
We now insert another complete set of states in between $\tilde{B}^{(2)}_{1\dots L}(\mu_{L})$ and $\mathcal{F}_{1\dots L}$ in Eq.(\ref{mixedB2}) to obtain,
\begin{equation}\begin{array}{lll}
 Z^{(B)}_{L,M}(\{\mu\},\{\xi\})
&=& \sum_{1 \le p_1 < \dots < p_M \le L}  \,_{1 \dots L}\langle 3 | \tilde{B}^{(1)}_{1\dots L}(\mu_1)\dots \tilde{B}^{(1)}_{1\dots L}(\mu_M)|1\rangle_{p_1 \dots p_M} | 3 \rangle_{\genfrac{}{}{0pt}{}{1\dots L}{\ne p_1 \dots p_M}}\\
&&\times \,_{p_1 \dots p_M}\langle 1 |\,_{\genfrac{}{}{0pt}{}{1\dots L}{\ne p_1 \dots p_M}}\langle 3 | \tilde{B}^{(2)}_{1\dots L}(\mu_{M+1})\dots \tilde{B}^{(2)}_{1\dots L}(\mu_{L})|1\rangle_{p_1 \dots p_M} | 2 \rangle_{\genfrac{}{}{0pt}{}{1\dots L}{\ne p_1 \dots p_M}} \\
&&\times \,_{p_1 \dots p_M}\langle 1 |\,_{\genfrac{}{}{0pt}{}{1\dots L}{\ne p_1 \dots p_M}}\langle 2 |\mathcal{F}_{1\dots L} |1\rangle_{q_1 \dots q_M} | 2 \rangle_{\genfrac{}{}{0pt}{}{1\dots L}{\ne q_1 \dots q_M}}.
\end{array}\label{mixedB3}\end{equation}
We now decrease the number of relevant vector spaces in each of the above \textit{pseudo} basic DWPF expressions.
\subsubsection{Reducing the number of relevant vector spaces - I}
Beginning with $\,_{p_1 \dots p_M}\langle 1 |\,_{\genfrac{}{}{0pt}{}{1\dots L}{\ne p_1 \dots p_M}}\langle 3 | \tilde{B}^{(2)}_{1\dots L}(\mu_{M+1})\dots \tilde{B}^{(2)}_{1\dots L}(\mu_{L})|1\rangle_{p_1 \dots p_M} | 2 \rangle_{\genfrac{}{}{0pt}{}{1\dots L}{\ne p_1 \dots p_M}}$, we use elementary matrix algebra and the fact that $\,_{p}\langle 1 |e^{(32)}_p=0$ to discard the values $p_1,\dots, p_M$ in the summations of the $\tilde{B}^{(2)}_{1\dots L}(\mu)$ operators to obtain,
\begin{equation}\begin{array}{ll}
&\,_{p_1 \dots p_M}\langle 1 |\,_{\genfrac{}{}{0pt}{}{1\dots L}{\ne p_1 \dots p_M}}\langle 3 | \tilde{B}^{(2)}_{1\dots L}(\mu_{M+1})\dots \tilde{B}^{(2)}_{1\dots L}(\mu_{L})|1\rangle_{p_1 \dots p_M} | 2 \rangle_{\genfrac{}{}{0pt}{}{1\dots L}{\ne p_1 \dots p_M}}\\
=& \prod^L_{i=M+1}\prod^{M}_{j=1}b_{31}(\mu_i,\xi_{p_j}) \,_{p_1 \dots p_M}\langle 1 |\,_{\genfrac{}{}{0pt}{}{1\dots L}{\ne p_1 \dots p_M}}\langle 3 | \tilde{B}^{(2)}_{\genfrac{}{}{0pt}{}{1\dots L}{\ne p_1 \dots p_M}}(\mu_{M+1})\dots \tilde{B}^{(2)}_{\genfrac{}{}{0pt}{}{1\dots L}{\ne p_1 \dots p_M}}(\mu_{L})|1\rangle_{p_1 \dots p_M} | 2 \rangle_{\genfrac{}{}{0pt}{}{1\dots L}{\ne p_1 \dots p_M}}.
\end{array}
\label{sssa}\end{equation}
Since the operators $\tilde{B}^{(2)}_{\genfrac{}{}{0pt}{}{ 1 \dots L}{\ne p_1 \dots p_M}}(\mu)$ 
act trivially on the vector space $V_{p_i}$, $i=1,\dots,M$ we decrease the number of relevant vector spaces in the reference states by M - hence Eq.(\ref{sssa}) becomes, 
\begin{equation*}\begin{array}{ll}
&\,_{p_1 \dots p_M}\langle 1 |\,_{\genfrac{}{}{0pt}{}{1\dots L}{\ne p_1 \dots p_M}}\langle 3 | \tilde{B}^{(2)}_{1\dots L}(\mu_{M+1})\dots \tilde{B}^{(2)}_{1\dots L}(\mu_{L})|1\rangle_{p_1 \dots p_M} | 2 \rangle_{\genfrac{}{}{0pt}{}{1\dots L}{\ne p_1 \dots p_M}}\\
=& \prod^L_{i=M+1}\prod^{M}_{j=1}b_{31}(\mu_i,\xi_{p_j})\,_{\genfrac{}{}{0pt}{}{1\dots L}{\ne p_1 \dots p_M}}\langle 3 | \tilde{B}^{(2)}_{\genfrac{}{}{0pt}{}{1\dots L}{\ne p_1 \dots p_M}}(\mu_{M+1})\dots \tilde{B}^{(2)}_{\genfrac{}{}{0pt}{}{1\dots L}{\ne p_1 \dots p_M}}(\mu_{L}) | 2 \rangle_{\genfrac{}{}{0pt}{}{1\dots L}{\ne p_1 \dots p_M}}\\
=& \prod^L_{i=M+1}\prod^{M}_{j=1}b_{31}(\mu_i,\xi_{p_j}) Z^{(B,2)}_{L-M}(\{\mu_{k}\}_{k=M+1,\dots,L}|\{ \xi_l\}_{\genfrac{}{}{0pt}{}{l=1,\dots, L}{\ne p_1, \dots, p_M}}).
\end{array}
\end{equation*}
\subsubsection{Reducing the number of relevant vector spaces - II}
We now focus on the term $\,_{1 \dots L}\langle 3 | \tilde{B}^{(1)}_{1\dots L}(\mu_1)\dots \tilde{B}^{(1)}_{1\dots L}(\mu_M)|1\rangle_{p_1 \dots p_M} | 3 \rangle_{\genfrac{}{}{0pt}{}{1\dots L}{\ne p_1 \dots p_M}}$. We reiterate that due to the elementary relations in Eq.(\ref{action}), there is no possibility of the Weyl matrices $e^{(21)}_{l_1} \otimes e^{(32)}_{l_2}$ in the expressions for the $\tilde{B}^{(1)}_{1\dots L}$ operators producing anything but zero, hence they can be discarded from the calculations. Additionally, using the fact that $e^{(31)}_p|3 \rangle_p =0$, we can discard the values $l \ne p_1,\dots,p_M$ in the summations of the $\tilde{B}^{(1)}_{1\dots L}$ operators. Applying the above facts, and elementary matrix multiplication, the expression now becomes,
\begin{equation}\begin{array}{rr}
&\,_{1 \dots L}\langle 3 | \tilde{B}^{(1)}_{1\dots L}(\mu_1)\dots \tilde{B}^{(1)}_{1\dots L}(\mu_M)|1\rangle_{p_1 \dots p_M} | 3 \rangle_{\genfrac{}{}{0pt}{}{1\dots L}{\ne p_1 \dots p_M}}\\
=&\sum^M_{\genfrac{}{}{0pt}{}{ l_1, \dots, l_M=1}{l_1 \ne \dots \ne l_M } } \prod^M_{i=1} \prod^{L}_{\genfrac{}{}{0pt}{}{ j=1}{\ne p_1, \dots, p_M}} \frac{a_3(\mu_i,\xi_j)}{b_{31}(\xi_j,\xi_{p_{l_i}}) \theta_3(\xi_{p_{l_i}},\xi_j)}\\
& \times \,_{1 \dots L}\langle 3 | \tilde{B}^{(1,p_{l_1})}_{p_1\dots p_M}(\mu_1)\dots \tilde{B}^{(1,p_{l_M})}_{p_1\dots p_M}(\mu_M)|1\rangle_{p_1 \dots p_M} | 3 \rangle_{\genfrac{}{}{0pt}{}{1\dots L}{\ne p_1 \dots p_M}},
\end{array}
\label{qqww}\end{equation}
where $\tilde{B}^{(1,p_j)}_{p_1\dots p_M}(\mu)$ is given by Eq. (\ref{H11}).

Since the product 
$\prod^M_{i=1} \prod^{L}_{\genfrac{}{}{0pt}{}{ j=1}{\ne p_1, \dots, p_M}} \frac{a_3(\mu_i,\xi_j)}{b_{31}(\xi_j,\xi_{p_{l_i}}) \theta_3(\xi_{p_{l_i}},\xi_j)}$ is independent of the value of $l_1, \dots, l_M$, and 
the operators $\tilde{B}^{(1,p_{l})}_{p_1\dots p_M}(\mu)$ act trivially on the vector 
space $V_{i}$, $i \ne p_1,\dots,p_M$, we can decrease the number of 
relevant vector spaces in the reference states by $L-M$ and perform the following simplification to Eq.(\ref{qqww}),
\begin{equation*}\begin{array}{ll}
& \,_{1 \dots L}\langle 3 | \tilde{B}^{(1)}_{1\dots L}(\mu_1)\dots \tilde{B}^{(1)}_{1\dots L}(\mu_M)|1\rangle_{p_1 \dots p_M} | 3 \rangle_{\genfrac{}{}{0pt}{}{1\dots L}{\ne p_1 \dots p_M}}\\
=&\prod^M_{i=1} \prod^{L}_{\genfrac{}{}{0pt}{}{ j=1}{\ne p_1, \dots, p_M}} \frac{a_3(\mu_i,\xi_j)}{b_{31}(\xi_j,\xi_{p_{i}}) \theta_3(\xi_{p_{i}},\xi_j)} \sum^M_{\genfrac{}{}{0pt}{}{ l_1, \dots, l_M=1}{l_1 \ne \dots \ne l_M } } \,_{p_1 \dots p_M}\langle 3 | \tilde{B}^{(1,p_{l_1})}_{p_1\dots p_M}(\mu_1)\dots \tilde{B}^{(1,p_{l_M})}_{p_1\dots p_M}(\mu_M)|1\rangle_{p_1 \dots p_M} \\
=&\prod^M_{i=1} \prod^{L}_{\genfrac{}{}{0pt}{}{ j=1}{\ne p_1, \dots, p_M}} \frac{a_3(\mu_i,\xi_j)}{b_{31}(\xi_j,\xi_{p_{i}}) \theta_3(\xi_{p_{i}},\xi_j)}\,_{p_1 \dots p_M}\langle 3 | \tilde{B}^{(1)}_{p_1\dots p_M}(\mu_1)\dots \tilde{B}^{(1)}_{p_1\dots p_M}(\mu_M)|1\rangle_{p_1 \dots p_M}\\
=&\prod^M_{i=1} \prod^{L}_{\genfrac{}{}{0pt}{}{ j=1}{\ne p_1, \dots, p_M}} \frac{a_3(\mu_i,\xi_j)}{b_{31}(\xi_j,\xi_{p_{i}}) \theta_3(\xi_{p_{i}},\xi_j)}Z^{(B,1)}_M(\{\mu_i\}_{i=1,\dots,M}|\{\xi_{p_j}\}_{j=1,\dots,M} ).
\end{array}
\end{equation*} 
Applying all the results from this section we obtain the following form for the mixed DWPF of type $B$ completely in terms of Boltzmann weights and the $\mathcal{F}$-matrix sandwiched between reference states,
\begin{equation}\begin{array}{lll}
 Z^{(B)}_{L,M}(\{\mu\},\{\xi\})&=& \sum_{1 \le p_1 < \dots < p_M \le L}\,_{p_1 \dots p_M}\langle 1 |\,_{\genfrac{}{}{0pt}{}{1\dots L}{\ne p_1 \dots p_M}}\langle 2 |\mathcal{F}_{1\dots L} |1\rangle_{q_1 \dots q_M} | 2 \rangle_{\genfrac{}{}{0pt}{}{1\dots L}{\ne q_1 \dots q_M}}\\
&& \times \prod^L_{i=M+1}\prod^{M}_{j=1}b_{31}(\mu_i,\xi_{p_j}) \prod^M_{k=1} \prod^{L}_{\genfrac{}{}{0pt}{}{ l=1}{\ne p_1, \dots, p_M}} \frac{a_3(\mu_k,\xi_l)}{b_{31}(\xi_l,\xi_{p_{k}}) \theta_3(\xi_{p_{k}},\xi_l)} \\
&&\times Z^{(B,1)}_M(\{\mu_m\}_{m=1,\dots,M}|\{\xi_{p_n}\}_{n=1,\dots,M}) Z^{(B,2)}_{L-M}(\{\mu_{r}\}_{r=M+1,\dots,L}|\{ \xi_s\}_{\genfrac{}{}{0pt}{}{s=1,\dots, L}{\ne p_1, \dots, p_M}}).
\end{array}\label{completeB}\end{equation}
\subsection{Mixed DWPF for C}
We now apply the $\mathcal{F}$-matrix to Eq.(\ref{mixedC}) to twist the monodromy operators to obtain,
\begin{equation}
 Z^{(C)}_{L,M}(\{\nu\},\{\xi\}) = \,_{q_1 \dots q_M}\langle 1 |\,_{\genfrac{}{}{0pt}{}{1\dots L}{\ne q_1 \dots q_M}}\langle 2 |\mathcal{F}^{-1}_{1\dots L} \tilde{C}^{(2)}_{1\dots L}(\nu_L)\dots \tilde{C}^{(2)}_{1\dots L}(\nu_{M+1}) \tilde{C}^{(1)}_{1\dots L}(\nu_{M})\dots \tilde{C}^{(1)}_{1\dots L}(\nu_{1}) |3\rangle_{1\dots L}.
\label{mixedC1}\end{equation}
Similarly to the previous mixed DWPF we note that there is an $\mathcal{F}^{-1}_{1\dots L}$ operator on the far left of Eq.(\ref{mixedC1}) as there is no equivalent expression such as Eq.(\ref{actref}) for mixed transpose reference states.
As in the previous calculation for $Z^{(B)}_{L,M}$ we proceed by inserting two complete sets of states into Eq.(\ref{mixedC1}) in separate stages for the sake of clarity. Consider first placing a complete set of states in between the operators $\tilde{C}^{(2)}_{1\dots L}(\nu_{M+1})$ and $ \tilde{C}^{(1)}_{1\dots L}(\nu_{M})$ to obtain,
\begin{equation}\begin{array}{lll}
 Z^{(C)}_{L,M}(\{\nu\},\{\xi\}) &=& \sum_{1 \le p_1 < \dots < p_M \le L} \,_{p_1 \dots p_M}\langle 1 |\,_{\genfrac{}{}{0pt}{}{1\dots L}{\ne p_1 \dots p_M}}\langle 3 | \tilde{C}^{(1)}_{1\dots L}(\nu_{M})\dots \tilde{C}^{(1)}_{1\dots L}(\nu_{1}) |3\rangle_{1\dots L}\\
&&\times  \,_{q_1 \dots q_M}\langle 1 |\,_{\genfrac{}{}{0pt}{}{1\dots L}{\ne q_1 \dots q_M}}\langle 2 |\mathcal{F}^{-1}_{1\dots L} \tilde{C}^{(2)}_{1\dots L}(\nu_L)\dots \tilde{C}^{(2)}_{1\dots L}(\nu_{M+1})|1\rangle_{p_1 \dots p_M} |3\rangle_{\genfrac{}{}{0pt}{}{1\dots L}{\ne p_1 \dots p_M}}.
\end{array}\label{mixedC2}\end{equation}
To verify that the above terms of the complete set of states are the only ones which are non zero, consider the expression, 
\begin{equation*}
 \tilde{C}^{(1)}_{1\dots L}(\nu_{M})\dots \tilde{C}^{(1)}_{1\dots L}(\nu_{1}) |3\rangle_{1\dots L},
\end{equation*}
and recall from Section \ref{arcolad} that only the Weyl matrix $e^{13}_l$ (as opposed to $e^{23}_{l_1}\otimes e^{12}_{l_2}$) gives non zero terms when applied to the reference state $|3\rangle_{1\dots L}$. Hence since $e^{13}_l |3 \rangle_l = |1 \rangle_l$, this means the only terms in the complete set of states which are non zero are,
\begin{equation*} 
 \sum_{1 \le p_1 < \dots < p_M \le L} |1\rangle_{p_1 \dots p_M} | 3 \rangle_{\genfrac{}{}{0pt}{}{1\dots L}{\ne p_1 \dots p_M}} \,_{p_1 \dots p_M}\langle 1 |\,_{\genfrac{}{}{0pt}{}{1\dots L}{\ne p_1 \dots p_M}}\langle 3 |.
\end{equation*}
We now insert another complete set of states in between $\mathcal{F}^{-1}_{1\dots L}$ and $\tilde{C}^{(2)}_{1\dots L}(\nu_L)$ in Eq.(\ref{mixedC2}) to obtain,
\begin{equation}\begin{array}{lll}
 Z^{(C)}_{L,M}(\{\nu\},\{\xi\})
&=& \sum_{1 \le p_1 < \dots < p_M \le L} \,_{q_1 \dots q_M}\langle 1 |\,_{\genfrac{}{}{0pt}{}{1\dots L}{\ne q_1 \dots q_M}}\langle 2 |\mathcal{F}^{-1}_{1\dots L}|1\rangle_{p_1 \dots p_M}|2\rangle_{\genfrac{}{}{0pt}{}{1\dots L}{\ne p_1 \dots p_M}} \\
&& \times \,_{p_1 \dots p_M}\langle 1 |\,_{\genfrac{}{}{0pt}{}{1\dots L}{\ne p_1 \dots p_M}}\langle 3 | \tilde{C}^{(1)}_{1\dots L}(\nu_{M})\dots \tilde{C}^{(1)}_{1\dots L}(\nu_{1}) |3\rangle_{1\dots L}\\
&&\times   \,_{p_1 \dots p_M}\langle 1 |\,_{\genfrac{}{}{0pt}{}{1\dots L}{\ne p_1 \dots p_M}}\langle 2 | \tilde{C}^{(2)}_{1\dots L}(\nu_L)\dots \tilde{C}^{(2)}_{1\dots L}(\nu_{M+1})|1\rangle_{p_1 \dots p_M} |3\rangle_{\genfrac{}{}{0pt}{}{1\dots L}{\ne p_1 \dots p_M}},
\end{array}\label{mixedC3}\end{equation}
and proceed to decrease the number of relevant vector spaces in each of the above pseudo basic DWPF expressions.
\subsubsection{Reducing the number of relevant vector spaces - I}
Beginning with $ \,_{p_1 \dots p_M}\langle 1 |\,_{\genfrac{}{}{0pt}{}{1\dots L}{\ne p_1 \dots p_M}}\langle 2 | \tilde{C}^{(2)}_{1\dots L}(\nu_L)\dots \tilde{C}^{(2)}_{1\dots L}(\nu_{M+1})|1\rangle_{p_1 \dots p_M} |3\rangle_{\genfrac{}{}{0pt}{}{1\dots L}{\ne p_1 \dots p_M}}$, we use the fact that $e^{(23)}_p |1 \rangle_p =0$ to discard the values $(p_1,\dots, p_M)$ in the summations of the $\tilde{C}^{(2)}_{1\dots L}(\nu)$ operators. Hence applying this fact, and elementary matrix multiplication, the expression simplifies to,
\begin{equation}\begin{array}{ll}
&\,_{p_1 \dots p_M}\langle 1 |\,_{\genfrac{}{}{0pt}{}{1\dots L}{\ne p_1 \dots p_M}}\langle 2 | \tilde{C}^{(2)}_{1\dots L}(\nu_L)\dots \tilde{C}^{(2)}_{1\dots L}(\nu_{M+1})|1\rangle_{p_1 \dots p_M} |3\rangle_{\genfrac{}{}{0pt}{}{1\dots L}{\ne p_1 \dots p_M}}\\
=& \prod^L_{i=M+1}\prod^{M}_{j=1}b_{21}(\nu_i,\xi_{p_j}) \,_{p_1 \dots p_M}\langle 1 |\,_{\genfrac{}{}{0pt}{}{1\dots L}{\ne p_1 \dots p_M}}\langle 2 | \tilde{C}^{(2)}_{\genfrac{}{}{0pt}{}{1\dots L}{\ne p_1 \dots p_M}}(\nu_L)\dots \tilde{C}^{(2)}_{\genfrac{}{}{0pt}{}{1\dots L}{\ne p_1 \dots p_M}}(\nu_{M+1})|1\rangle_{p_1 \dots p_M} |3\rangle_{\genfrac{}{}{0pt}{}{1\dots L}{\ne p_1 \dots p_M}}.
\end{array}
\label{sssx}\end{equation}
Since the operators $\tilde{C}^{(2)}_{\genfrac{}{}{0pt}{}{ 1 \dots L}{\ne p_1 \dots p_M}}(\nu)$ act 
trivially on the vector space $V_{p_i}$, $i=1,\dots,M$, we 
can decrease the number of relevant vector spaces in the reference states by M - hence Eq.(\ref{sssx}) becomes, 
\begin{equation*}\begin{array}{ll}
&\,_{p_1 \dots p_M}\langle 1 |\,_{\genfrac{}{}{0pt}{}{1\dots L}{\ne p_1 \dots p_M}}\langle 2 | \tilde{C}^{(2)}_{1\dots L}(\nu_L)\dots \tilde{C}^{(2)}_{1\dots L}(\nu_{M+1})|1\rangle_{p_1 \dots p_M} |3\rangle_{\genfrac{}{}{0pt}{}{1\dots L}{\ne p_1 \dots p_M}}\\
=& \prod^L_{i=M+1}\prod^{M}_{j=1}b_{21}(\nu_i,\xi_{p_j}) \,\,_{\genfrac{}{}{0pt}{}{1\dots L}{\ne p_1 \dots p_M}}\langle 2 | \tilde{C}^{(2)}_{\genfrac{}{}{0pt}{}{1\dots L}{\ne p_1 \dots p_M}}(\nu_L)\dots \tilde{C}^{(2)}_{\genfrac{}{}{0pt}{}{1\dots L}{\ne p_1 \dots p_M}}(\nu_{M+1}) |3\rangle_{\genfrac{}{}{0pt}{}{1\dots L}{\ne p_1 \dots p_M}}\\
=& \prod^L_{i=M+1}\prod^{M}_{j=1}b_{21}(\nu_i,\xi_{p_j}) Z^{(C,2)}_{L-M} (\{\nu_{k}\}_{k=M+1,\dots,L}|\{ \xi_l\}_{\genfrac{}{}{0pt}{}{l=1,\dots, L}{\ne p_1, \dots, p_M}}).
\end{array}
\end{equation*}
\subsubsection{Reducing the number of relevant vector spaces - II}
We now focus on the term $ \,_{p_1 \dots p_M}\langle 1 |\,_{\genfrac{}{}{0pt}{}{1\dots L}{\ne p_1 \dots p_M}}\langle 3 | \tilde{C}^{(1)}_{1\dots L}(\nu_{M})\dots \tilde{C}^{(1)}_{1\dots L}(\nu_{1}) |3\rangle_{1\dots L}$. Recall
that due to the elementary relations in Eq.(\ref{action}), there is no possibility of the Weyl matrices $e^{(12)}_{l_1} \otimes e^{(23)}_{l_2}$ in the expressions for the $\tilde{C}^{(1)}_{1\dots L}$ operators to produce anything but zero, hence they can be discarded from the calculations. Additionally, using the fact that $\,_p \langle 3| e^{(13)}_p =0$, we can discard the values $l \ne p_1,\dots,p_M$ in the summations of the $\tilde{C}^{(1)}_{1\dots L}$ operators. Applying the above facts, and elementary matrix multiplication, the expression now becomes,
\begin{equation}\begin{array}{rr}
& \,_{p_1 \dots p_M}\langle 1 |\,_{\genfrac{}{}{0pt}{}{1\dots L}{\ne p_1 \dots p_M}}\langle 3 | \tilde{C}^{(1)}_{1\dots L}(\nu_{M})\dots \tilde{C}^{(1)}_{1\dots L}(\nu_{1}) |3\rangle_{1\dots L}\\
=&\sum^M_{\genfrac{}{}{0pt}{}{ l_1, \dots, l_M=1}{l_1 \ne \dots \ne l_M } } \prod^M_{i=1} \prod^{L}_{\genfrac{}{}{0pt}{}{ j=1}{\ne p_1, \dots, p_M}} a_3(\nu_i,\xi_j) \theta_3(\xi_{j},\xi_{p_{l_i}} )\\
& \times\,_{p_1 \dots p_M}\langle 1 |\,_{\genfrac{}{}{0pt}{}{1\dots L}{\ne p_1 \dots p_M}}\langle 3 | \tilde{C}^{(1,p_{l_M})}_{p_1\dots p_M}(\nu_M)\dots \tilde{C}^{(1,p_{l_1})}_{p_1\dots p_M}(\nu_1) |3\rangle_{1\dots L},
\end{array}
\label{qqxx}\end{equation}
where $\tilde{C}^{(1,p_j)}_{p_1\dots p_M}(\nu)$is given by Eq. (\ref{H22}).

Since the product
$\prod^M_{i=1} \prod^{L}_{\genfrac{}{}{0pt}{}{ j=1}{\ne p_1, \dots, p_M}} a_3(\nu_i,\xi_j) \theta_3(\xi_{j},\xi_{p_{l_i}} )$ is independent of the value of $l_1, \dots, l_M$, and the operators $\tilde{C}^{(1,p_{l})}_{p_1\dots p_M}(\nu)$ 
act trivially on the vector spaces $V_{i}$, $i \ne p_1,\dots,p_M$, we can decrease the 
number of relevant vector spaces in the reference states by $L-M$ and simplify Eq. (\ref{qqxx}) as follows,
\begin{equation*}\begin{array}{ll}
& \,_{p_1 \dots p_M}\langle 1 |\,_{\genfrac{}{}{0pt}{}{1\dots L}{\ne p_1 \dots p_M}}\langle 3 | \tilde{C}^{(1)}_{1\dots L}(\nu_{M})\dots \tilde{C}^{(1)}_{1\dots L}(\nu_{1}) |3\rangle_{1\dots L}\\
=&\prod^M_{i=1} \prod^{L}_{\genfrac{}{}{0pt}{}{ j=1}{\ne p_1, \dots, p_M}} a_3(\nu_i,\xi_j) \theta_3(\xi_{j},\xi_{p_{i}} ) \sum^M_{\genfrac{}{}{0pt}{}{ l_1, \dots, l_M=1}{l_1 \ne \dots \ne l_M } } \,_{p_1 \dots p_M}\langle 1 | \tilde{C}^{(1,p_{l_M})}_{p_1\dots p_M}(\nu_M)\dots \tilde{C}^{(1,p_{l_1})}_{p_1\dots p_M}(\nu_1) |3\rangle_{p_1\dots p_M} \\
=& \prod^M_{i=1} \prod^{L}_{\genfrac{}{}{0pt}{}{ j=1}{\ne p_1, \dots, p_M}} a_3(\nu_i,\xi_j) \theta_3(\xi_{j},\xi_{p_{i}} )  \,_{p_1 \dots p_M}\langle 1 | \tilde{C}^{(1)}_{p_1\dots p_M}(\nu_M)\dots \tilde{C}^{(1)}_{p_1\dots p_M}(\nu_1) |3\rangle_{p_1\dots p_M}\\
=& \prod^M_{i=1} \prod^{L}_{\genfrac{}{}{0pt}{}{ j=1}{\ne p_1, \dots, p_M}} a_3(\nu_i,\xi_j) \theta_3(\xi_{j},\xi_{p_{i}} )  Z^{(C,1)}_M(\{\nu_i\}_{i=1,\dots,M}|\{\xi_{p_j}\}_{j=1,\dots,M} ).
\end{array}
\end{equation*}
Hence we obtain the 
following form for the mixed DWPF of type $C$ completely in terms of Boltzmann weights and the 
inverse $\mathcal{F}$-matrix sandwiched between reference states,
\begin{equation}\begin{array}{lll}
 Z^{(C)}_{L,M}(\{\nu\},\{\xi\})&=& \sum_{1 \le p_1 < \dots < p_M \le L} \,_{q_1 \dots q_M}\langle 1 |\,_{\genfrac{}{}{0pt}{}{1\dots L}{\ne q_1 \dots q_M}}\langle 2 |\mathcal{F}^{-1}_{1\dots L}|1\rangle_{p_1 \dots p_M}|2\rangle_{\genfrac{}{}{0pt}{}{1\dots L}{\ne p_1 \dots p_M}}\\
&& \times \prod^L_{i=M+1}\prod^{M}_{j=1}b_{21}(\nu_i,\xi_{p_j}) \prod^M_{k=1} \prod^{L}_{\genfrac{}{}{0pt}{}{ l=1}{\ne p_1, \dots, p_M}} a_3(\nu_k,\xi_l) \theta_3(\xi_{l},\xi_{p_{k}} ) \\
&&\times Z^{(C,1)}_M(\{\nu_m\}_{m=1,\dots,M}|\{\xi_{p_n}\}_{n=1,\dots,M} )Z^{(C,2)}_{L-M}(\{\nu_{r}\}_{r=M+1,\dots,L}|\{ \xi_s\}_{\genfrac{}{}{0pt}{}{s=1,\dots, L}{\ne p_1, \dots, p_M}}).
\end{array}\label{completeC}\end{equation}

\section{Conclusions}\label{CONCLU}

In this article we have argued that the factorized $F$-matrices associated to the $R$-matrix
of the $U(1)^{(N-1)}$ vertex model can be constructed for arbitrary Boltzmann weights.
Our analysis is purely algebraic relying only on the structure of the $R$-matrix
as well as on the corresponding unitarity and Yang-Baxter relations. We have applied
this formulation to the $N=3$ fifteen-vertex model which is the simplest extension
of the asymmetric $N=2$ six-vertex model in such family of integrable models. For $N=3$ we
have exhibited the algebraic expressions of relevant monodromy matrix elements in the $F$-basis. This
allowed us to compute the domain wall partition functions related to the creation and
annihilation fields for arbitrary weights. 

We have motived our approach by showing that the Yang-Baxter relations for $N=3$ hide
a general structure of Boltzmann weights. The underlying algebraic variety is at least
governed by the  intersection
of two quadrics in the projective $\mathbb{P}^4$ space leading 
us to a surface of del Pezzo type. Interestingly enough, this type of variety also governs the
integrability of the 
$N=2$ vertex model. In fact, one can show that the weights of the asymmetric
six-vertex model lie on a cubic del Pezzo surface. A natural question to investigate is
whether the del Pezzo structure persists for general $N$ or even higher dimension algebraic
variety emerges when $N > 3$. In any case, this observation emphasizes 
the importance of attempts to
establish results for integrable vertex models that are independent of 
any specific
parameterization of Boltzmann weights.

\section*{Acknowledgments}
The authors thank the Brazilian Research Agencies FAPESP and CNPq for financial support.

\addcontentsline{toc}{section}{Appendix A}
\section*{\bf Appendix A: The Yang-Baxter for $N=3$ } \setcounter{equation}{0}
\renewcommand{\theequation}{A.\arabic{equation}}
Here we describe some details entering the
general solution of the Yang-Baxter equation for $N=3$ exhibited in Section
\ref{VERTEX}. The algebraic solution consists in the elimination of the weights
dependent on the variables $\xi_1$ and $\xi_2$ leading to determine the algebraic invariants
constraining the remaining Boltzmann weights. We
start by solving the relations involving only two triple products,
see Eqs.(\ref{YB1}-\ref{YB3}). After eliminating
the weights $c_{21}(\xi_1,\xi_2)$, $c_{32}(\xi_1,\xi_2)$, $b_{31}(\xi_1,\xi_2)$,
$b_{13}(\xi_1,\xi_2)$, $c_{13}(\xi_1,\xi_2)$ and $b_{23}(\xi_1,\xi_2)$ one finds that
there exists only three independent relations. They are easily
separable providing us the following invariants,
\begin{equation}
\frac{b_{32}(\xi_i,\xi_3)}{b_{12}(\xi_i,\xi_3)}=\delta_1,~~~~~
\frac{b_{31}(\xi_i,\xi_3)}{b_{21}(\xi_i,\xi_3)}=\delta_2,~~~~~
\frac{b_{13}(\xi_i,\xi_3)}{b_{23}(\xi_i,\xi_3)}=\delta_3~~\text{for}~~i=1,2,
\label{inv2termos}\end{equation}
where $\delta_1$, $\delta_2$ and $\delta_3$ are free parameters.

Taking into account this solution the number of relations with three triple products
(\ref{YB4}-\ref{YB11})
reduce to thirty independent functional equations. Among
of them there exists eight relations which are suitable
for carrying out further elimination of weights.
Their explicit forms are,
\begin{eqnarray}
c_{12}(\xi_1,\xi_2)a_{1}(\xi_1,\xi_3)b_{21}(\xi_2,\xi_3) &+&
b_{21}(\xi_1,\xi_2)c_{12}(\xi_1,\xi_3)c_{21}(\xi_2,\xi_3)\nonumber\\&=&
c_{12}(\xi_1,\xi_2)b_{21}(\xi_1,\xi_3)a_{1}(\xi_2,\xi_3),
\label{neweq2}\end{eqnarray}
\begin{eqnarray}
b_{21}(\xi_1,\xi_2)c_{12}(\xi_1,\xi_3)b_{12}(\xi_2,\xi_3) &+&
c_{12}(\xi_1,\xi_2)a_{1}(\xi_1,\xi_3)c_{12}(\xi_2,\xi_3)\nonumber\\&=&
a_{1}(\xi_1,\xi_2)c_{12}(\xi_1,\xi_3)a_{1}(\xi_2,\xi_3),
\label{neweq1}\end{eqnarray}
\begin{eqnarray}
b_{12}(\xi_1,\xi_2)a_{1}(\xi_1,\xi_3)c_{21}(\xi_2,\xi_3)&+&
c_{12}(\xi_1,\xi_2)c_{21}(\xi_1,\xi_3)b_{12}(\xi_2,\xi_3)\nonumber\\&=&
a_{1}(\xi_1,\xi_2)b_{12}(\xi_1,\xi_3)c_{21}(\xi_2,\xi_3),
\label{neweq5}\end{eqnarray}
\begin{eqnarray}
c_{12}(\xi_1,\xi_2)a_{2}(\xi_1,\xi_3)b_{12}(\xi_2,\xi_3)&+&
b_{12}(\xi_1,\xi_2)c_{12}(\xi_1,\xi_3)c_{21}(\xi_2,\xi_3)\nonumber\\&=&
c_{12}(\xi_1,\xi_2)b_{12}(\xi_1,\xi_3)a_{2}(\xi_2,\xi_3),
\label{neweq27}\end{eqnarray}
\begin{eqnarray}
c_{23}(\xi_1,\xi_2)a_{3}(\xi_1,\xi_3)b_{21}(\xi_1,\xi_3)b_{23}(\xi_2,\xi_3)&+&
b_{21}(\xi_1,\xi_2)c_{23}(\xi_1,\xi_3)b_{23}(\xi_1,\xi_3)c_{32}(\xi_2,\xi_3)\nonumber\\&=&
c_{23}(\xi_1,\xi_2)b_{23}(\xi_1,\xi_3)b_{21}(\xi_1,\xi_3)a_{3}(\xi_2,\xi_3),
\label{neweq66}\end{eqnarray}
\begin{eqnarray}
\delta_{1}b_{21}(\xi_1,\xi_2)c_{23}(\xi_1,\xi_3)b_{23}(\xi_1,\xi_3)b_{12}(\xi_2,\xi_3)&+&
c_{23}(\xi_1,\xi_2)a_{3}(\xi_1,\xi_3)b_{21}(\xi_1,\xi_3)c_{23}(\xi_2,\xi_3)\nonumber\\&=&
a_{3}(\xi_1,\xi_2)c_{23}(\xi_1,\xi_3)b_{21}(\xi_1,\xi_3)a_{3}(\xi_2,\xi_3),
\label{neweq47}\end{eqnarray}
\begin{eqnarray}
\delta_{1}c_{23}(\xi_1,\xi_2)c_{32}(\xi_1,\xi_3)b_{12}(\xi_2,\xi_3) &+&
b_{32}(\xi_1,\xi_2)a_{3}(\xi_1,\xi_3)c_{32}(\xi_2,\xi_3)
\nonumber\\&=&
\delta_{1}a_{3}(\xi_1,\xi_2)b_{12}(\xi_1,\xi_3)c_{32}(\xi_2,\xi_3),
\label{neweq48}\end{eqnarray}
\begin{eqnarray}
\delta_{1}c_{23}(\xi_1,\xi_2)a_{2}(\xi_1,\xi_3)b_{12}(\xi_2,\xi_3)
 &+&
b_{32}(\xi_1,\xi_2)c_{23}(\xi_1,\xi_3)c_{32}(\xi_2,\xi_3)\nonumber\\&=&
\delta_{1}c_{23}(\xi_1,\xi_2)b_{12}(\xi_1,\xi_3)a_{2}(\xi_2,\xi_3).
\label{neweq38}\end{eqnarray}

From Eqs.(\ref{neweq2}-\ref{neweq5}) we are able to eliminate the
weights $b_{21}(\xi_1,\xi_2)$, $a_1(\xi_1,\xi_2)$, $b_{12}(\xi_1,\xi_2)$ and
by substituting them in Eq.(\ref{neweq27}) we find that it leads to the following
constraint,
\begin{equation}
\frac{a_1(\xi_i,\xi_3)a_2(\xi_i,\xi_3) + b_{12}(\xi_i,\xi_3)b_{21}(\xi_i,\xi_3) - c_{12}(\xi_i,\xi_3)c_{21}(\xi_i,\xi_3)}
{a_{1}(\xi_i,\xi_3)b_{12}(\xi_i,\xi_3)}=\delta_4~~\text{for}~~i=1,2,
\label{inv4}\end{equation}
where $\delta_4$ is a constant.

The same procedure can be implemented for Eqs.(\ref{neweq66}-\ref{neweq38}). By eliminating
the weights $a_3(\xi_1,\xi_2)$, $b_{32}(\xi_1,\xi_2)$ and $c_{23}(\xi_1,\xi_2)$ one
obtains the additional constraint,
\begin{equation}
\frac{a_{3}(\xi_i,\xi_3)a_{2}(\xi_i,\xi_3) + \delta_{1}b_{12}(\xi_i,\xi_3)b_{23}(\xi_i,\xi_3) - c_{23}(\xi_i,\xi_3)c_{32}(\xi_i,\xi_3)}
{a_{3}(\xi_i,\xi_3)b_{12}(\xi_i,\xi_3)}=\delta_5~~\text{for}~~i=1,2,
\label{inv5}\end{equation}
where $\delta_5$ is a free parameter.

At this point we are left to eliminate only the Boltzmann weights $c_{13}(\xi_1,\xi_2)$ and
$a_2(\xi_1,\xi_2)$. This is done by performing linear combinations among certain
remaining three terms relations coming from Eqs.(\ref{YB4}-\ref{YB11}). Remarkably enough all the consistency conditions
are solved by means of the following extra invariants,
\begin{equation}
\frac{\delta_{1}a_{1}(\xi_i,\xi_3)b_{23}(\xi_i,\xi_3) - a_{3}(\xi_i,\xi_3)b_{21}(\xi_i,\xi_3)}
{b_{23}(\xi_i,\xi_3)b_{21}(\xi_i,\xi_3)}=\delta_6~~\text{for}~~i=1,2,
\label{inv6}\end{equation}
\begin{eqnarray}
&&\frac{\delta_{1}\left[a_{1}(\xi_1,\xi_3)c_{23}(\xi_1,\xi_3) - c_{13}(\xi_1,\xi_3)c_{21}(\xi_1,\xi_3)\right]}
{c_{23}(\xi_1,\xi_3)b_{21}(\xi_1,\xi_3)}=\nonumber\\&&
\frac{\delta_{1}a_{1}(\xi_2,\xi_3)c_{12}(\xi_2,\xi_3)b_{23}(\xi_2,\xi_3) - c_{13}(\xi_2,\xi_3)b_{21}(\xi_2,\xi_3)c_{32}(\xi_2,\xi_3)}
{c_{12}(\xi_2,\xi_3)b_{23}(\xi_2,\xi_3)b_{21}(\xi_2,\xi_3)}=\delta_7,
\label{inv7}\end{eqnarray}
\begin{eqnarray}
&&\frac{\delta_{1}a_{1}(\xi_1,\xi_3)c_{12}(\xi_1,\xi_3)b_{23}(\xi_1,\xi_3) - c_{13}(\xi_1,\xi_3)b_{21}(\xi_1,\xi_3)c_{32}(\xi_1,\xi_3)}
{c_{12}(\xi_1,\xi_3)b_{23}(\xi_1,\xi_3)b_{21}(\xi_1,\xi_3)}=\nonumber\\&&
\frac{\delta_{1}\left[a_{1}(\xi_2,\xi_3)c_{23}(\xi_2,\xi_3) - c_{13}(\xi_2,\xi_3)c_{21}(\xi_2,\xi_3)\right]}
{c_{23}(\xi_2,\xi_3)b_{21}(\xi_2,\xi_3)}=\delta_8,
\label{inv8}\end{eqnarray}
\begin{eqnarray}
&&\frac{\delta_{1}\left[a_{1}(\xi_1,\xi_3)c_{32}(\xi_1,\xi_3) - c_{12}(\xi_1,\xi_3)c_{31}(\xi_1,\xi_3)\right]}
{b_{21}(\xi_1,\xi_3)c_{32}(\xi_1,\xi_3)}=\nonumber\\&&
\frac{\delta_{1}a_{1}(\xi_2,\xi_3)b_{23}(\xi_2,\xi_3)c_{21}(\xi_2,\xi_3) - c_{23}(\xi_2,\xi_3)b_{21}(\xi_2,\xi_3)c_{31}(\xi_2,\xi_3)}
{b_{23}(\xi_2,\xi_3)b_{21}(\xi_2,\xi_3)c_{21}(\xi_2,\xi_3)}=\delta_9,
\label{inv9}\end{eqnarray}
\begin{eqnarray}
&&\frac{\delta_{1}a_{1}(\xi_1,\xi_3)b_{23}(\xi_1,\xi_3)c_{21}(\xi_1,\xi_3) - c_{23}(\xi_1,\xi_3)b_{21}(\xi_1,\xi_3)c_{31}(\xi_1,\xi_3)}
{b_{23}(\xi_1,\xi_3)b_{21}(\xi_1,\xi_3)c_{21}(\xi_1,\xi_3)}=\nonumber\\&&
\frac{\delta_{1}\left[a_{1}(\xi_2,\xi_3)c_{32}(\xi_2,\xi_3) - c_{12}(\xi_2,\xi_3)c_{31}(\xi_2,\xi_3)\right]}
{b_{21}(\xi_2,\xi_3)c_{32}(\xi_2,\xi_3)}=\delta_{10},
\label{inv10}\end{eqnarray}
where $\delta_6$, $\delta_7$, $\delta_8$, $\delta_9$ and $\delta_{10}$ are yet
new free parameters.

It turns out that the remaining Yang-Baxter functional relations
lead us to branches that impose
further constraints among certain
weights and the invariants
obtained so far. We find that one such possible branch is,
\begin{equation}
 \frac{a_{2}(\xi_i,\xi_3)b_{21}(\xi_i,\xi_3)}{a_1(\xi_i,\xi_3)b_{12}(\xi_i,\xi_3)} =
\delta_1\frac{a_{2}(\xi_i,\xi_3)b_{23}(\xi_i,\xi_3)}{a_3(\xi_i,\xi_3)b_{12}(\xi_i,\xi_3)} =
\frac{\delta_1(\delta_4\delta_7-\delta_1)}{\delta_7^2}~~\text{for}~~i=1,2,
\label{branch1}\end{equation}
while the invariants $\delta_3, \delta_5,\delta_6, \delta_8, \delta_9, \delta_{10}$
are fixed by,
\begin{equation}
\delta_3=\frac{\delta_7^2}{\delta_2(\delta_4\delta_7-\delta_1)},~~~~\delta_5=\delta_4,~~~~\delta_6=0,~~~~\delta_8=\delta_7,~~~~
\delta_9=\delta_{10}=\frac{\delta_1 \delta_7}{\delta_4 \delta_7-\delta_1}.
\label{invtravas}\end{equation}

We have now reached a point where all the weights
in the variables $\xi_1$ and $\xi_2$ have been
eliminated, while the weights in the variables
$\xi_{1,2}$ and $\xi_3$ are constrained by the
algebraic invariants (\ref{inv2termos},\ref{inv4}-\ref{branch1}).
The final step of our analysis consists 
to make the intersection of these algebraic invariants.
The procedure for performing such intersection
is as follows. We first note that the weights $a_2(\xi_i,\xi_3)$,
$b_{23}(\xi_i,\xi_3)$, $b_{31}(\xi_i,\xi_3)$, $b_{32}(\xi_i,\xi_3)$,
$a_3(\xi_i,\xi_3)$, $c_{13}(\xi_i,\xi_3)$ and $c_{32}(\xi_i,\xi_3)$
can be linearly extracted from (\ref{inv2termos},\ref{inv5}-\ref{branch1}),
leading us to,
\begin{equation}
a_2(\xi_i,\xi_3) = \frac{\delta_{1} (\delta_{1} - \delta_{4} \delta_{7}) a_{1}(\xi_i,\xi_3) c_{12}(\xi_i,\xi_3) c_{21}(\xi_i,\xi_3)}{\left[\delta_{1} a_{1}(\xi_i,\xi_3) - \delta_{7} b_{21}(\xi_i,\xi_3)\right]\left[(\delta_{1} - \delta_{4} \delta_{7}) a_{1}(\xi_i,\xi_3) + \delta_{7} b_{21}(\xi_i,\xi_3)\right]},
\label{expa2}\end{equation}
\begin{equation}
b_{23}(\xi_i,\xi_3) = \frac{(\delta_{1} - \delta_{4} \delta_{7}) b_{21}(\xi_i,\xi_3) c_{23}(\xi_i,\xi_3) c_{31}(\xi_i,\xi_3)}{c_{21}(\xi_i,\xi_3)\delta_{1} \left[(\delta_{1} - \delta_{4} \delta_{7}) a_{1}(\xi_i,\xi_3) +\delta_{7} b_{21}(\xi_i,\xi_3)\right]},
\label{expb23}\end{equation}
\begin{equation}
b_{31}(\xi_i,\xi_3) = \delta_{2} b_{21}(\xi_i,\xi_3),
\label{expb31}\end{equation}
\begin{equation}
b_{32}(\xi_i,\xi_3) = \frac{\delta_{1} \delta_{7}^2 b_{21}(\xi_i,\xi_3) c_{12}(\xi_i,\xi_3) c_{21}(\xi_i,\xi_3)}{\left[\delta_{1} a_{1}(\xi_i,\xi_3) - \delta_{7} b_{21}(\xi_i,\xi_3)\right]\left[(-\delta_{1} + \delta_{4} \delta_{7}) a_{1}(\xi_i,\xi_3) - \delta_{7} b_{21}(\xi_i,\xi_3)\right]},
\label{expb32}\end{equation}
\begin{equation}
a_3(\xi_i,\xi_3) = \frac{(\delta_{1} - \delta_{4} \delta_{7}) a_{1}(\xi_i,\xi_3) c_{23}(\xi_i,\xi_3) c_{31}(\xi_i,\xi_3)}{c_{21}(\xi_i,\xi_3)\left[(\delta_{1} - \delta_{4} \delta_{7}) a_{1}(\xi_i,\xi_3) + \delta_{7} b_{21}(\xi_i,\xi_3)\right]},
\label{expa3}\end{equation}
\begin{equation}
c_{13}(\xi_i,\xi_3) = \frac{c_{23}(\xi_i,\xi_3)\left[\delta_{1} a_{1}(\xi_i,\xi_3) - \delta_{7} b_{21}(\xi_i,\xi_3)\right] }{\delta_{1} c_{21}(\xi_i,\xi_3)},
\label{expc13}\end{equation}
\begin{equation}
c_{32}(\xi_i,\xi_3) = \frac{(\delta_{1} - \delta_{4} \delta_{7}) c_{12}(\xi_i,\xi_3) c_{31}(\xi_i,\xi_3)}{(\delta_{1} - \delta_{4} \delta_{7}) a_{1}(\xi_i,\xi_3) + \delta_{7} b_{21}(\xi_i,\xi_3)},
\label{expc32}\end{equation}
where $i=1,2$.

We next substitute the
weight (\ref{expa2}) in Eq.(\ref{inv4})
and as result we find that the weights $a_1(\xi_i,\xi_3)$,
$b_{12}(\xi_i,\xi_3)$, $b_{21}(\xi_i,\xi_3)$,
$c_{12}(\xi_i,\xi_3)$ and $c_{21}(\xi_i,\xi_3)$
are constrained by,
\begin{eqnarray}
&&\left[\frac{\delta_1(\delta_4\delta_7-\delta_1)}{\delta_7^2}\right]a_1(\xi_i,\xi_3)^2b_{12}(\xi_i,\xi_3)-
\delta_4a_1(\xi_i,\xi_3)b_{12}(\xi_i,\xi_3)b_{21}(\xi_i,\xi_3)
\nonumber\\&&+
b_{12}(\xi_i,\xi_3)b_{21}(\xi_i,\xi_3)^2-b_{21}(\xi_i,\xi_3)c_{12}(\xi_i,\xi_3)c_{21}(\xi_i,\xi_3)=0,~~\textrm{for}~~i=1,2.
\label{hyperapp}\end{eqnarray}
By performing the definition $\delta_7=\delta_1 \Delta_1$
and $\delta_4 = \Delta_2$ we see that the form
of Eq.(\ref{hyperapp}) is the same as that of
the hypersurface (\ref{HYPER}) given in the main text.
This is the case because the spectral parameter
$\xi_3$ is a common variable for all the weights entering
Eq.(\ref{hyperapp}). Therefore, through the identification,
\begin{eqnarray}
&&a_2(\xi_i,\xi_3)=a(\xi_i),~~b_{12}(\xi_i,\xi_3)=b(\xi_i),~~
b_{21}(\xi_i,\xi_3)=\bar{b}(\xi_i),~~c_{12}(\xi_i,\xi_3)=c(\xi_i),~~
c_{21}(\xi_i,\xi_3)=\bar{c}(\xi_i),\nonumber\\
&&c_{23}(\xi_i,\xi_3)=h_1(\xi_i),~~c_{31}(\xi_i,\xi_3)=h_2(\xi_i),
\label{notacao1}\end{eqnarray}
we see that Eq.(\ref{hyperapp}) becomes exactly Eq.(\ref{HYPER}).

We conclude by observing that the earlier $U(1)\otimes U(1)$
Yang-Baxter solution presented in the literature
\cite{DEV,PK} is indeed a particular case of the $R-$matrix
given in the text.
In fact, the so called Perk-Schultz solution associated to the $U_q[Su(3)]$ quantum
algebra is obtained
by setting,
\begin{eqnarray}
&&\delta_1=\delta_2=1,~~\Delta_1=q,~~\Delta_4=q+1/q,~~a(\xi)=1,~~\bar{b}(\xi)=b(\xi)=q\frac{\xi^2-1}{\xi^2-q^2},\nonumber\\&&
h_1(\xi)=\xi h_2(\xi)=\bar{c}(\xi)=c(\xi)=\xi\frac{q^2-1}{q^2-\xi^2},
\end{eqnarray}
where $q$ is a free constant and $\xi$ is the spectral
parameter. In this special case
we see that the $R-$matrix is of the difference form, i.e., $R=R(\xi_1/\xi_2)$.

\addcontentsline{toc}{section}{Appendix B}
\section*{\bf Appendix B: The analysis for $L=2$ } \setcounter{equation}{0}
\renewcommand{\theequation}{B.\arabic{equation}}
Here we present the explicit expressions of the entries (\ref{kappaL21},\ref{kappaL22})  
together with the corresponding simplifications using
the Yang-Baxter and unitarity relations;

\begin{equation*}\begin{array}{lll}
\kappa^{(D)}_1&=& c_{21}(\xi_1,\xi_2)\underbrace{\{ b_{31}(\mu,\xi_1)b_{32}(\mu,\xi_2)-b_{32}(\mu,\xi_1)b_{31}(\mu,\xi_2)\}}_{\textrm{apply Eq.(\ref{YB2})-\{3,1,2\}}}=0\\
\kappa^{(D)}_2&=&  \underbrace{b_{31}(\mu,\xi_1)a_{3}(\mu,\xi_2)c_{31}(\xi_1,\xi_2)+c_{13}(\mu,\xi_1)c_{31}(\mu,\xi_2)b_{31}(\xi_1,\xi_2)}_{\textrm{apply Eq.(\ref{YB4})-\{3,1\}}}\\
&& -a_{3}(\mu,\xi_1)b_{31}(\mu,\xi_2)c_{31}(\xi_1,\xi_2)=0\\
\kappa^{(D)}_3&=&  \underbrace{b_{32}(\mu,\xi_1)a_{3}(\mu,\xi_2)c_{32}(\xi_1,\xi_2)+c_{23}(\mu,\xi_1)c_{32}(\mu,\xi_2)b_{32}(\xi_1,\xi_2)}_{\textrm{apply Eq.(\ref{YB4})-\{3,2\}}}
\\
&&-a_{3}(\mu,\xi_1)b_{32}(\mu,\xi_2)c_{32}(\xi_1,\xi_2)=0
\end{array}\end{equation*}
\begin{equation*}\begin{array}{lll}
\kappa^{(C_2)}_1 &=& b_{21}(\mu,\xi_1)c_{32}(\mu,\xi_2)\\
\kappa^{(C_2)}_2 &=& a_{2}(\mu,\xi_1)c_{32}(\mu,\xi_2)-\frac{c_{32}(\mu,\xi_1)b_{32}(\mu,\xi_2)\overbrace{c_{32}(\xi_1,\xi_2)}^{\textrm{apply Eq.(\ref{bigc})-\{3,2\}}}}{b_{32}(\xi_1,\xi_2)}\\
&=& \frac{1}{b_{32}(\xi_2,\xi_1)}\underbrace{\left\{ c_{32}(\mu,\xi_2)a_{2}(\mu,\xi_1)b_{32}(\xi_2,\xi_1)+b_{32}(\mu,\xi_1)c_{32}(\mu,\xi_1)c_{23}(\xi_2,\xi_1) \right\}}_{\textrm{apply Eq.(\ref{YB9})-\{3,2\}}}\\
&=&\frac{ c_{32}(\mu,\xi_2)b_{32}(\mu,\xi_1)a_{2}(\xi_2,\xi_1)}{b_{32}(\xi_2,\xi_1)}\\
\kappa^{(C_2)}_3 &=&\underbrace{c_{32}(\mu,\xi_1)a_{3}(\mu,\xi_2)c_{32}(\xi_1,\xi_2)+b_{23}(\mu,\xi_1)c_{32}(\mu,\xi_2)b_{32}(\xi_1,\xi_2)}_{\textrm{apply Eq.(\ref{YB6})-\{3,2\}}}\\
&=&a_{3}(\mu,\xi_1)c_{32}(\mu,\xi_2)a_{3}(\xi_1,\xi_2)\\
\kappa^{(C_2)}_4 &=&\frac{c_{32}(\mu,\xi_1)b_{31}(\mu,\xi_2)b_{21}(\xi_1,\xi_2)}{b_{31}(\xi_1,\xi_2)}\\
\kappa^{(C_2)}_5 &=&\frac{c_{32}(\mu,\xi_1)b_{32}(\mu,\xi_2)}{b_{32}(\xi_1,\xi_2)}\\
\kappa^{(C_2)}_6 &=&c_{32}(\mu,\xi_1)a_{3}(\mu,\xi_2)\\
\kappa^{(C_2)}_7 &=& b_{21}(\mu,\xi_1)c_{32}(\mu,\xi_2)c_{21}(\xi_1,\xi_2)+c_{12}(\mu,\xi_1)c_{31}(\mu,\xi_2)b_{21}(\xi_1,\xi_2)\\
&&-\frac{c_{32}(\mu,\xi_1)\overbrace{b_{31}(\mu,\xi_2)b_{21}(\xi_1,\xi_2)}^{\textrm{apply Eq.(\ref{YB3})-\{2,3,1\}}}c_{31}(\xi_1,\xi_2)}{b_{31}(\xi_1,\xi_2)}\\
&=& \underbrace{b_{21}(\mu,\xi_1)c_{32}(\mu,\xi_2)c_{21}(\xi_1,\xi_2)+c_{12}(\mu,\xi_1)c_{31}(\mu,\xi_2)b_{21}(\xi_1,\xi_2)}_{\textrm{apply Eq.(\ref{YB11})-\{3,2,1\}}}\\ &&-c_{32}(\mu,\xi_1)b_{21}(\mu,\xi_2)c_{31}(\xi_1,\xi_2) = 0
\end{array}\end{equation*}
\begin{equation*}\begin{array}{lll}
 \kappa^{(B_2)}_1 &=&b_{31}(\mu,\xi_1)c_{23}(\mu,\xi_2) \\
 \kappa^{(B_2)}_2 &=& b_{32}(\mu,\xi_1)c_{23}(\mu,\xi_2)\\
\kappa^{(B_2)}_3 &=& \frac{a_{3}(\mu,\xi_1)c_{23}(\mu,\xi_2)}{b_{32}(\xi_1,\xi_2)}\\
\kappa^{(B_2)}_4 &=& \frac{c_{23}(\mu,\xi_1)b_{21}(\mu,\xi_2)b_{31}(\xi_1,\xi_2)}{b_{21}(\xi_1,\xi_2)}\\
\kappa^{(B_2)}_5 &=&\underbrace{c_{23}(\mu,\xi_1)a_{2}(\mu,\xi_2)b_{32}(\xi_1,\xi_2)+b_{32}(\mu,\xi_1)c_{23}(\mu,\xi_2)c_{32}(\xi_1,\xi_2)}_{\textrm{apply Eq.(\ref{YB8})-\{2,3\}}}\\
&=&c_{23}(\mu,\xi_1)b_{32}(\mu,\xi_2)a_{2}(\xi_1,\xi_2)\\
\kappa^{(B_2)}_6 &=& c_{23}(\mu,\xi_1)b_{23}(\mu,\xi_2)-\frac{a_{3}(\mu,\xi_1)c_{23}(\mu,\xi_2)\overbrace{c_{32}(\xi_1,\xi_2)}^{\textrm{apply Eq.(\ref{bigc})-\{2,3\}}}}{b_{32}(\xi_1,\xi_2)}\\
&=& \frac{1}{b_{32}(\xi_2,\xi_1)}\underbrace{\left\{ b_{23}(\mu,\xi_2)c_{23}(\mu,\xi_1)b_{32}(\xi_2,\xi_1)+c_{23}(\mu,\xi_2)a_{3}(\mu,\xi_1)c_{23}(\xi_2,\xi_1) \right\}}_{\textrm{apply Eq.(\ref{YB7})-\{2,3\}}}\\
&=&\frac{ a_{3}(\mu,\xi_2)c_{23}(\mu,\xi_1)a_{3}(\xi_2,\xi_1)}{b_{32}(\xi_2,\xi_1)}\\
\kappa^{(B_2)}_7 &=& c_{13}(\mu,\xi_1)c_{21}(\mu,\xi_2)b_{31}(\xi_1,\xi_2)+b_{31}(\mu,\xi_1)c_{23}(\mu,\xi_2)c_{31}(\xi_1,\xi_2)\\
&&-\frac{c_{23}(\mu,\xi_1)\overbrace{b_{21}(\mu,\xi_2)b_{31}(\xi_1,\xi_2)}^{\textrm{apply Eq.(\ref{YB3})-\{3,2,1\}}}c_{21}(\xi_1,\xi_2)}{b_{21}(\xi_1,\xi_2)}\\
&=& \underbrace{c_{13}(\mu,\xi_1)c_{21}(\mu,\xi_2)b_{31}(\xi_1,\xi_2)+b_{31}(\mu,\xi_1)c_{23}(\mu,\xi_2)c_{31}(\xi_1,\xi_2)}_{\textrm{apply Eq.(\ref{YB11})-\{2,3,1\}}}\\
&&-c_{23}(\mu,\xi_1)b_{31}(\mu,\xi_2)c_{21}(\xi_1,\xi_2) = 0
\end{array}\end{equation*}
\begin{equation*}\begin{array}{lll}
 \kappa^{(C_1)}_1 &=&a_{1}(\mu,\xi_1)c_{31}(\mu,\xi_2) - \frac{c_{31}(\mu,\xi_1)b_{31}(\mu,\xi_2)\overbrace{c_{31}(\xi_1,\xi_2)}^{\textrm{apply Eq.(\ref{bigc})-\{3,1\}}}}{b_{31}(\xi_1,\xi_2)}\\
&=&  \frac{1}{b_{31}(\xi_2,\xi_1)}\underbrace{\left\{ c_{31}(\mu,\xi_2)a_{1}(\mu,\xi_1)b_{31}(\xi_2,\xi_1)+b_{31}(\mu,\xi_2)c_{31}(\mu,\xi_1)c_{13}(\xi_2,\xi_1) \right\}}_{\textrm{apply Eq.(\ref{YB9})-\{3,1\}}}\\
&=&\frac{ c_{31}(\mu,\xi_2)b_{31}(\mu,\xi_1)a_{1}(\xi_2,\xi_1)}{b_{31}(\xi_2,\xi_1)}\\
 \kappa^{(C_1)}_2 &=& c_{21}(\mu,\xi_1)c_{32}(\mu,\xi_2)c_{21}(\xi_1,\xi_2)+b_{12}(\mu,\xi_1)c_{31}(\mu,\xi_2)b_{21}(\xi_1,\xi_2)\\
&&-\frac{c_{31}(\mu,\xi_1)b_{32}(\mu,\xi_2)c_{21}(\xi_1,\xi_2)\overbrace{c_{32}(\xi_1,\xi_2)}^{\textrm{apply Eq.(\ref{bigc})-\{3,2\}}}}{b_{32}(\xi_1,\xi_2)}\\
&=& \frac{c_{21}(\xi_1,\xi_2)}{b_{32}(\xi_2,\xi_1)} \underbrace{\{c_{32}(\mu,\xi_2)c_{21}(\mu,\xi_1)b_{32}(\xi_2,\xi_1)+b_{32}(\mu,\xi_2)c_{31}(\mu,\xi_1)c_{23}(\xi_2,\xi_1)\}}_{\textrm{apply Eq.(\ref{YB10})-\{3,2,1\}}}\\
&&+b_{12}(\mu,\xi_1)c_{31}(\mu,\xi_2)b_{21}(\xi_1,\xi_2)\\
&=&\frac{c_{31}(\mu,\xi_2)}{b_{32}(\xi_2,\xi_1)} \{\underbrace{c_{21}(\xi_1,\xi_2)c_{21}(\xi_2,\xi_1)}_{\textrm{apply Eq.(\ref{bigb})-\{2,1\}}} b_{32}(\mu,\xi_1)+b_{12}(\mu,\xi_1)b_{21}(\xi_1,\xi_2)b_{32}(\xi_2,\xi_1)\}\\
&=&\frac{c_{31}(\mu,\xi_2)}{b_{32}(\xi_2,\xi_1)} \{b_{32}(\mu,\xi_1) + b_{21}(\xi_1,\xi_2)\underbrace{[b_{12}(\mu,\xi_1)b_{32}(\xi_2,\xi_1) -b_{32}(\mu,\xi_1)b_{12}(\xi_2,\xi_1)]}_{\textrm{apply Eq.(\ref{YB3})-\{1,3,2\}}}\}\\
&=& \frac{b_{32}(\mu,\xi_1)c_{31}(\mu,\xi_2)}{b_{32}(\xi_2,\xi_1)}\\
\kappa^{(C_1)}_3 &=& \underbrace{c_{31}(\mu,\xi_1)a_{3}(\mu,\xi_2)c_{31}(\xi_1,\xi_2)+b_{13}(\mu,\xi_1)c_{31}(\mu,\xi_2)b_{31}(\xi_1,\xi_2)}_{\textrm{apply Eq.(\ref{YB6})-\{3,1\}}}\\
&=&a_{3}(\mu,\xi_1)c_{31}(\mu,\xi_2)a_{3}(\xi_1,\xi_2)\\
\kappa^{(C_1)}_4 &=& \frac{c_{31}(\mu,\xi_1)b_{31}(\mu,\xi_2)}{b_{31}(\xi_1,\xi_2)}\\
\kappa^{(C_1)}_5 &=&\frac{c_{31}(\mu,\xi_1)b_{32}(\mu,\xi_2)}{b_{32}(\xi_1,\xi_2)}\\
\kappa^{(C_1)}_6 &=& c_{31}(\mu,\xi_1)a_{3}(\mu,\xi_2)\\
\kappa^{(C_1)}_7 &=& c_{21}(\mu,\xi_1)c_{32}(\mu,\xi_2) - \frac{c_{31}(\mu,\xi_1)b_{32}(\mu,\xi_2)\overbrace{c_{32}(\xi_1,\xi_2)}^{\textrm{apply Eq.(\ref{bigc})-\{3,2\}}}}{b_{32}(\xi_1,\xi_2)}\\
&=&  \frac{1}{b_{32}(\xi_2,\xi_1)}\underbrace{\left\{ c_{32}(\mu,\xi_2)c_{21}(\mu,\xi_1)b_{32}(\xi_2,\xi_1)+ b_{32}(\mu,\xi_2)c_{31}(\mu,\xi_1)c_{23}(\xi_2,\xi_1) \right\}}_{\textrm{apply Eq.(\ref{YB10})-\{3,2,1\}}}\\
&=&\frac{ c_{31}(\mu,\xi_2)b_{32}(\mu,\xi_1)c_{21}(\xi_2,\xi_1)}{b_{32}(\xi_2,\xi_1)}\\
\kappa^{(C_1)}_8 &=& \frac{c_{31}(\mu,\xi_1)b_{32}(\mu,\xi_2)c_{21}(\xi_1,\xi_2)}{b_{32}(\xi_1,\xi_2)}
\end{array}\end{equation*}
\begin{equation*}\begin{array}{lll}
 \kappa^{(B_1)}_1 &=&b_{31}(\mu,\xi_1)c_{13}(\mu,\xi_2)\\
 \kappa^{(B_1)}_2 &=&\frac{b_{32}(\mu,\xi_1)c_{13}(\mu,\xi_2)}{b_{21}(\xi_1,\xi_2)} \\
\kappa^{(B_1)}_3 &=& \frac{a_{3}(\mu,\xi_1)c_{13}(\mu,\xi_2)}{b_{31}(\xi_1,\xi_2)}\\
\kappa^{(B_1)}_4 &=&  \underbrace{c_{13}(\mu,\xi_1)a_{1}(\mu,\xi_2)b_{31}(\xi_1,\xi_2)+b_{31}(\mu,\xi_1)c_{13}(\mu,\xi_2)c_{31}(\xi_1,\xi_2)}_{\textrm{apply Eq.(\ref{YB8})-\{1,3\}}}\\
&=&c_{13}(\mu,\xi_1)b_{31}(\mu,\xi_2)a_{1}(\xi_1,\xi_2)\\
\kappa^{(B_1)}_5 &=&
 -\underbrace{\frac{c_{21}(\xi_1,\xi_2)}{b_{21}(\xi_1,\xi_2)}}_{\textrm{apply Eq.(\ref{bigc})-\{2,1\}}} \underbrace{\{c_{23}(\mu,\xi_1)c_{12}(\mu,\xi_2)b_{32}(\xi_1,\xi_2)+b_{32}(\mu,\xi_1)c_{13}(\mu,\xi_2)c_{32}(\xi_1,\xi_2)\}}_{\textrm{apply Eq.(\ref{YB11})-\{1,3,2\}}}\\
&&+c_{13}(\mu,\xi_1)b_{12}(\mu,\xi_2)b_{32}(\xi_1,\xi_2)\\
&=&\frac{c_{13}(\mu,\xi_1)}{b_{21}(\xi_2,\xi_1)} \{\underbrace{c_{12}(\xi_1,\xi_2)c_{12}(\xi_2,\xi_1)}_{\textrm{apply Eq.(\ref{bigb})-\{1,2\}}} b_{32}(\mu,\xi_2)+b_{12}(\mu,\xi_2)b_{32}(\xi_1,\xi_2)b_{21}(\xi_2,\xi_1)\}\\
&=&\frac{c_{13}(\mu,\xi_1)}{b_{21}(\xi_2,\xi_1)} \{b_{32}(\mu,\xi_2) + b_{21}(\xi_2,\xi_1)\underbrace{[b_{12}(\mu,\xi_2)b_{32}(\xi_1,\xi_2) -b_{32}(\mu,\xi_2)b_{12}(\xi_1,\xi_2)]}_{\textrm{apply Eq.(\ref{YB3})-\{1,3,2\}}}\}\\
&=& \frac{c_{12}(\mu,\xi_1)b_{32}(\mu,\xi_2)}{b_{21}(\xi_2,\xi_1)}\\
\kappa^{(B_1)}_6 &=& c_{13}(\mu,\xi_1)b_{13}(\mu,\xi_2) - \frac{a_{3}(\mu,\xi_1)c_{13}(\mu,\xi_2)\overbrace{c_{31}(\xi_1,\xi_2)}^{\textrm{apply Eq.(\ref{bigc})-\{3,1\}}}}{b_{31}(\xi_1,\xi_2)}\\
&=&  \frac{1}{b_{31}(\xi_2,\xi_1)}\underbrace{\left\{ b_{13}(\mu,\xi_2)c_{13}(\mu,\xi_1)b_{31}(\xi_2,\xi_1)+ c_{13}(\mu,\xi_2)a_{3}(\mu,\xi_1)c_{13}(\xi_2,\xi_1) \right\}}_{\textrm{apply Eq.(\ref{YB7})-\{1,3\}}}\\
&=&\frac{ a_{3}(\mu,\xi_2)c_{13}(\mu,\xi_1)a_{3}(\xi_2,\xi_1)}{b_{31}(\xi_2,\xi_1)}\\
\kappa^{(B_1)}_7 &=& -\frac{b_{32}(\mu,\xi_1)c_{13}(\mu,\xi_2)\overbrace{c_{21}(\xi_1,\xi_2)}^{\textrm{apply Eq.(\ref{bigc})-\{2,1\}}}}{b_{21}(\xi_1,\xi_2)}\\
&=&\frac{b_{23}(\mu,\xi_1)c_{13}(\mu,\xi_2)c_{12}(\xi_2,\xi_1)}{b_{21}(\xi_2,\xi_1)}\\
\kappa^{(B_1)}_8 &=&  \frac{1}{b_{21}(\xi_1,\xi_2)}\underbrace{\{c_{23}(\mu,\xi_1)c_{12}(\mu,\xi_2)b_{32}(\xi_1,\xi_2)+b_{32}(\mu,\xi_1)c_{13}(\mu,\xi_2)c_{32}(\xi_1,\xi_2)\}}_{\textrm{apply Eq.(\ref{YB11})-\{1,3,2\}}}\\
&=&\frac{c_{13}(\mu,\xi_1)b_{32}(\mu,\xi_2)c_{12}(\xi_1,\xi_2)}{b_{21}(\xi_1,\xi_2)}
\end{array}\end{equation*}

\addcontentsline{toc}{section}{Appendix C}
\section*{\bf Appendix C: The analysis for $L=3$ } \setcounter{equation}{0}
\renewcommand{\theequation}{C.\arabic{equation}}
In what follows we present the explicit expressions of 
the entries (\ref{gammaL3})  
together with the corresponding simplifications using
the Yang-Baxter and unitarity relations;
\begin{equation*}\begin{array}{lll}
 \alpha^{(C_1)}_1 &=& \frac{c_{21}(\mu,\xi_1)c_{32}(\mu,\xi_2)\overbrace{b_{31}(\mu,\xi_3)b_{21}(\xi_2,\xi_3)}^{\textrm{apply Eq.(\ref{YB3})-\{3,2,1\}}} }{b_{21}(\xi_1,\xi_3)b_{31}(\xi_2,\xi_3)} -\frac{c_{31}(\mu,\xi_1)b_{32}(\mu,\xi_2)b_{31}(\mu,\xi_3)\overbrace{c_{32}(\xi_1,\xi_2)}^{\textrm{apply Eq.(\ref{bigc})-\{3,2\}}} }{b_{31}(\xi_1,\xi_3)b_{32}(\xi_1,\xi_2)}\\
&=&\frac{c_{21}(\mu,\xi_1)c_{32}(\mu,\xi_2)b_{21}(\mu,\xi_3) }{b_{21}(\xi_1,\xi_3)} + \frac{c_{31}(\mu,\xi_1)b_{32}(\mu,\xi_2)\overbrace{b_{31}(\mu,\xi_3)b_{21}(\xi_1,\xi_3)}^{\textrm{apply Eq.(\ref{YB3})-\{3,2,1\}} }c_{23}(\xi_2,\xi_1) }{b_{32}(\xi_2,\xi_1)b_{31}(\xi_1,\xi_3)b_{21}(\xi_1,\xi_3)}\\
&=& \frac{b_{21}(\mu,\xi_3)}{b_{32}(\xi_2,\xi_1)b_{21}(\xi_1,\xi_3)}\underbrace{\{c_{32}(\mu,\xi_2)c_{21}(\mu,\xi_1)b_{32}(\xi_2,\xi_1)+b_{32}(\mu,\xi_2)c_{31}(\mu,\xi_1)c_{23}(\xi_2,\xi_1) \}}_{\textrm{apply Eq.(\ref{YB10})-\{1,3,2\}}}\\
&=&\frac{b_{21}(\mu,\xi_3)c_{31}(\mu,\xi_2)b_{32}(\mu,\xi_1)c_{21}(\xi_2,\xi_1)}{b_{32}(\xi_2,\xi_1)b_{21}(\xi_1,\xi_3)}\\
 \alpha^{(C_1)}_2 &=& \frac{c_{21}(\mu,\xi_1)c_{32}(\mu,\xi_2)b_{32}(\mu,\xi_3) }{b_{32}(\xi_2,\xi_3)} + \frac{c_{31}(\mu,\xi_1)b_{32}(\mu,\xi_2)b_{32}(\mu,\xi_3)\overbrace{c_{32}(\xi_1,\xi_2)}^{\textrm{apply Eq.(\ref{bigc})-\{3,2\}} } }{b_{32}(\xi_2,\xi_3)b_{32}(\xi_1,\xi_2) }\\
&=& \frac{b_{32}(\mu,\xi_3)}{b_{32}(\xi_2,\xi_1)b_{32}(\xi_2,\xi_3)}\underbrace{\{c_{32}(\mu,\xi_2)c_{21}(\mu,\xi_1)b_{32}(\xi_2,\xi_1)+b_{32}(\mu,\xi_2)c_{31}(\mu,\xi_1)c_{23}(\xi_2,\xi_1) \}}_{\textrm{apply Eq.(\ref{YB10})-\{1,3,2\}}}\\
&=&\frac{b_{32}(\mu,\xi_3)c_{31}(\mu,\xi_2)b_{32}(\mu,\xi_1)c_{21}(\xi_2,\xi_1)}{b_{32}(\xi_2,\xi_1)b_{32}(\xi_2,\xi_3)}\\
 \alpha^{(C_1)}_3 &=& c_{21}(\mu,\xi_1)c_{32}(\mu,\xi_2)a_{3}(\mu,\xi_3) - \frac{c_{31}(\mu,\xi_1)b_{32}(\mu,\xi_2)a_{3}(\mu,\xi_3)\overbrace{c_{32}(\xi_1,\xi_2)}^{\textrm{apply Eq.(\ref{bigc})-\{3,2\}} } }{b_{32}(\xi_1,\xi_2) }\\
&=& \frac{a_{3}(\mu,\xi_3)}{b_{32}(\xi_2,\xi_1)}\underbrace{\{c_{32}(\mu,\xi_2)c_{21}(\mu,\xi_1)b_{32}(\xi_2,\xi_1)+b_{32}(\mu,\xi_2)c_{31}(\mu,\xi_1)c_{23}(\xi_2,\xi_1) \}}_{\textrm{apply Eq.(\ref{YB10})-\{1,3,2\}}}\\
&=&\frac{a_{3}(\mu,\xi_3)c_{31}(\mu,\xi_2)b_{32}(\mu,\xi_1)c_{21}(\xi_2,\xi_1)}{b_{32}(\xi_2,\xi_1)}
\end{array}\end{equation*}
\begin{equation*}\begin{array}{lll}
 \beta^{(C_1)}_1 &=& \frac{c_{21}(\mu,\xi_1)b_{21}(\mu,\xi_2)c_{32}(\mu,\xi_3) }{b_{21}(\xi_1,\xi_2)} -\frac{c_{31}(\mu,\xi_1)\overbrace{b_{31}(\mu,\xi_2)b_{21}(\xi_1,\xi_2)}^{\textrm{apply Eq.(\ref{YB3})-\{3,2,1\}}} b_{32}(\mu,\xi_3) \overbrace{c_{32}(\xi_1,\xi_3)}^{\textrm{apply Eq.(\ref{bigc})-\{3,2\}}} }{b_{31}(\xi_1,\xi_2)b_{21}(\xi_1,\xi_2)b_{32}(\xi_1,\xi_3)}\\
&=& \frac{b_{21}(\mu,\xi_2)}{b_{21}(\xi_1,\xi_2)b_{32}(\xi_3,\xi_1)} \underbrace{\{c_{32}(\mu,\xi_3)c_{21}(\mu,\xi_1)b_{32}(\xi_3,\xi_1)+b_{32}(\mu,\xi_3)c_{31}(\mu,\xi_1)c_{23}(\xi_3,\xi_1) \}}_{\textrm{apply Eq.(\ref{YB10})-\{1,3,2\}}}\\
&=&\frac{b_{21}(\mu,\xi_2)c_{31}(\mu,\xi_3)b_{32}(\mu,\xi_1)c_{21}(\xi_3,\xi_1)}{b_{21}(\xi_1,\xi_2)b_{32}(\xi_3,\xi_1)}\\
\end{array}\end{equation*}
\begin{equation*}\begin{array}{lll}
 \beta^{(C_1)}_2 &=& c_{21}(\mu,\xi_1)a_{2}(\mu,\xi_2)c_{32}(\mu,\xi_3)- \frac{c_{21}(\mu,\xi_1)c_{32}(\mu,\xi_2)b_{32}(\mu,\xi_3)\overbrace{c_{32}(\xi_2,\xi_3)}^{\textrm{apply Eq.(\ref{bigc})-\{3,2\}} } }{b_{32}(\xi_2,\xi_3) }\\
&& + \frac{c_{31}(\mu,\xi_1)b_{32}(\mu,\xi_2)b_{32}(\mu,\xi_3)c_{32}(\xi_1,\xi_2)\overbrace{c_{32}(\xi_2,\xi_3)}^{\textrm{apply Eq.(\ref{bigc})-\{3,2\}} } }{b_{32}(\xi_1,\xi_2)b_{32}(\xi_2,\xi_3) }-  \frac{c_{31}(\mu,\xi_1)b_{32}(\mu,\xi_2)b_{32}(\mu,\xi_3)a_{2}(\xi_1,\xi_2)c_{32}(\xi_1,\xi_3) }{ b_{32}(\xi_1,\xi_2) b_{32}(\xi_1,\xi_3) }\\
&=&  \frac{c_{21}(\mu,\xi_1)}{b_{32}(\xi_3,\xi_2)}\underbrace{\{c_{32}(\mu,\xi_3)a_{2}(\mu,\xi_2)b_{32}(\xi_3,\xi_2)+b_{32}(\mu,\xi_3)c_{32}(\mu,\xi_2)c_{23}(\xi_3,\xi_2) \}}_{\textrm{apply Eq.(\ref{YB9})-\{3,2\}}}\\
&&-\frac{c_{31}(\mu,\xi_1)b_{32}(\mu,\xi_2)b_{32}(\mu,\xi_3)}{b_{32}(\xi_1,\xi_2)b_{32}(\xi_3,\xi_2)b_{32}(\xi_1,\xi_3)}\underbrace{\{b_{32}(\xi_1,\xi_3)c_{32}(\xi_1,\xi_2)c_{23}(\xi_3,\xi_2)+c_{32}(\xi_1,\xi_3)a_{2}(\xi_1,\xi_2)b_{32}(\xi_3,\xi_2) \}}_{\textrm{apply Eq.(\ref{YB9})-\{3,2\}}}
\end{array}\end{equation*}
\begin{equation*}\begin{array}{lll}
&=&\frac{c_{21}(\mu,\xi_1)c_{32}(\mu,\xi_3)b_{32}(\mu,\xi_2)a_{2}(\xi_3,\xi_2)}{b_{32}(\xi_3,\xi_2)}-\frac{c_{31}(\mu,\xi_1)b_{32}(\mu,\xi_2)b_{32}(\mu,\xi_3)\overbrace{c_{32}(\xi_1,\xi_3)}^{\textrm{apply Eq.(\ref{bigc})-\{3,2\}}}a_{2}(\xi_3,\xi_2)}{b_{32}(\xi_1,\xi_3)b_{32}(\xi_3,\xi_2)}\\
&=& \frac{b_{32}(\mu,\xi_2)a_{2}(\xi_3,\xi_2)}{b_{32}(\xi_3,\xi_2)b_{32}(\xi_3,\xi_1)}\underbrace{\{c_{32}(\mu,\xi_3)c_{21}(\mu,\xi_1)b_{32}(\xi_3,\xi_1)+b_{32}(\mu,\xi_3)c_{31}(\mu,\xi_1)c_{23}(\xi_3,\xi_1) \}}_{\textrm{apply Eq.(\ref{YB10})-\{1,3,2\}}}\\
&=&\frac{b_{32}(\mu,\xi_2)a_{2}(\xi_3,\xi_2)c_{31}(\mu,\xi_3)b_{32}(\mu,\xi_1)c_{21}(\xi_3,\xi_1)}{b_{32}(\xi_3,\xi_2)b_{32}(\xi_3,\xi_1)}
\end{array}\end{equation*}
\begin{equation*}\begin{array}{lll}
 \beta^{(C_1)}_3 &=& c_{21}(\mu,\xi_1)\underbrace{\{c_{32}(\mu,\xi_2)a_{3}(\mu,\xi_3)c_{32}(\xi_2,\xi_3)+b_{23}(\mu,\xi_2)c_{32}(\mu,\xi_3)b_{32}(\xi_2,\xi_3) \}}_{\textrm{apply Eq.(\ref{YB6})-\{3,2\}}}\\
&&-\frac{c_{31}(\mu,\xi_1)c_{32}(\xi_1,\xi_2)}{b_{32}(\xi_1,\xi_2)}\underbrace{\{b_{32}(\mu,\xi_2)a_{3}(\mu,\xi_3)c_{32}(\xi_2,\xi_3)+c_{23}(\mu,\xi_2)c_{32}(\mu,\xi_3)b_{32}(\xi_2,\xi_3) \}}_{\textrm{apply Eq.(\ref{YB4})-\{3,2\}}}\\
&&+\frac{c_{31}(\mu,\xi_1)a_{3}(\mu,\xi_2)b_{32}(\mu,\xi_3)\overbrace{c_{32}(\xi_1,\xi_2)}^{\textrm{apply Eq.(\ref{bigc})-\{3,2\}}} c_{32}(\xi_2,\xi_3)}{b_{32}(\xi_1,\xi_2)}-\frac{c_{31}(\mu,\xi_1)a_{3}(\mu,\xi_2)b_{32}(\mu,\xi_3)c_{32}(\xi_1,\xi_3)a_{3}(\xi_2,\xi_3)}{b_{32}(\xi_1,\xi_3)}\\
&=& c_{21}(\mu,\xi_1)a_{3}(\mu,\xi_2)c_{32}(\mu,\xi_3)a_{3}(\xi_2,\xi_3) -\frac{c_{31}(\mu,\xi_1)c_{32}(\xi_1,\xi_2)a_{3}(\mu,\xi_2)b_{32}(\mu,\xi_3)c_{32}(\xi_1,\xi_3)}{b_{32}(\xi_1,\xi_2)}\\
&&-\frac{c_{31}(\mu,\xi_1)a_{3}(\mu,\xi_2)b_{32}(\mu,\xi_3)}{b_{32}(\xi_1,\xi_3)b_{32}(\xi_2,\xi_1)}\underbrace{\{c_{23}(\xi_2,\xi_1)c_{32}(\xi_2,\xi_3)b_{32}(\xi_1,\xi_3)+b_{32}(\xi_2,\xi_1)a_{3}(\xi_2,\xi_3)c_{32}(\xi_1,\xi_3) \}}_{\textrm{apply Eq.(\ref{YB4})-\{3,2\}}}\\
&=& c_{21}(\mu,\xi_1)a_{3}(\mu,\xi_2)c_{32}(\mu,\xi_3)a_{3}(\xi_2,\xi_3) -\frac{ c_{31}(\mu,\xi_1) \overbrace{c_{32}(\xi_1,\xi_2)}^{\textrm{apply Eq.(\ref{bigc})-\{3,2\}}} a_{3}(\mu,\xi_2) b_{32}(\mu,\xi_3) c_{32}(\xi_1,\xi_3) }{ b_{32}(\xi_1,\xi_2) }\\
&&-\frac{c_{31}(\mu,\xi_1)a_{3}(\mu,\xi_2)b_{32}(\mu,\xi_3)a_{3}(\xi_2,\xi_1)b_{32}(\xi_2,\xi_3)\overbrace{c_{32}(\xi_1,\xi_3)}^{\textrm{apply Eq.(\ref{bigc})-\{3,2\}}}}{b_{32}(\xi_1,\xi_3)b_{32}(\xi_2,\xi_1)} \\
&=& c_{21}(\mu,\xi_1)a_{3}(\mu,\xi_2)c_{32}(\mu,\xi_3)a_{3}(\xi_2,\xi_3) \\
&&+ \frac{c_{31}(\mu,\xi_1)a_{3}(\mu,\xi_2)b_{32}(\mu,\xi_3)}{b_{32}(\xi_3,\xi_1)b_{32}(\xi_2,\xi_1)}\underbrace{\{c_{32}(\xi_2,\xi_3)c_{23}(\xi_2,\xi_1)b_{32}(\xi_3,\xi_1)+b_{32}(\xi_2,\xi_3)a_{3}(\xi_2,\xi_1)c_{23}(\xi_3,\xi_1) \}}_{\textrm{apply Eq.(\ref{YB5})-\{3,2\}}}
\end{array}\end{equation*}
\begin{equation*}\begin{array}{lll}
&=& \frac{a_{3}(\mu,\xi_2)a_{3}(\xi_2,\xi_3)}{b_{32}(\xi_3,\xi_1)}\underbrace{\{c_{32}(\mu,\xi_3)c_{21}(\mu,\xi_1)b_{32}(\xi_3,\xi_1) + b_{32}(\mu,\xi_3)c_{31}(\mu,\xi_1)c_{23}(\xi_3,\xi_1) \}}_{\textrm{apply Eq.(\ref{YB10})-\{1,3,2\}}}\\
&=& \frac{a_{3}(\mu,\xi_2)a_{3}(\xi_2,\xi_3)c_{31}(\mu,\xi_3)b_{32}(\mu,\xi_1)c_{21}(\xi_3,\xi_1)}{b_{32}(\xi_3,\xi_1)}
\end{array}\end{equation*}
\begin{equation*}\begin{array}{lll}
 \gamma^{(C_1)}_1 &=& a_{1}(\mu,\xi_1)c_{21}(\mu,\xi_2)c_{32}(\mu,\xi_3)-\frac{c_{21}(\mu,\xi_1)b_{21}(\mu,\xi_2) c_{32}(\mu,\xi_3) \overbrace{c_{21}(\xi_1,\xi_2)}^{\textrm{apply Eq.(\ref{bigc})-\{2,1\}}} }{b_{21}(\xi_1,\xi_2)} \\
&&+\frac{c_{31}(\mu,\xi_1)b_{31}(\mu,\xi_2) b_{32}(\mu,\xi_3) \overbrace{c_{31}(\xi_1,\xi_2)}^{\textrm{apply Eq.(\ref{bigc})-\{3,1\}}} \overbrace{c_{32}(\xi_2,\xi_3)}^{\textrm{apply Eq.(\ref{bigc})-\{3,2\}}} }{b_{31}(\xi_1,\xi_2)b_{32}(\xi_1,\xi_2)} -\frac{a_{1}(\mu,\xi_1) c_{31}(\mu,\xi_2) b_{32}(\mu,\xi_3) \overbrace{c_{32}(\xi_2,\xi_3)}^{\textrm{apply Eq.(\ref{bigc})-\{3,2\}}} }{b_{32}(\xi_2,\xi_3)} 
\end{array}\end{equation*}
\begin{equation*}\begin{array}{lll}
&=& \frac{c_{32}(\mu,\xi_3)}{b_{21}(\xi_2,\xi_1)}\underbrace{\{c_{21}(\mu,\xi_2)a_{1}(\mu,\xi_1)b_{21}(\xi_2,\xi_1) +b_{21}(\mu,\xi_2)c_{21}(\mu,\xi_1)c_{12}(\xi_2,\xi_1) \}}_{\textrm{apply Eq.(\ref{YB9})-\{2,1\}}}\\
&&+ \frac{b_{32}(\mu,\xi_3)c_{23}(\xi_3,\xi_2)}{b_{31}(\xi_2,\xi_1)b_{32}(\xi_3,\xi_2)}\underbrace{\{b_{31}(\mu,\xi_2)c_{31}(\mu,\xi_1)c_{31}(\xi_2,\xi_1) +c_{31}(\mu,\xi_2)a_{1}(\mu,\xi_1)b_{31}(\xi_2,\xi_1) \}}_{\textrm{apply Eq.(\ref{YB9})-\{3,1\}}}\\
&=&\frac{c_{32}(\mu,\xi_3)c_{21}(\mu,\xi_2)b_{21}(\mu,\xi_1)a_{1}(\xi_2,\xi_1) }{b_{21}(\xi_2,\xi_1)} + \frac{b_{32}(\mu,\xi_3)c_{23}(\xi_3,\xi_2)c_{31}(\mu,\xi_2)a_{1}(\xi_2,\xi_1)\overbrace{b_{31}(\mu,\xi_1)}^{\textrm{apply Eq.(\ref{YB3})-\{3,2,1\}}} }{b_{32}(\xi_3,\xi_2)b_{31}(\xi_2,\xi_1)} \\
&=& \frac{b_{21}(\mu,\xi_1)a_{1}(\xi_2,\xi_1)}{b_{21}(\xi_2,\xi_1)b_{32}(\xi_3,\xi_2)}\underbrace{\{c_{32}(\mu,\xi_3)c_{21}(\mu,\xi_2)b_{32}(\xi_3,\xi_2)+b_{32}(\mu,\xi_3)c_{31}(\mu,\xi_2)c_{23}(\xi_3,\xi_2) \}}_{\textrm{apply Eq.(\ref{YB10})-\{1,3,2\}}}\\
&=& \frac{b_{21}(\mu,\xi_1)a_{1}(\xi_2,\xi_1)c_{31}(\mu,\xi_3)b_{32}(\mu,\xi_2)c_{21}(\xi_3,\xi_2)}{b_{21}(\xi_2,\xi_1)b_{32}(\xi_3,\xi_2)}
\end{array}\end{equation*}
\begin{equation*}\begin{array}{lll}
 \gamma^{(C_1)}_2 &=& b_{12}(\mu,\xi_1)c_{21}(\mu,\xi_2)c_{32}(\mu,\xi_3)c_{21}(\xi_1,\xi_2)-\frac{b_{12}(\mu,\xi_1)c_{31}(\mu,\xi_2) b_{32}(\mu,\xi_3) b_{21}(\xi_1,\xi_2) \overbrace{c_{32}(\xi_1,\xi_2)}^{\textrm{apply Eq.(\ref{bigc})-\{3,2\}}} }{b_{32}(\xi_1,\xi_2)} \\
&&+c_{21}(\mu,\xi_1) a_{2}(\mu,\xi_2) c_{32}(\mu,\xi_3) c_{21}(\xi_1,\xi_2) -\frac{c_{21}(\mu,\xi_1) c_{32}(\mu,\xi_2) b_{32}(\mu,\xi_3) c_{21}(\xi_1,\xi_2) \overbrace{c_{32}(\xi_2,\xi_3)}^{\textrm{apply Eq.(\ref{bigc})-\{3,2\}}} }{b_{32}(\xi_2,\xi_3)} \\
&&+\frac{c_{31}(\mu,\xi_1)b_{32}(\mu,\xi_2) b_{32}(\mu,\xi_3)c_{32}(\xi_1,\xi_2) c_{21}(\xi_1,\xi_2) \overbrace{c_{32}(\xi_2,\xi_3)}^{\textrm{apply Eq.(\ref{bigc})-\{3,2\}}} }{b_{32}(\xi_1,\xi_2)b_{32}(\xi_2,\xi_3)} -\frac{c_{31}(\mu,\xi_1) b_{32}(\mu,\xi_2) b_{32}(\mu,\xi_3) a_{2}(\xi_1,\xi_2) c_{21}(\xi_1,\xi_2) c_{32}(\xi_1,\xi_3) }{b_{32}(\xi_1,\xi_2)b_{32}(\xi_1,\xi_3)}\\
&=& \frac{b_{12}(\mu,\xi_1)b_{21}(\xi_1,\xi_2)}{b_{32}(\xi_3,\xi_2)}\underbrace{\{c_{32}(\mu,\xi_3)c_{21}(\mu,\xi_2)b_{32}(\xi_3,\xi_2) +b_{32}(\mu,\xi_3)c_{31}(\mu,\xi_2)c_{23}(\xi_3,\xi_2) \}}_{\textrm{apply Eq.(\ref{YB10})-\{1,3,2\}}}\\
&&+ \frac{c_{21}(\mu,\xi_1)c_{21}(\xi_1,\xi_2)}{b_{32}(\xi_3,\xi_2)}\underbrace{\{c_{32}(\mu,\xi_3)a_{2}(\mu,\xi_2)b_{32}(\xi_3,\xi_2) +b_{32}(\mu,\xi_3)c_{32}(\mu,\xi_2)c_{23}(\xi_3,\xi_2) \}}_{\textrm{apply Eq.(\ref{YB9})-\{3,2\}}}\\
&&- \frac{c_{31}(\mu,\xi_1)b_{32}(\mu,\xi_2)b_{32}(\mu,\xi_3)c_{21}(\xi_1,\xi_2)}{b_{32}(\xi_1,\xi_2)b_{32}(\xi_1,\xi_2)b_{32}(\xi_3,\xi_2)} \underbrace{\{b_{32}(\xi_1,\xi_3)c_{32}(\xi_1,\xi_2)c_{23}(\xi_3,\xi_2) + c_{32}(\xi_1,\xi_3)a_{2}(\xi_1,\xi_2)b_{32}(\xi_3,\xi_2) \}}_{\textrm{apply Eq.(\ref{YB9})-\{3,2\}}}
\end{array}\end{equation*}
\begin{equation*}\begin{array}{lll}
&=&\frac{b_{12}(\mu,\xi_1)b_{21}(\xi_1,\xi_2)c_{31}(\mu,\xi_3) b_{32}(\mu,\xi_2)c_{21}(\xi_3,\xi_2) }{b_{21}(\xi_3,\xi_2)} + \frac{c_{21}(\mu,\xi_1)c_{21}(\xi_1,\xi_2) c_{32}(\mu,\xi_3)b_{32}(\mu,\xi_2)a_{2}(\xi_3,\xi_2) }{b_{32}(\xi_3,\xi_2)} \\ &&-\frac{c_{31}(\mu,\xi_1)b_{32}(\mu,\xi_2)b_{32}(\mu,\xi_3)c_{21}(\xi_1,\xi_2)\overbrace{c_{32}(\xi_1,\xi_3)}^{\textrm{apply Eq.(\ref{bigc})-\{3,2\}}} a_{2}(\xi_3,\xi_2) }{b_{32}(\xi_1,\xi_3)b_{32}(\xi_3,\xi_2)} \\
&=&\frac{b_{12}(\mu,\xi_1)b_{21}(\xi_1,\xi_2)c_{31}(\mu,\xi_3) b_{32}(\mu,\xi_2)c_{21}(\xi_3,\xi_2) }{b_{21}(\xi_3,\xi_2)}\\
&&+ \frac{c_{21}(\xi_1,\xi_2)a_{2}(\xi_3,\xi_2)b_{32}(\mu,\xi_2)}{b_{32}(\xi_3,\xi_2)b_{32}(\xi_3,\xi_1)}\underbrace{\{c_{32}(\mu,\xi_3)c_{21}(\mu,\xi_1)b_{32}(\xi_3,\xi_1)+b_{32}(\mu,\xi_3)c_{31}(\mu,\xi_1)c_{23}(\xi_3,\xi_1) \}}_{\textrm{apply Eq.(\ref{YB10})-\{1,3,2\}}}\\
&=& \frac{\overbrace{b_{12}(\mu,\xi_1) b_{32}(\xi_3,\xi_1)}^{\textrm{apply Eq.(\ref{YB3})-\{1,3,2\}}} b_{21}(\xi_1,\xi_2)c_{31}(\mu,\xi_3) b_{32}(\mu,\xi_2)c_{21}(\xi_3,\xi_2) }{ b_{32}(\xi_3,\xi_1) b_{32}(\xi_3,\xi_2)} + \frac{c_{31}(\mu,\xi_3)b_{32}(\mu,\xi_2)c_{21}(\xi_1,\xi_2) a_{2}(\xi_3,\xi_2) b_{32}(\mu,\xi_1) c_{21}(\xi_3,\xi_1)}{b_{32}(\xi_3,\xi_2)b_{32}(\xi_3,\xi_1)}\\
&=& \frac{c_{31}(\mu,\xi_3)b_{32}(\mu,\xi_2)b_{32}(\mu,\xi_1)}{b_{32}(\xi_3,\xi_2)b_{32}(\xi_3,\xi_1)} \underbrace{\{b_{12}(\xi_3,\xi_1)c_{21}(\xi_3,\xi_2)b_{21}(\xi_1,\xi_2) +c_{21}(\xi_3,\xi_1)a_{2}(\xi_3,\xi_2)c_{21}(\xi_1,\xi_2) \}}_{\textrm{apply Eq.(\ref{YB6})-\{2,1\}}}\\
&=& \frac{c_{31}(\mu,\xi_3)b_{32}(\mu,\xi_2)b_{32}(\mu,\xi_1)a_{2}(\xi_3,\xi_1)a_{2}(\xi_1,\xi_2)c_{21}(\xi_3,\xi_2)}{b_{21}(\xi_3,\xi_1)b_{32}(\xi_3,\xi_2)}
\end{array}\end{equation*}
\begin{equation*}\begin{array}{lll}
 \gamma^{(C_1)}_3 &=& \frac{b_{13}(\mu,\xi_1) c_{21}(\mu,\xi_2) c_{32}(\mu,\xi_3) b_{31}(\xi_1,\xi_2) b_{32}(\xi_1,\xi_2) }{ b_{32}(\xi_1,\xi_2) } -\frac{b_{13}(\mu,\xi_1)c_{31}(\mu,\xi_2) b_{32}(\mu,\xi_3) b_{31}(\xi_1,\xi_2)b_{32}(\xi_1,\xi_3) \overbrace{c_{32}(\xi_2,\xi_3)}^{\textrm{apply Eq.(\ref{bigc})-\{3,2\}}} }{b_{32}(\xi_1,\xi_2)b_{32}(\xi_2,\xi_3)} \\
&&+\frac{c_{31}(\mu,\xi_1) c_{23}(\mu,\xi_2) c_{32}(\mu,\xi_3) c_{31}(\xi_1,\xi_2) b_{32}(\xi_1,\xi_3) }{ b_{32}(\xi_1,\xi_2) } - \frac{c_{31}(\mu,\xi_1)a_{3}(\mu,\xi_2) b_{32}(\mu,\xi_3) c_{31}(\xi_1,\xi_2) b_{32}(\xi_1,\xi_3) \overbrace{c_{32}(\xi_2,\xi_3)}^{\textrm{apply Eq.(\ref{bigc})-\{3,2\}}} }{b_{32}(\xi_1,\xi_2)b_{32}(\xi_2,\xi_3)} \\
&&+ \frac{c_{31}(\mu,\xi_1) b_{32}(\mu,\xi_2) a_{3}(\mu,\xi_3) c_{21}(\xi_1,\xi_2) c_{32}(\xi_1,\xi_3)}{b_{32}(\xi_1,\xi_2)}\\
&=& \frac{b_{13}(\mu,\xi_1)b_{31}(\xi_1,\xi_2)b_{32}(\xi_1,\xi_3)}{b_{32}(\xi_1,\xi_2)b_{12}(\xi_3,\xi_2)} \underbrace{\{c_{32}(\mu,\xi_3)c_{21}(\mu,\xi_2)b_{32}(\xi_3,\xi_2) + b_{32}(\mu,\xi_3)c_{31}(\mu,\xi_2)c_{23}(\xi_3,\xi_2) \}}_{\textrm{apply Eq.(\ref{YB10})-\{1,3,2\}}}\\
&&+ \frac{c_{31}(\mu,\xi_1)c_{31}(\xi_1,\xi_2)b_{32}(\xi_1,\xi_3)}{b_{32}(\xi_1,\xi_2)b_{32}(\xi_3,\xi_2)} \underbrace{\{c_{32}(\mu,\xi_3)c_{23}(\mu,\xi_2)b_{32}(\xi_3,\xi_2) +b_{32}(\mu,\xi_3)a_{3}(\mu,\xi_2)c_{23}(\xi_3,\xi_2) \}}_{\textrm{apply Eq.(\ref{YB5})-\{3,2\}}}\\
&&+ \frac{c_{31}(\mu,\xi_1) b_{32}(\mu,\xi_2) a_{3}(\mu,\xi_3) c_{21}(\xi_1,\xi_2) c_{32}(\xi_1,\xi_3)}{b_{32}(\xi_1,\xi_2)}\\
&=&\frac{b_{13}(\mu,\xi_1)b_{31}(\xi_1,\xi_2)b_{32}(\xi_1,\xi_3) c_{31}(\mu,\xi_3)b_{32}(\mu,\xi_2)c_{21}(\xi_3,\xi_2) }{b_{32}(\xi_1,\xi_2)b_{32}(\xi_3,\xi_2)}  \\
&&+\frac{c_{31}(\mu,\xi_1)b_{32}(\mu,\xi_2)a_{3}(\mu,\xi_3)}{b_{32}(\xi_1,\xi_2)b_{32}(\xi_3,\xi_2)} \underbrace{\{b_{32}(\xi_1,\xi_3)c_{31}(\xi_1,\xi_2)c_{23}(\xi_3,\xi_2) + c_{32}(\xi_1,\xi_3)c_{21}(\xi_1,\xi_2)b_{32}(\xi_3,\xi_2) \}}_{\textrm{apply Eq.(\ref{YB10})-\{1,3,2\}}} \\
&=&\frac{b_{13}(\mu,\xi_1)b_{32}(\mu,\xi_2)c_{31}(\mu,\xi_3)\overbrace{ b_{31}(\xi_1,\xi_2) }^{\textrm{apply Eq.(\ref{YB2})-\{3,2,1\}}}b_{32}(\xi_1,\xi_3) c_{21}(\xi_3,\xi_2) }{b_{32}(\xi_3,\xi_2)b_{32}(\xi_1,\xi_2)}+ \frac{c_{31}(\mu,\xi_1)b_{32}(\mu,\xi_2)a_{3}(\mu,\xi_3) c_{31}(\xi_1,\xi_3)c_{21}(\xi_3,\xi_2) }{b_{32}(\xi_3,\xi_2)}\\
&=& \frac{b_{32}(\mu,\xi_2)c_{21}(\xi_3,\xi_2)}{b_{32}(\xi_3,\xi_2)} \underbrace{\{b_{13}(\mu,\xi_1)c_{31}(\mu,\xi_3)b_{31}(\xi_1,\xi_3) + c_{31}(\mu,\xi_1) a_{3}(\mu,\xi_3)c_{31}(\xi_1,\xi_3) \}}_{\textrm{apply Eq.(\ref{YB6})-\{3,1\}}}\\
&=& \frac{c_{32}(\mu,\xi_2)c_{21}(\xi_3,\xi_2)a_{3}(\mu,\xi_1)c_{31}(\mu,\xi_3)a_{3}(\xi_1,\xi_3)}{b_{32}(\xi_3,\xi_2)}
\end{array}\end{equation*}
\begin{equation*}\begin{array}{lll}
 \delta^{(C_1)}_1 &=& \frac{c_{31}(\mu,\xi_1)b_{32}(\mu,\xi_2)\overbrace{b_{31}(\mu,\xi_3)b_{21}(\xi_1,\xi_3)}^{\textrm{apply Eq.(\ref{YB3})-\{3,2,1\}}} c_{21}(\xi_1,\xi_2)}{b_{32}(\xi_1,\xi_2)b_{31}(\xi_1,\xi_3)b_{21}(\xi_2,\xi_3)} \\
&=& \frac{c_{31}(\mu,\xi_1)b_{32}(\mu,\xi_2)b_{21}(\mu,\xi_3) c_{21}(\xi_1,\xi_2) }{ b_{32}(\xi_1,\xi_2) b_{21}(\xi_2,\xi_3)} \\
 \delta^{(C_1)}_2 &=&\frac{c_{31}(\mu,\xi_1)b_{32}(\mu,\xi_2)b_{32}(\mu,\xi_3)c_{21}(\xi_1,\xi_2)}{b_{32}(\xi_1,\xi_2)b_{32}(\xi_1,\xi_3)}\\
 \delta^{(C_1)}_3 &=& \frac{c_{31}(\mu,\xi_1)b_{32}(\mu,\xi_2)a_{3}(\mu,\xi_3)c_{21}(\xi_1,\xi_2)}{b_{32}(\xi_1,\xi_2)}\\
 \phi^{(C_1)}_1 &=&\frac{c_{31}(\mu,\xi_1)b_{31}(\mu,\xi_2)b_{32}(\mu,\xi_3)a_{1}(\xi_1,\xi_2)c_{21}(\xi_1,\xi_3)}{ b_{31}(\xi_1,\xi_2) b_{32}(\xi_1,\xi_3)}- \frac{c_{31}(\mu,\xi_1)\overbrace{b_{32}(\mu,\xi_2)b_{31}(\mu,\xi_3)}^{\textrm{apply Eq.(\ref{YB2})-\{3,2,1\}}} c_{21}(\xi_1,\xi_2)b_{21}(\xi_1,\xi_3 )\overbrace{ c_{21}(\xi_2,\xi_3)}^{\textrm{apply Eq.(\ref{bigc})-\{2,1\}}} }{b_{31}(\xi_1,\xi_2)b_{32}(\xi_1,\xi_3)b_{21}(\xi_2,\xi_3)} \\
&=&\frac{c_{31}(\mu,\xi_1)b_{31}(\mu,\xi_2)b_{32}(\mu,\xi_3)}{b_{31}(\xi_1,\xi_2)b_{32}(\xi_1,\xi_3)b_{21}(\xi_3,\xi_2)} \underbrace{\{c_{21}(\xi_1,\xi_3)a_{1}(\xi_1,\xi_2)b_{21}(\xi_3,\xi_2) + b_{21}(\xi_1,\xi_3)c_{21}(\xi_1,\xi_2)c_{12}(\xi_3,\xi_2) \}}_{\textrm{apply Eq.(\ref{YB9})-\{2,1\}}}\\
&=& \frac{c_{31}(\mu,\xi_1)\overbrace{b_{31}(\mu,\xi_2)b_{21}(\xi_1,\xi_2)}^{\textrm{apply Eq.(\ref{YB3})-\{3,2,1\}}}b_{32}(\mu,\xi_3) c_{21}(\xi_1,\xi_3)a_{1}(\xi_3,\xi_2) }{ b_{31}(\xi_1,\xi_2) b_{32}(\xi_1,\xi_3)b_{21}(\xi_3,\xi_2)} \\
&=& \frac{c_{31}(\mu,\xi_1)b_{21}(\mu,\xi_2) b_{32}(\mu,\xi_3) c_{21}(\xi_1,\xi_3)a_{1}(\xi_3,\xi_2) }{ b_{32}(\xi_1,\xi_3)b_{21}(\xi_3,\xi_2)}
\end{array}\end{equation*}
\begin{equation*}\begin{array}{lll}
 \phi^{(C_1)}_2 &=&\frac{c_{31}(\mu,\xi_1)b_{32}(\mu,\xi_2) b_{32}(\mu,\xi_3) c_{21}(\xi_1,\xi_3) a_{2}(\xi_2,\xi_3) }{ b_{32}(\xi_1,\xi_2)b_{32}(\xi_1,\xi_3)}\\
 \phi^{(C_1)}_3 &=& \frac{c_{31}(\mu,\xi_1)a_{3}(\mu,\xi_2)b_{32}(\mu,\xi_3)c_{21}(\xi_1,\xi_3)}{b_{32}(\xi_1,\xi_3)}\\
 \omega^{(C_1)}_1 &=& \frac{a_{1}(\mu,\xi_1) c_{31}(\mu,\xi_2) b_{32}(\mu,\xi_3) c_{21}(\xi_2,\xi_3)  }{ b_{32}(\xi_2,\xi_3) } -\frac{c_{31}(\mu,\xi_1)b_{31}(\mu,\xi_2) b_{32}(\mu,\xi_3) \overbrace{c_{31}(\xi_1,\xi_2)}^{\textrm{apply Eq.(\ref{bigc})-\{3,1\}}} c_{21}(\xi_2,\xi_3) }{b_{31}(\xi_1,\xi_2)b_{32}(\xi_2,\xi_3)} \\
&&+\frac{c_{31}(\mu,\xi_1) b_{32}(\mu,\xi_2) b_{31}(\mu,\xi_3) \overbrace{c_{31}(\xi_1,\xi_2)}^{\textrm{apply Eq.(\ref{bigc})-\{3,1\}}} c_{21}(\xi_2,\xi_3)}{ b_{31}(\xi_1,\xi_2) b_{32}(\xi_2,\xi_3)} - \frac{c_{31}(\mu,\xi_1)b_{32}(\mu,\xi_2) b_{31}(\mu,\xi_3) c_{31}(\xi_1,\xi_3) c_{23}(\xi_2,\xi_3) }{ b_{31}(\xi_1,\xi_3) b_{32}(\xi_2,\xi_3)} \\
&&- \frac{c_{21}(\mu,\xi_1) c_{32}(\mu,\xi_2) b_{31}(\mu,\xi_3) c_{21}(\xi_1,\xi_3) b_{21}(\xi_2,\xi_3)}{b_{21}(\xi_1,\xi_3)b_{31}(\xi_2,\xi_3)}\\
&=& \frac{b_{32}(\mu,\xi_3)c_{21}(\xi_2,\xi_3)}{b_{32}(\xi_2,\xi_3)b_{31}(\xi_2,\xi_1)} \underbrace{\{c_{31}(\mu,\xi_2)a_{1}(\mu,\xi_1)b_{31}(\xi_2,\xi_1) + b_{31}(\mu,\xi_2)c_{31}(\mu,\xi_1)c_{13}(\xi_2,\xi_1) \}}_{\textrm{apply Eq.(\ref{YB9})-\{3,1\}}}\\
&&- \frac{c_{31}(\mu,\xi_1)b_{32}(\mu,\xi_2)b_{31}(\mu,\xi_3)}{b_{31}(\xi_2,\xi_1)b_{32}(\xi_2,\xi_3)b_{31}(\xi_1,\xi_3)} \underbrace{\{c_{13}(\xi_2,\xi_1)c_{21}(\xi_2,\xi_3)b_{31}(\xi_1,\xi_3) +b_{31}(\xi_2,\xi_1)c_{23}(\xi_2,\xi_3)c_{31}(\xi_1,\xi_3) \}}_{\textrm{apply Eq.(\ref{YB11})-\{2,3,1\}}}\\
&&- \frac{c_{21}(\mu,\xi_1) c_{32}(\mu,\xi_2) b_{31}(\mu,\xi_3) c_{21}(\xi_1,\xi_3) b_{21}(\xi_2,\xi_3)}{b_{21}(\xi_1,\xi_3)b_{31}(\xi_2,\xi_3)}\\
&=&\frac{b_{32}(\mu,\xi_3)c_{21}(\xi_2,\xi_3)c_{31}(\mu,\xi_2)b_{31}(\mu,\xi_1)a_{1}(\xi_2,\xi_1) }{b_{32}(\xi_2,\xi_3)b_{31}(\xi_2,\xi_1)} -  \frac{c_{31}(\mu,\xi_1)b_{32}(\mu,\xi_2)b_{31}(\mu,\xi_3)c_{32}(\xi_2,\xi_1)b_{31}(\xi_2,\xi_3)c_{21}(\xi_1,\xi_3)}{\underbrace{b_{31}(\xi_2,\xi_1)b_{32}(\xi_2,\xi_3)}_{\textrm{apply Eq.(\ref{YB2})-\{3,1,2\}}}b_{31}(\xi_1,\xi_3)}  \\
&& - \frac{c_{21}(\mu,\xi_1) c_{32}(\mu,\xi_2) b_{31}(\mu,\xi_3) c_{21}(\xi_1,\xi_3) b_{21}(\xi_2,\xi_3)}{\underbrace{b_{21}(\xi_1,\xi_3)b_{31}(\xi_2,\xi_3)}_{\textrm{apply Eq.(\ref{YB3})-\{2,3,1\}}}}
\end{array}\end{equation*}
\begin{equation*}\begin{array}{lll}
&=&\frac{b_{32}(\mu,\xi_3)c_{21}(\xi_2,\xi_3)c_{31}(\mu,\xi_2)b_{31}(\mu,\xi_1)a_{1}(\xi_2,\xi_1) }{b_{32}(\xi_2,\xi_3)b_{31}(\xi_2,\xi_1)}\\
&&-\frac{c_{21}(\xi_1,\xi_3)b_{31}(\mu,\xi_3)}{b_{31}(\xi_1,\xi_3)b_{32}(\xi_2,\xi_1)} \underbrace{\{b_{32}(\mu,\xi_2)c_{31}(\mu,\xi_1)c_{23}(\xi_2,\xi_1) + c_{32}(\mu,\xi_2)c_{21}(\mu,\xi_1)b_{32}(\xi_2,\xi_1) \}}_{\textrm{apply Eq.(\ref{YB10})-\{1,3,2\}}} \\
&=& \frac{b_{32}(\mu,\xi_3)c_{21}(\xi_2,\xi_3)c_{31}(\mu,\xi_2)b_{31}(\mu,\xi_1)a_{1}(\xi_2,\xi_1) }{b_{32}(\xi_2,\xi_3)b_{31}(\xi_2,\xi_1)} - \frac{c_{21}(\xi_1,\xi_3)\overbrace{b_{31}(\mu,\xi_3)b_{32}(\mu,\xi_1)}^{\textrm{apply Eq.(\ref{YB2})-\{3,1,2\}}} c_{31}(\mu,\xi_2) c_{21}(\xi_2,\xi_1)}{ b_{31}(\xi_1,\xi_3)b_{32}(\xi_2,\xi_1)}\\
&=& \frac{b_{32}(\mu,\xi_3)c_{21}(\xi_2,\xi_3)c_{31}(\mu,\xi_2)b_{31}(\mu,\xi_1)a_{1}(\xi_2,\xi_1) }{b_{32}(\xi_2,\xi_3)b_{31}(\xi_2,\xi_1)} - \frac{b_{32}(\mu,\xi_3)c_{31}(\mu,\xi_2)c_{21}(\xi_1,\xi_3) b_{31}(\mu,\xi_1)c_{21}(\xi_2,\xi_1) \overbrace{b_{32}(\xi_2,\xi_3) b_{31}(\xi_2,\xi_1) }^{\textrm{apply Eq.(\ref{YB2})-\{3,1,2\}}}}{ b_{32}(\xi_2,\xi_3)b_{31}(\xi_1,\xi_3)b_{32}(\xi_2,\xi_1)b_{31}(\xi_2,\xi_1)}\\
&=& \frac{b_{32}(\mu,\xi_3)c_{21}(\xi_2,\xi_3)c_{31}(\mu,\xi_2)b_{31}(\mu,\xi_1)a_{1}(\xi_2,\xi_1) }{b_{32}(\xi_2,\xi_3)b_{31}(\xi_2,\xi_1)} - \frac{b_{32}(\mu,\xi_3)c_{31}(\mu,\xi_2)b_{31}(\mu,\xi_1)c_{21}(\xi_1,\xi_3) c_{21}(\xi_2,\xi_1) \overbrace{b_{31}(\xi_2,\xi_3) b_{21}(\xi_1,\xi_3) }^{\textrm{apply Eq.(\ref{YB3})-\{2,3,1\}}}}{ b_{32}(\xi_2,\xi_3)b_{31}(\xi_1,\xi_3)b_{32}(\xi_2,\xi_1)b_{21}(\xi_1,\xi_3)}\\
&=& \frac{b_{32}(\mu,\xi_3)c_{21}(\xi_2,\xi_3)c_{31}(\mu,\xi_2)b_{31}(\mu,\xi_1)a_{1}(\xi_2,\xi_1) }{b_{32}(\xi_2,\xi_3)b_{31}(\xi_2,\xi_1)} - \frac{b_{32}(\mu,\xi_3)c_{31}(\mu,\xi_2)b_{31}(\mu,\xi_1)\overbrace{c_{21}(\xi_1,\xi_3)}^{\textrm{apply Eq.(\ref{bigc})-\{2,1\}}} c_{21}(\xi_2,\xi_1)b_{21}(\xi_2,\xi_3) }{ b_{32}(\xi_2,\xi_3)b_{31}(\xi_1,\xi_3)b_{21}(\xi_1,\xi_3)}\\
&=& \frac{b_{32}(\mu,\xi_3)c_{31}(\mu,\xi_2)b_{31}(\mu,\xi_1)}{b_{21}(\xi_3,\xi_1)b_{32}(\xi_2,\xi_3)b_{31}(\xi_2,\xi_1)} \underbrace{\{c_{21}(\xi_2,\xi_3)a_{1}(\xi_2,\xi_1)b_{21}(\xi_3,\xi_1) + b_{21}(\xi_2,\xi_3) c_{21}(\xi_2,\xi_1)c_{12}(\xi_3,\xi_1) \}}_{\textrm{apply Eq.(\ref{YB9})-\{2,1\}}}\\
&=& \frac{b_{32}(\mu,\xi_3)c_{31}(\mu,\xi_2)\overbrace{b_{31}(\mu,\xi_1)b_{21}(\xi_2,\xi_1)}^{\textrm{apply Eq.(\ref{YB3})-\{3,2,1\}}} c_{21}(\xi_2,\xi_3)a_{1}(\xi_3,\xi_1)}{b_{21}(\xi_3,\xi_1)b_{32}(\xi_2,\xi_3)b_{31}(\xi_2,\xi_1)} \\
&=& \frac{b_{32}(\mu,\xi_3)c_{31}(\mu,\xi_2)b_{21}(\mu,\xi_1)c_{21}(\xi_2,\xi_3) a_{1}(\xi_3,\xi_1)}{ b_{21}(\xi_3,\xi_1)b_{32}(\xi_2,\xi_3)} \\
 \omega^{(C_1)}_2 &=& \frac{c_{21}(\mu,\xi_1)c_{32}(\mu,\xi_2)b_{32}(\mu,\xi_3)c_{21}(\xi_1,\xi_3)a_{2}(\xi_2,\xi_3)}{b_{32}(\xi_2,\xi_3)}-\frac{c_{31}(\mu,\xi_1) b_{32}(\mu,\xi_2)b_{32}(\mu,\xi_3)\overbrace{c_{32}(\xi_1,\xi_2)}^{\textrm{apply Eq.(\ref{bigc})-\{3,2\}}} c_{21}(\xi_1,\xi_3) a_{2}(\xi_2,\xi_3)}{ b_{32}(\xi_1,\xi_2)b_{32}(\xi_2,\xi_3)} \\
&&+\frac{b_{12}(\mu,\xi_1)c_{31}(\mu,\xi_2)b_{32}(\mu,\xi_3)b_{21}(\xi_1,\xi_3)c_{21}(\xi_2,\xi_3)}{b_{32}(\xi_2,\xi_3)} \\
&=& \frac{b_{32}(\mu,\xi_3)c_{21}(\xi_1,\xi_3)a_{2}(\xi_2,\xi_3)}{b_{32}(\xi_2,\xi_1)b_{32}(\xi_2,\xi_3)} \underbrace{\{c_{32}(\mu,\xi_2)c_{21}(\mu,\xi_1)b_{32}(\xi_2,\xi_1) +b_{32}(\mu,\xi_2)c_{31}(\mu,\xi_1)c_{23}(\xi_2,\xi_1) \}}_{\textrm{apply Eq.(\ref{YB10})-\{1,3,2\}}}\\
&&+ \frac{b_{12}(\mu,\xi_1)c_{31}(\mu,\xi_2)b_{32}(\mu,\xi_3)b_{21}(\xi_1,\xi_3)c_{21}(\xi_2,\xi_3)}{b_{32}(\xi_2,\xi_3)}\\
&=&\overbrace{\frac{b_{32}(\mu,\xi_1)}{b_{32}(\xi_2,\xi_1)}}^{\textrm{apply Eq.(\ref{YB3})-\{3,1,2\}}} \frac{b_{32}(\mu,\xi_3)c_{21}(\xi_1,\xi_3)a_{2}(\xi_2,\xi_3)c_{31}(\mu,\xi_2)c_{21}(\xi_2,\xi_1) }{b_{32}(\xi_2,\xi_3)} + \frac{b_{12}(\mu,\xi_1)c_{31}(\mu,\xi_2)b_{32}(\mu,\xi_3)b_{21}(\xi_1,\xi_3)c_{21}(\xi_2,\xi_3)}{b_{32}(\xi_2,\xi_3)}\\
&=&\overbrace{\frac{b_{12}(\mu,\xi_1)}{b_{12}(\xi_2,\xi_1)}}^{\textrm{apply Eq.(\ref{YB3})-\{3,1,2\}}} \frac{c_{31}(\mu,\xi_2)b_{32}(\mu,\xi_3)}{b_{32}(\xi_2,\xi_3)} \underbrace{\{c_{21}(\xi_2,\xi_1)a_{2}(\xi_2,\xi_3)c_{21}(\xi_1,\xi_3)+b_{12}(\xi_2,\xi_1)c_{21}(\xi_2,\xi_3)b_{21}(\xi_1,\xi_3) \}}_{\textrm{apply Y-B (13,5)}}\\
&=& \frac{c_{31}(\mu,\xi_1)b_{32}(\mu,\xi_3)b_{32}(\mu,\xi_1)a_{2}(\xi_2,\xi_)c_{21}(\xi_2,\xi_3)a_{2}(\xi_1,\xi_3)}{b_{32}(\xi_2,\xi_1)b_{32}(\xi_2,\xi_3)}
\end{array}\end{equation*}
\begin{equation*}\begin{array}{lll}
 \omega^{(C_1)}_3 &=& \frac{c_{31}(\mu,\xi_1)a_{3}(\mu,\xi_2)b_{32}(\mu,\xi_3)}{b_{32}(\xi_1,\xi_3)b_{32}(\xi_2,\xi_3)} \underbrace{\{c_{32}(\xi_1,\xi_2)c_{21}(\xi_1,\xi_3)b_{32}(\xi_2,\xi_3) +b_{32}(\xi_1,\xi_2)c_{31}(\xi_1,\xi_3)c_{23}(\xi_2,\xi_3) \}}_{\textrm{apply Eq.(\ref{YB10})-\{1,3,2\}}}\\
&&+ \frac{b_{13}(\mu,\xi_1)c_{31}(\mu,\xi_2)b_{32}(\mu,\xi_3)\overbrace{b_{32}(\xi_1,\xi_2)b_{31}(\xi_1,\xi_3)}^{\textrm{apply Eq.(\ref{YB2})-\{3,1,2\}}} c_{21}(\xi_2,\xi_3)}{ b_{32}(\xi_1,\xi_3)b_{32}(\xi_2,\xi_3)}\\
&=&\frac{b_{32}(\mu,\xi_3)c_{21}(\xi_2,\xi_3)b_{32}(\xi_1,\xi_3)}{b_{32}(\xi_1,\xi_3)b_{32}(\xi_2,\xi_3)} \underbrace{\{c_{31}(\mu,\xi_1)a_{3}(\mu,\xi_2)c_{31}(\xi_1,\xi_2) +b_{13}(\mu,\xi_1)c_{31}(\mu,\xi_2)b_{31}(\xi_1,\xi_2) \}}_{\textrm{apply Eq.(\ref{YB6})-\{3,1\}}}\\
&=& \frac{b_{32}(\mu,\xi_3)c_{21}(\xi_2,\xi_3)a_{3}(\mu,\xi_1)c_{31}(\mu,\xi_2)a_{3}(\xi_1,\xi_2)}{b_{32}(\xi_2,\xi_3)}
\end{array}\end{equation*}

\end{document}